\documentclass[twocolumn,a4paper]{svjour3} 
\smartqed  
\makeatletter
\expandafter\let\csname opt@amsmath.sty\endcsname\relax
\makeatother

\usepackage{mathtools}
\newcommand*\phantomrel[1]{\mathrel{\phantom{#1}}}

\usepackage{graphicx}

\usepackage{subfig}
\usepackage{algorithm,algpseudocode}
\usepackage{tabularx}
\usepackage{amsfonts} 
\usepackage[numbers]{natbib}
\usepackage{hyperref}
\hypersetup{
colorlinks=true,
linkcolor=blue,
citecolor=blue,
urlcolor=blue
}
\usepackage{esvect}
\usepackage{graphicx}
\usepackage{amsmath}

\usepackage{xcolor}
 \journalname{Preprint submitted to Computer Methods and Programs in Biomedicine}
\begin{document}
\raggedbottom

\title{Prediction of the Position of External Markers Using a Recurrent Neural Network Trained With Unbiased Online Recurrent Optimization for Safe Lung Cancer Radiotherapy
}

\titlerunning{Prediction of the Position of External Markers With UORO for Safe Radiotherapy}        

\author{Michel Pohl         \and
        Mitsuru Uesaka \and
        Hiroyuki Takahashi \and
        Kazuyuki Demachi \and
        Ritu Bhusal Chhatkuli
}


\institute{Michel Pohl \at
              The University of Tokyo, 113-8654 Tokyo, Japan\\
              \email{michel.pohl@centrale-marseille.fr}
           \and
           Mitsuru Uesaka \at
              Japan Atomic Energy Commission, 100-8914 Tokyo, Japan
           \and
           Hiroyuki Takahashi \and Kazuyuki Demachi \at
              The University of Tokyo, 113-8654 Tokyo, Japan
           \and
           Ritu Bhusal Chhatkuli \at
              National Institutes for Quantum and Radiological Science and Technology, 263-8555 Chiba, Japan              
}

\date{ }

\maketitle

\begin{abstract}
\setlength{\parindent}{0pt}

\emph{Background and Objective}: During lung cancer radiotherapy, the position of infrared reflective objects on the chest can be recorded to estimate the tumor location. However, radiotherapy systems have a latency inherent to robot control limitations that impedes the radiation delivery precision. Prediction with online learning of recurrent neural networks (RNN) allows for adaptation to non-stationary respiratory signals, but classical methods such as real-time recurrent learning (RTRL) and truncated backpropagation through time are respectively slow and biased. This study investigates the capabilities of unbiased online recurrent optimization (UORO) to forecast respiratory motion and enhance safety in lung radiotherapy.

\emph{Methods}: We used nine observation records of the three-dimensional (3D) position of three external markers on the chest and abdomen of healthy individuals breathing during intervals from 73s to 222s. The sampling frequency was 10Hz, and the amplitudes of the recorded trajectories range from 6mm to 40mm in the superior-inferior direction. We forecast the 3D location of each marker simultaneously with a horizon value (the time interval in advance for which the prediction is made) between 0.1s and 2.0s, using an RNN trained with UORO. We compare its performance with an RNN trained with RTRL, least mean squares (LMS), and offline linear regression. We provide closed-form expressions for quantities involved in the loss gradient calculation in UORO, thereby making its implementation efficient. Training and cross-validation were performed during the first minute of each sequence.

\emph{Results}: On average over the horizon values considered and the nine sequences, UORO achieves the lowest root-mean-square (RMS) error and maximum error among the compared algorithms. These errors are respectively equal to 1.3mm and 8.8mm, and the prediction time per time step was lower than 2.8ms (Dell Intel core i9-9900K 3.60 GHz). Linear regression has the lowest RMS error for the horizon values 0.1s and 0.2s, followed by LMS for horizon values between 0.3s and 0.5s, and UORO for horizon values greater than 0.6s.

\emph{Conclusions}: UORO can accurately predict the 3D position of external markers for intermediate to high response times with an acceptable time performance. This will help limit unwanted damage to healthy tissues caused by radiotherapy.

\end{abstract}

\section{Introduction}

\subsection{External markers in lung cancer radiotherapy} 
\label{subs:external markers}

The National Cancer Institute estimates that 236,000 new cases of lung and bronchus cancer appeared in the United States in 2021. Furthermore, it estimates that 132,000 deaths occurred in 2021, making up 21.7\% of all cancer deaths \cite{NIC2020LungBronchusCancer}.   

\begin{figure}[thb!]
\centering
\includegraphics[width=0.8\columnwidth]{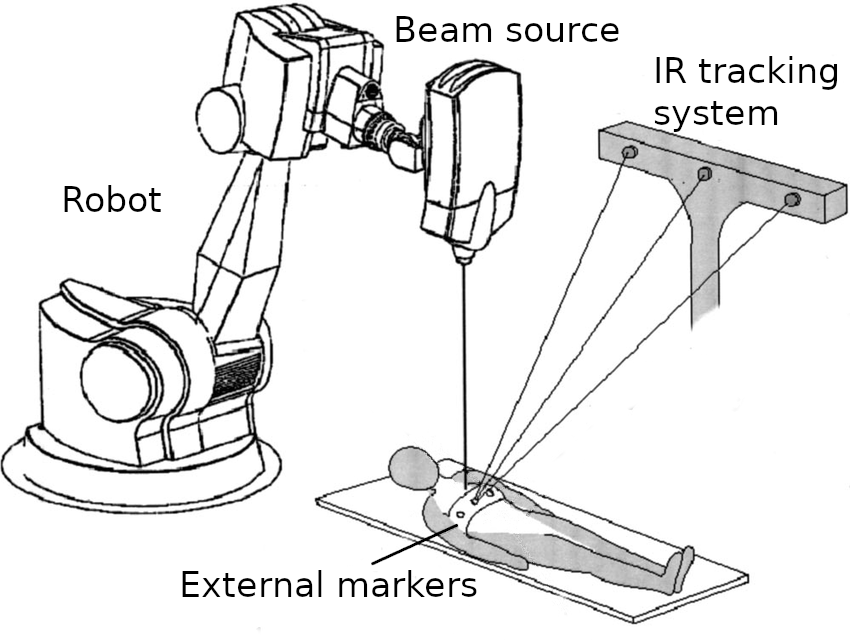}
\caption{Radiotherapy treatment system (Cyberknife) using external markers to guide the irradiation beam\protect\footnotemark.} \label{fig:treatment system}
\end{figure}

During lung cancer radiotherapy, respiratory motion makes tumor targeting difficult. Indeed, the amplitude of lung tumor motion due to breathing can exceed 5cm in the superior-inferior (SI) direction \cite{sarudis2017systematic}. Respiratory motion is largely cyclic but exhibits changes in frequency and amplitude, shifts and drifts, and varies across patients and fractions \cite{verma2010survey, ehrhardt20134d}. The term "shift" designates abrupt changes of the respiratory signal, whereas "drift" designates continuous variations of the mean tumor position. Baseline drifts of $1.65 \pm 5.95$ mm (mean position $\pm$ standard deviation) in the craniocaudal direction have been observed in \cite{takao2016intrafractional}. To overcome this problem, one can record the position of external markers placed on the chest and abdomen with infrared cameras (e.g., Cyberknife system \cite{khankan2017demystifying} in Fig. \ref{fig:treatment system}). By using an appropriate correspondence model, these positions may be correlated to the three-dimensional (3D) position and shape of the tumor for accurate irradiation \cite{ehrhardt20134d, mcclelland2013respiratory}.

\footnotetext{Adapted from \cite{schweikard2004respiration} with permission from Wiley, Copyright 2004 American Association of Physicists in Medicine.}

\subsection{Compensation of treatment system latency via prediction}
\label{section:intro pred in radiotherapy}

\begin{figure*}[htb!]
    \centering
    \subfloat[\normalsize Previous ANN models proposed for respiratory motion prediction in radiotherapy. The term "RNN" here designates a vanilla RNN, as opposed to LSTMs and gated recurrent units (GRU).\protect\footnotemark]{ {\includegraphics[width=.95\textwidth]{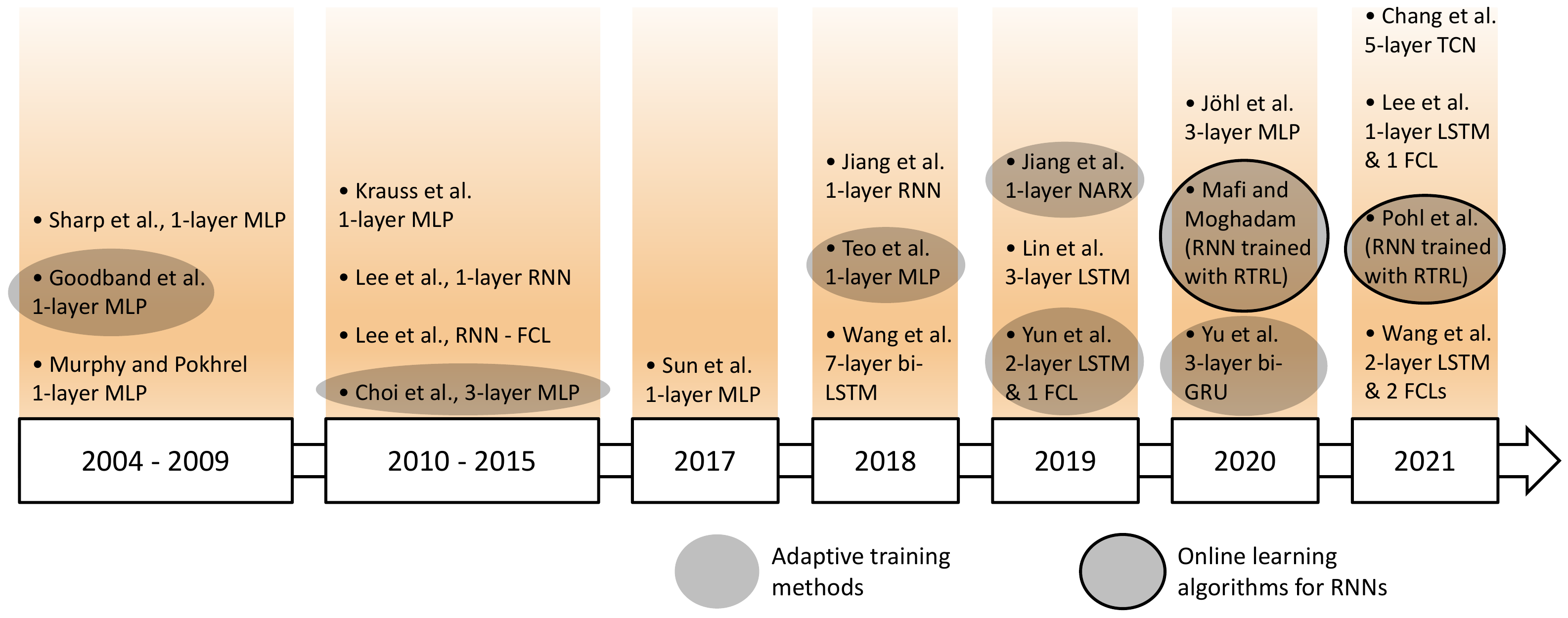} \label{fig:ANNs in radiotherapy}}}%
    \quad
    \subfloat[\normalsize Evolution and development of algorithms for online learning of RNNs. Our study is the first to assess the potential of UORO for respiratory motion compensation in radiotherapy.]{{\includegraphics[width=.95\textwidth]{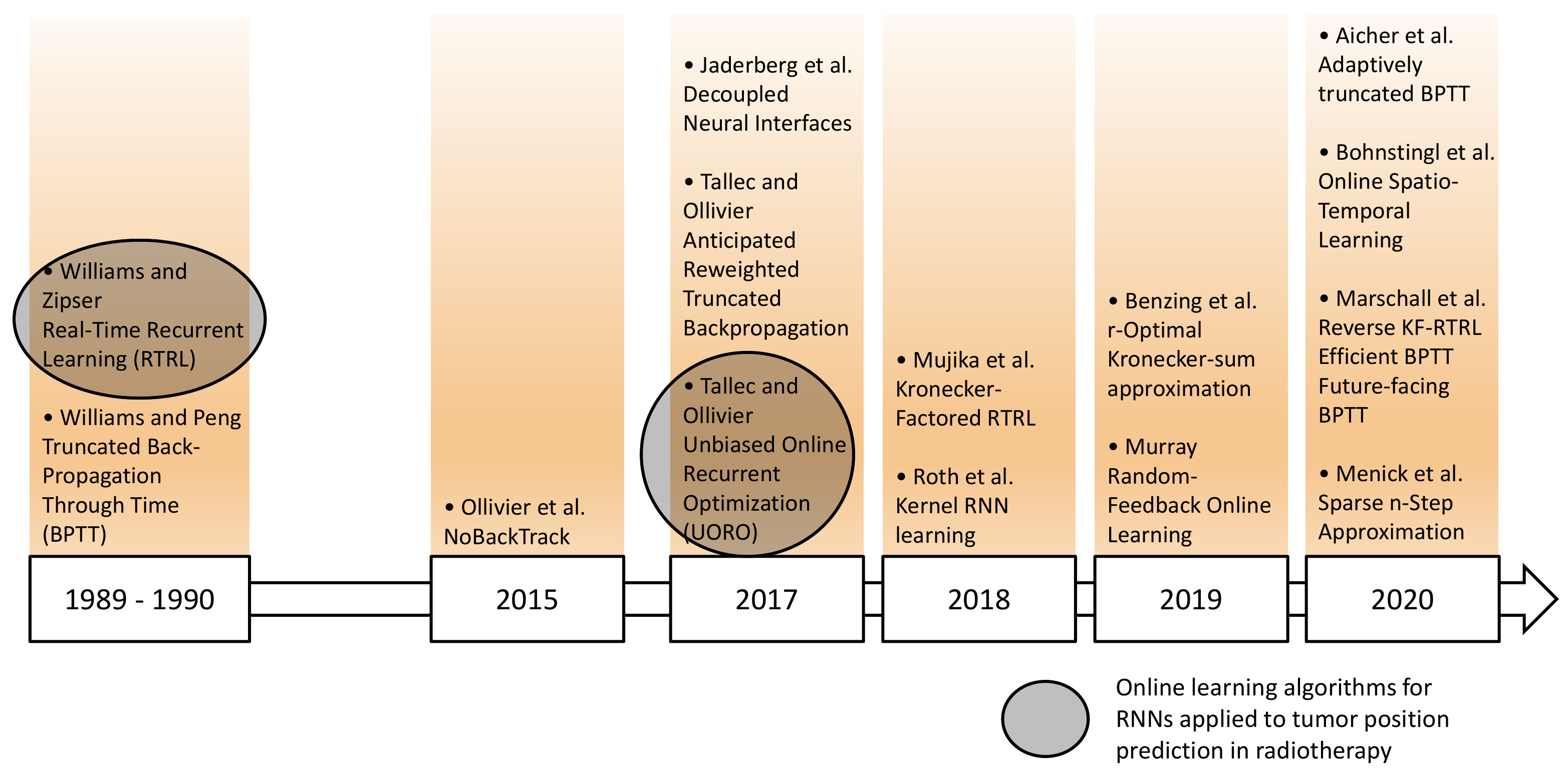} \label{fig:online learning of RNNs}}}%
    \caption{Time scope of our study, both from the radiotherapy application perspective and the algorithmic research perspective}%
\end{figure*}

\begin{figure}[htb!]
\centering
\includegraphics[width=0.6\columnwidth]{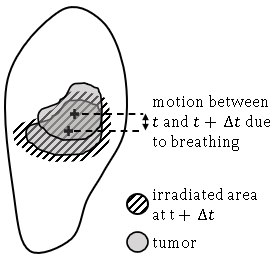}
\caption{Excessive irradiation of healthy lung tissue due to an overall system delay $ \Delta t$ not compensated. The area irradiated, represented here using diagonal stripes, is larger than the tumor size, to take into consideration effects such as the variation of the tumor shape during the treatment\protect\footnotemark .} \label{fig:irradiation delay}
\end{figure}

Radiotherapy treatment machines are subject to a time latency due to communication delays, robot control, and radiation system delivery preparation. Verma et al. reported that "for most radiation treatments, the latency will be more than 100ms, and can be up to two seconds" \cite{verma2010survey}. Delay compensation is necessary to minimize excessive damage to healthy tissues (Fig. \ref{fig:irradiation delay}). To achieve that, various prediction methods,  including Bayesian filtering based on Kalman theory \cite{remy2021potential}, relevance vector machines \cite{fan2020respiratory}, and online sequential forecasting random convolution nodes \cite{wang2020fast}, have been proposed. Reviews and comparisons of the classical methods can be found in \cite{verma2010survey, lee2014prediction, johl2020performance, ehrhardt20134d}. Among the approaches studied, artificial neural networks (ANNs) form a class of algorithms that perform well at forecasting tasks. Different ANN architectures and training algorithms have been investigated extensively in the context of respiratory motion prediction, mostly for one-dimensional (1D) traces (Fig. \ref{fig:ANNs in radiotherapy}) \cite{sharp2004prediction, goodband2008comparison, murphy2009optimization, krauss2011comparative, lee2011respiratory, lee2013customized, choi2014performance, sun2017respiratory, kai2018prediction, teo2018feasibility, wang2018feasibility, jiang2019prediction, lin2019towards, yun2019deep, johl2020performance, mafi2020real, yu2020rapid, chang2021real, lee2021geometric, POHL2021101941, wang2021real}. ANNs are efficient for performing prediction with a high response time, which is the time interval in advance for which the prediction is made, also called the look-ahead time or horizon, and for non-stationary and complex signals. However, they are heavily computer resource intensive and have high processing times \cite{verma2010survey}. Furthermore, deep ANNs need large amounts of data for training which can be practically difficult because of patient data regulations, and the prediction results are strongly dependent on the database chosen. Tumor motion is essentially three-dimensional, but most previous works about ANNs applied to respiratory motion management in radiotherapy focused on univariate time-series forecasting.

\footnotetext{MLP, FCL, NARX, and TCN respectively stand for "multilayer perceptron", "fully connected layer", "nonlinear autoregressive exogenous model", and "temporal convolutional network". By abuse of language, the number of layers mentioned actually refers to the number of hidden layers. For instance, a "1-layer MLP" architecture refers to an MLP with 1 hidden layer. \label{footnote abbreviations}}

Recurrent neural networks (RNNs) are characterized by a feedback loop that acts as a memory and enables the retention of information over time. They are able to efficiently learn features and long-term dependencies from sequential and time-series data \cite{salehinejad2017recent}. As a result, almost all of the recent research about ANNs applied to time series forecasting for motion management in radiotherapy focuses on RNNs \cite{kai2018prediction, wang2018feasibility, yun2019deep, lin2019towards, mafi2020real, yu2020rapid, lee2021geometric, wang2021real, POHL2021101941}. It has experimentally been observed that forecasting respiratory signals with an RNN could improve radiation delivery accuracy \cite{lee2021geometric}. RNN models such as long short term memory (LSTM) networks have also been used in related medical data processing problems such as cardiorespiratory motion prediction from X-ray angiography sequences \cite{azizmohammadi2019model} and next-frame prediction in medical image sequences, including chest dynamic imaging \cite{nabavi2020respiratory, romaguera2020prediction}.

Our research investigates the feasibility of predicting breathing motion with online training algorithms for RNNs. In contrast to offline methods, online methods update the synaptic weights with each new training example. That enables the neural network to adapt to the continuously changing breathing patterns of the patient (section \ref{subs:external markers}), therefore providing robustness to complex motion. Because online learning enables adaptation to examples unseen in the training set, it can be viewed as a way to compensate for the difficulty of acquiring and using large training databases for medical applications. Adaptive or dynamic learning has been applied many times to radiotherapy, and several studies demonstrated the benefit of that approach in comparison with static models \cite{krauss2011comparative, teo2018feasibility, mafi2020real}. Real-time recurrent learning (RTRL) \cite{williams1989learning} is one of these dynamic approaches that has already been used for predicting tumor motion from the Cyberknife Synchrony system \cite{mafi2020real} and the SyncTraX system \cite{jiang2019prediction}, as well as the position of internal points in the chest \cite{POHL2021101941}.

Many techniques for online training of RNNs have recently emerged \cite{ollivier2015training, tallec2017unbiasing, jaderberg2017decoupled, mujika2018approximating, roth2018kernel, benzing2019optimal, murray2019local, aicher2020adaptively, menick2020practical, marschall2020unified, bohnstingl2020online}, such as unbiased online recurrent optimization (UORO) \cite{tallec2017unbiased} (Fig. \ref{fig:online learning of RNNs}). Most of these seek to approximate RTRL, which suffers from a large computational complexity. They also aim to provide an unbiased estimation of the loss gradient that truncated backpropagation through time (truncated BPTT) \cite{jaeger2002tutorial} cannot compute, guaranteeing an appropriate balance between short-term and long-term temporal dependencies. The theoretical convergence of RTRL and UORO, which could not be proved by standard stochastic gradient descent theory, has recently been established \cite{masse2020convergence}.


\footnotetext{Reprinted from \cite{POHL2021101941}, Copyright 2021, with permission from Elsevier.}

\subsection{Content of this study}

This is the first study evaluating the capabilities of RNNs trained online with UORO to predict the position of external markers on the chest and abdomen for safety in radiotherapy. In contrast to most of the studies mentioned in Section \ref{section:intro pred in radiotherapy} focusing on the prediction of 1D signals, we tackle the problem of multivariate forecasting of the 3D coordinates of the markers, as this will help estimate the tumor location more precisely during the treatment. The proposed RNN framework does not perform prediction for each marker separately but instead learns patterns about the correlation between their motion to potentially increase the forecasting accuracy. We provide closed-form expressions for some quantities involved in UORO in the specific case of vanilla RNNs, as the original article \citep{tallec2017unbiased} only describes UORO for a general RNN model. We compare UORO with different forecasting algorithms, namely RTRL, least mean squares (LMS), and linear regression, for different look-ahead values $h$, ranging from $h_{min} = 0.1s$ to $h_{max} = 2.0s$, by observing different prediction metrics as $h$ varies. We divide the subjects' data into two groups: regular and irregular breathing, to quantify the robustness of each prediction algorithm. We analyze the influence of the hyper-parameters on the prediction accuracy of UORO as the horizon value changes and discuss the selection of the best hyper-parameters.

\section{Material and Methods}

\subsection{Marker position data}

In this study, we use 9 records of the 3D position of 3 external markers on the chest and abdomen of individuals lying on a treatment couch (HexaPOD), acquired by an infrared camera (NDI Polaris). The duration of each sequence is between 73s and 320s and the sampling rate is 10Hz. The superior-inferior, left-right, and antero-posterior trajectories respectively range between 6mm and 40mm, between 2mm and 10mm, and between 18 mm and 45mm. In five of the sequences, the breathing motion is normal and in the four remaining sequences, the individuals were asked to perform actions such as talking or laughing. Additional details concerning the dataset can be found in \cite{krilavicius2016predicting}.

\subsection{The RTRL and UORO algorithms for training RNNs}

\begin{figure} [htb!]
	\centering
		\includegraphics[width=\columnwidth]{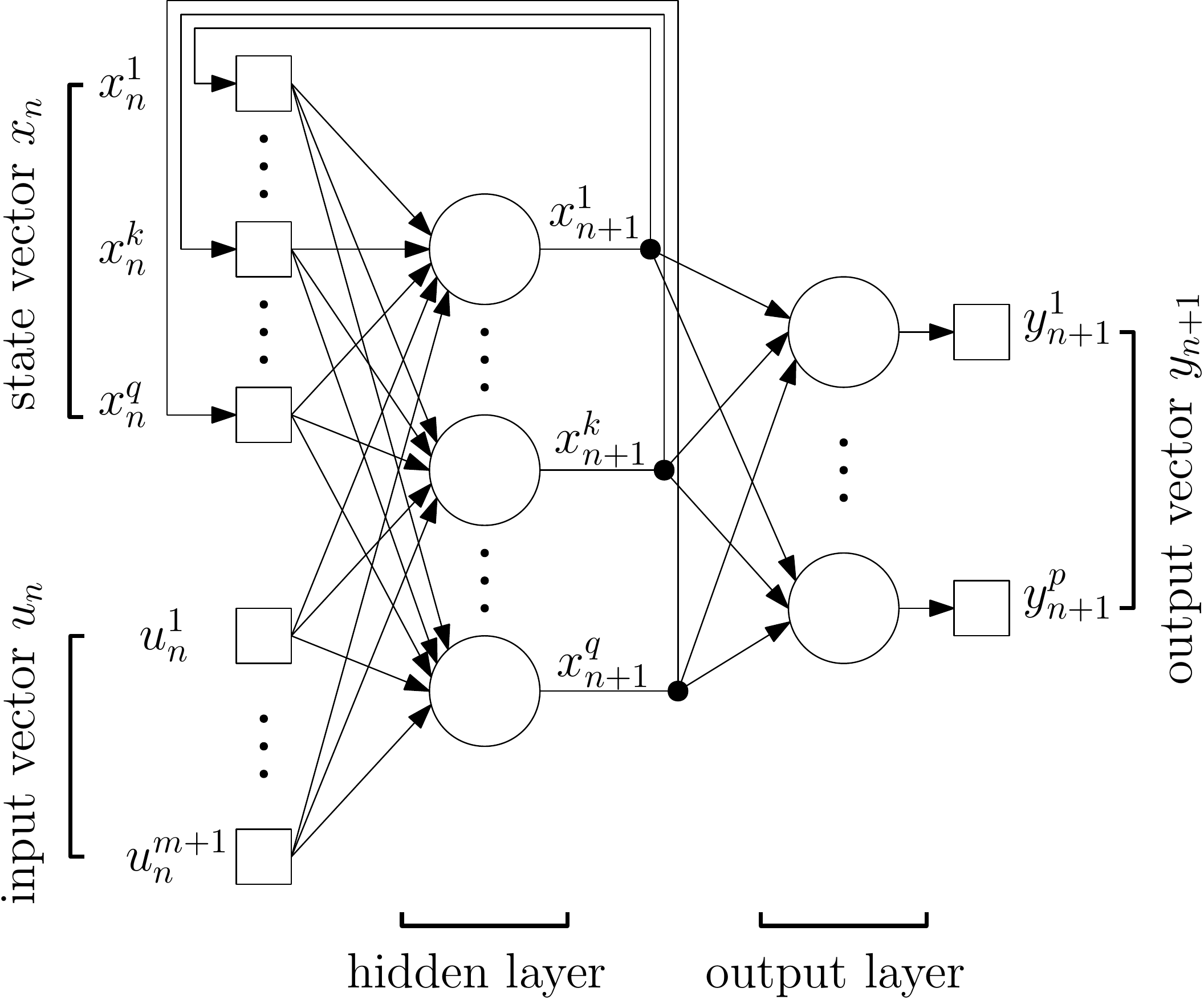}
	\caption{Structure of the RNN predicting the positions of the markers. The input vector \text{\normalsize $u_n$}  corresponds to the positions in the past and the output vector \text{\normalsize $y_{n+1}$} corresponds to the predicted positions\protect\footnotemark  .}
	\label{fig:rnn_structure}
\end{figure}

In this study, we train an RNN with one hidden layer to predict the position of 3 markers in the future. RNNs with one hidden layer are characterized by the state equation, which describes the dynamics of the internal states, and the measurement equation, which describes how the RNN output is influenced by the hidden states (Eq. \ref{eq:RNN_general_eqs}). In the following, we denote by $u_n \in \mathbb{R}^{m+1}$, $x_n \in \mathbb{R}^q$, $y_{n+1} \in \mathbb{R}^p$, and $\theta_n$ the input, state, output, and synaptic weight vectors at time $t_n$. Fig. \ref{fig:rnn_structure} gives a graphical representation of these two equations.

\begin{equation} \label{eq:RNN_general_eqs}
x_{n+1} = F_{st}(x_n, u_n, \theta_n)
\qquad
y_{n+1} = F_{out}(x_n, u_n, \theta_n)
\end{equation}

\footnotetext{Reprinted from \cite{POHL2021101941}, Copyright 2021, with permission from Elsevier.}

The instantaneous square loss $L_n$ of the network can be calculated from the instantaneous error $e_n$ between the vector $y_n^{*}$ containing the ground-truth positions and the output $y_n$ containing the predicted positions (Eq. \ref{eq:loss function}).    

\begin{equation} \label{eq:loss function}
 e_{n} = y_n^* - y_n
\qquad
 L_{n} = \frac{1}{2} \|e_n\|_2^2
\end{equation}

\begin{figure}[thb!]
\centering
\includegraphics[width=0.80\columnwidth]{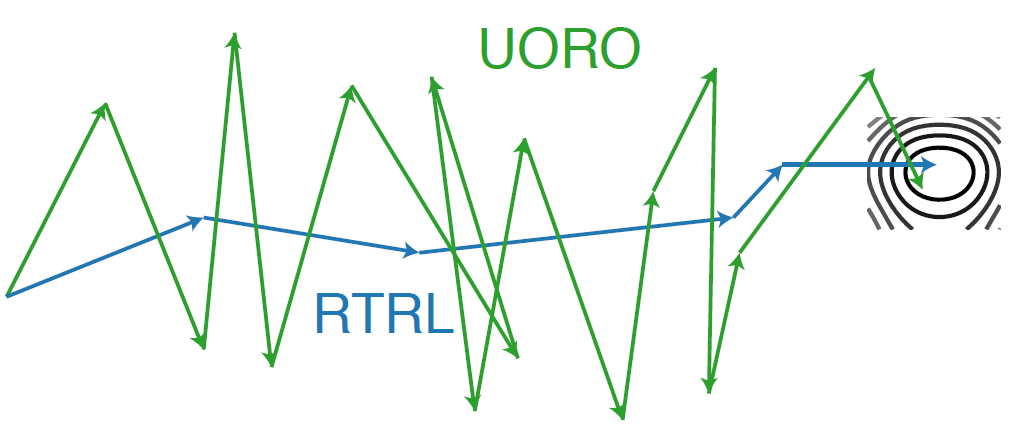}
\caption[Weak alignment of the gradients of UORO with those of RTRL from the loss optimization viewpoint]{Weak alignment of the gradients of UORO with those of RTRL from the loss optimization viewpoint.\protect \footnotemark} \label{fig:UORO vs RTRL optimization point of view}
\end{figure}

By using the chain rule, one can derive Eqs. \ref{eq:parameters_gradient} and \ref{eq:influence update}, which describe how changes of the parameter vector $\theta_n$ affect the instantaneous loss $L_{n+1}$ and state vector $x_{n+1}$. Computation of the gradient of $L_{n+1}$ with respect to the parameter vector using Eq. \ref{eq:parameters_gradient}, followed by recursive computation of the influence matrix $\partial{x_n}/ \partial \theta$ using Eq. \ref{eq:influence update} constitutes the RTRL algorithm. RTRL is computationally expensive, and UORO attempts to solve that problem by approximating the influence matrix with an unbiased rank-one estimator. In UORO, two random column vectors $\tilde{x}_n$ and $\tilde{\theta}_n$ are recursively updated at each time step so that $\mathbb{E}(\tilde{x}_n \tilde{\theta}_n^T) = \partial{x_n}/ \partial \theta$. It was reported that "UORO's noisy estimates of the true gradient are almost orthogonal with RTRL at each time point, but the errors average out over time and allow UORO \mbox{to find} the same solution" \cite{marschall2020unified} (Fig. \ref{fig:UORO vs RTRL optimization point of view}).

\begin{multline} \label{eq:parameters_gradient}
\frac{\partial{L_{n+1}}}{\partial \theta} = 
\frac{\partial{L_{n+1}}}{\partial y}(y_{n+1}) 
\left[ \frac{\partial{F_{out}}}{\partial x}(x_n, u_n, \theta_n)
\frac{\partial{x_n}}{\partial \theta} \right. \\
\left. + \frac{\partial{F_{out}}}{\partial \theta}(x_n, u_n, \theta_n) \right]
\end{multline}

\begin{equation} \label{eq:influence update}
\frac{\partial{x_{n+1}}}{\partial \theta} = 
\frac{\partial{F_{st}}}{\partial x}(x_n, u_n, \theta_n) 
\frac{\partial{x_n}}{\partial \theta} + 
\frac{\partial{F_{st}}}{\partial \theta}(x_n, u_n, \theta_n)
\end{equation}

\footnotetext{Reprinted from \citep{marschall2020unified}, Copyright JMLR 2020.}

\begin{figure*}[hbt!]
\centering
\begin{minipage}{.8\textwidth}
\begin{algorithm}[H]
\small
\caption{UORO}
\label{alg:RNN-UORO}
\begin{algorithmic}[1]
\State \textbf{Parameters} :
\State $L \in \mathbb{N}^* $ : signal history length, $n_M = 3 $ : number of external markers considered
\State $m = 3 n_M L$, $q \in \mathbb{N}^*$ and $p = 3 n_M$ : dimension of the RNN input, RNN state and RNN output
\State $\eta \in \mathbb{R}_{>0} $ and $\tau \in \mathbb{R}_{>0 }$ : learning rate and gradient threshold
\State $\sigma_{init} \in \mathbb{R}_{>0} $ : standard deviation of the Gaussian distribution of the initial weights
\State $\epsilon_{norm} = 1.10^{-7}, \epsilon_{prop} = 1.10^{-7} $
\State
\State \textbf{Initialization}
\State $W_{a,n=1}$, $W_{b,n=1}$, $W_{c,n=1}$ : synaptic weight matrices of respective sizes $q \times q$, $q \times (m+1)$ and $p \times q$, initialized randomly according to a Gaussian distribution with std. deviation $\sigma_{init}$. 
\State \textit{Notation :} $|W_{a}| = q^2$, $|W_{b}| = q(m+1)$, $|W_{c}| = p q$, and $|W| = q(p+q+m+1)$
\State $x_{n=1} := 0_{q \times 1}$ : state vector, $\tilde{x}_{n=1} := 0_{q \times 1}, \tilde{\theta}_{n=1} := 0_{1 \times |W|}$ : vectors such that $\partial{x_n}/\partial{\theta} \approx \tilde{x}_n \tilde{\theta}_n$
\State $\delta\theta := 0_{1 \times |W|}, \delta\theta_g = 0_{1 \times |W|}$ : vectors defined by 
$\delta\theta = \dfrac{\partial L_{n+1}}{\partial y}\dfrac{\partial F_{out}}{\partial \theta}$ and 
$\delta\theta_g = \nu^T \dfrac{\partial F_{st}}{\partial \theta}$
\State
\State \textbf{Learning and prediction}
\For{$n = 1,2,...$}
\State $z_n := W_{a,n}x_n + W_{b,n}u_n$, $x_{n+1} := \Phi(z_n)$ (hidden state update)
\State $y_{n+1} := W_{c,n}x_{n+1}$ (prediction), $e_{n+1} := y^*_{n+1} - y_{n+1}$ (error vector update)
\State $[\delta\theta_{1+|W_{a}|+|W_{b}|}, ..., \delta\theta_{|W|}] := -[{(e_{n+1} x_{n+1}^T)}_{1,1}, ..., (e_{n+1} x_{n+1}^T)_{p,q}]$
\State $\nabla_x L_{n+1} := - e_{n+1}^T W_{c,n}$, $\Delta\theta := \nabla_x L_{n+1} \tilde{x}_n \tilde{\theta}_n + \delta{\theta}$ (gradient estimate)
\State $\nu$ : column vector of size $q$ with random values in $\{-1, 1\}$
\State $\tilde{x}_{n+1} := \dfrac{\Phi[W_{a,n}(x_n+\epsilon_{prop}\tilde{x}_n) + W_{b,n}u_n]-x_{n+1}}{\epsilon_{prop}}$ (tangent forward propagation)
\State $\delta\theta_g^{aux} := \nu * \Phi'(z_n)$ (element-wise product) 
\State $[(\delta\theta_g)_1, ..., (\delta\theta_g)_{|W_{a}|}] := [{(\delta\theta_g^{aux} x_n^T)}_{1,1}, ..., {(\delta\theta_g^{aux} x_n^T)}_{q,q}]$
\State $[(\delta\theta_g)_{|W_{a}|+1}, ..., (\delta\theta_g)_{|W_{a}|+|W_{b}|}] := [{(\delta\theta_g^{aux} u_n^T)}_{1,1}, ..., {(\delta\theta_g^{aux} u_n^T)}_{q,m+1}]$
\State 
\State $\rho_0 := \sqrt{\dfrac{\|\tilde{\theta}\|_2}{\|\tilde{x}\|_2+\epsilon_{norm}}}+\epsilon_{norm}$, 
\quad $\rho_1 := \sqrt{\dfrac{\|\tilde{\theta_g}\|_2}{\|\nu\|_2+\epsilon_{norm}}}+\epsilon_{norm}
$
\State $ \tilde{x}_{n+1} := \rho_0 \tilde{x}_{n+1} + \rho_1 \nu$ 
\quad $ \tilde{\theta}_{n+1}:= \tilde{\theta}_n/\rho_0 + (\delta\theta_g)/\rho_1$
\State $\theta_n := [(W_{a,n})_{1,1}, ..., (W_{a,n})_{q,q}, (W_{b,n})_{1,1}, ..., (W_{b,n})_{q,m+1}, (W_{c,n})_{1,1}, ..., (W_{c,n})_{p,q}]$
\If{$ \|\Delta\theta\|_2 > \tau $} 
\State $\Delta\theta := \dfrac{\tau}{\|\Delta\theta\|_2} \Delta\theta $ (gradient clipping)
\EndIf 
\State $\theta_{n+1} := \theta_n - \eta \Delta\theta$ (weights update)
\State 
$ W_{a,n+1} := \begin{bmatrix} 
(\theta_{n+1})_1 &...& (\theta_{n+1})_{q(q-1)+1}\\
... &...& ...\\
(\theta_{n+1})_q &...& (\theta_{n+1})_{|W_a|}
\end{bmatrix} $ 
$ W_{b,n+1} := \begin{bmatrix} 
(\theta_{n+1})_{|W_a|+1} &...& (\theta_{n+1})_{|W_a|+qm+1}\\
... &...& ...\\
(\theta_{n+1})_{|W_a|+q} &...& (\theta_{n+1})_{|W_a|+|W_b|}
\end{bmatrix} $ 
\State $ W_{c,n+1} := \begin{bmatrix} 
(\theta_{n+1})_{|W_a|+|W_b|+1} & ... & (\theta_{n+1})_{|W_a|+|W_b|+p(q-1)+1}\\
... & ... & ...\\
(\theta_{n+1})_{|W_a|+|W_b|+p} & ... & (\theta_{n+1})_{|W_a|+|W_b|+|W_c|}
\end{bmatrix} $
\EndFor
\State
\State \textit{Convention : for} $A \in \mathbb{R}^{M} \times \mathbb{R}^{N}$ \textit{we note} $[A_{1,1}, ..., A_{M,N}] =
[A_{1,1}, ..., A_{M,1}, A_{1,2}, ..., A_{M,N}]$

\end{algorithmic}
\end{algorithm}

\end{minipage}

\end{figure*}

\subsection{Online prediction of the position of the markers with a vanilla RNN}

If we denote by $\vec{u}_j(t_k) = [u_j^x(t_k), u_j^y(t_k), u_j^z(t_k)]$ the normalized 3D displacement of marker $j$ at time $t_k$, the input $u_n$ of the RNN consists of the concatenation of the vectors $\vec{u}_j(t_n)$, ..., $\vec{u}_j(t_{n+L-1})$ for each marker $j$, where $L$ designates the signal history length (SHL), expressed here in number of time steps. The prediction of the displacement of the 3 markers is performed simultaneously to use information about the correlation between each of them. The output vector $y_{n+1}$ consists of the position of these 3 points at time $t_{n+L+h-1}$, where $h$ refers to the horizon value, expressed also in number of time steps (Eq. \ref{eq:RNN_in_out_def}).

\begin{equation} \label{eq:RNN_in_out_def}
u_n
=
\begin{pmatrix}
1 \\
u_1^x(t_n)\\
u_1^y(t_n)\\
u_1^z(t_n)\\
...\\
u_3^z(t_n)\\
u_1^x(t_{n+1})\\
...\\
u_3^z(t_{n+L-1})\\
\end{pmatrix}
\quad
y_{n+1}
=
\begin{pmatrix}
u_1^x(t_{n+L+h-1})\\
u_1^y(t_{n+L+h-1})\\
u_1^z(t_{n+L+h-1})\\
...\\
u_3^z(t_{n+L+h-1})\\
\end{pmatrix}
\end{equation}

\begin{figure} [thb!]
	\centering
		\includegraphics[width=0.8\columnwidth]{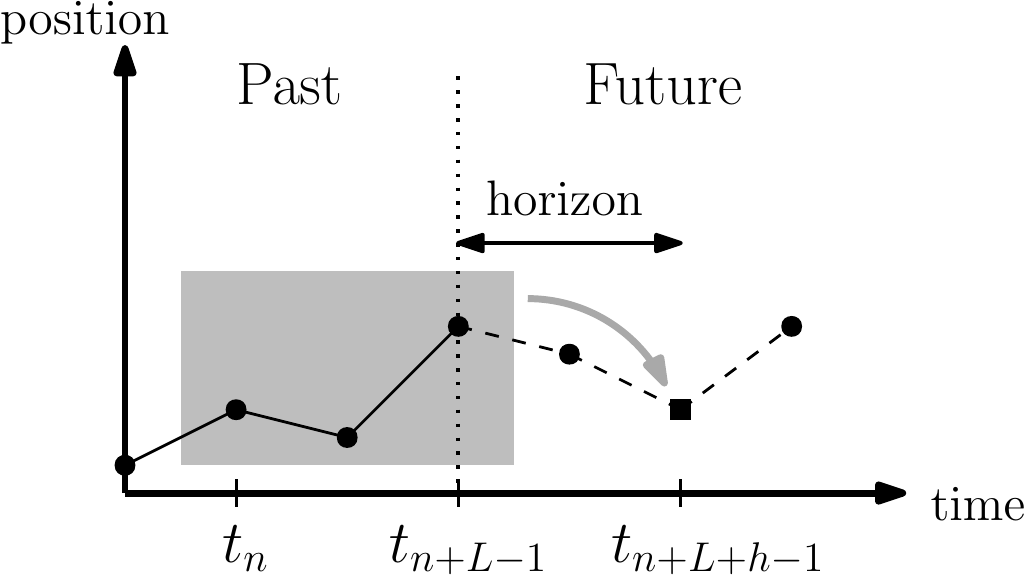}
	\caption{Forecasting a 1D position signal. The signal history length $L$ is the time interval in the past, the information of which is used for performing one prediction. The horizon $h$, also called response time or look-ahead time, is the time interval in advance for which the prediction is made (cf Eq. \ref{eq:RNN_in_out_def}).}
\end{figure}

In this work, we use a vanilla RNN structure, described by Eqs. \ref{eq:state_vanilla} and \ref{eq:measurement_vanilla}, where the parameter vector $\theta_n$ consists of the elements of the matrices $W_{a,n}$, $W_{b,n}$, and $W_{c,n}$ of respective size $q \times q$, $q \times (m+1)$, and $p \times q$. $\Phi$ is the nonlinear activation function and we use here the hyperbolic tangent function (Eq. \ref{eq:non_linearity}). 

\begin{equation} \label{eq:state_vanilla}
F_{st}(x_n, u_n, \theta_n) = \Phi(W_{a,n} x_n + W_{b,n} u_n)
\end{equation}

\begin{equation} \label{eq:measurement_vanilla}
F_{out}(x_n, u_n, \theta_n) = W_{c,n} F_{st}(x_n, u_n, \theta_n)
\end{equation}

\begin{equation}\label{eq:non_linearity}
\Phi \left(
\begin{matrix}
z_1\\
...\\
z_q
\end{matrix}
\right) 
=
\begin{pmatrix}
tanh (z_1)\\
...\\
tanh (z_q)
\end{pmatrix}
\end{equation}

RNNs updated by the gradient descent rule may be unstable. Therefore, we prevent large weight updates by clipping the gradient norm to avoid numerical instability \cite{pascanu2013difficulty}. The optimizer is stochastic gradient descent. The implementation of UORO in the case of the vanilla model described above is detailed in Algorithm \ref{alg:RNN-UORO}. The quantities $\nabla_x L_{n+1}$, $\delta \theta$, and $\delta \theta_g$ are calculated in Appendix \ref{appendix:UORO_alg}. The RNN characteristics are summarized in Table \ref{table:RNNs_configuration}.

\begin{table}[thb!]
\normalsize
\setlength{\tabcolsep}{5pt}
\begin{tabular}{ll}
\hline
RNN characteristic &  \\
\hline
Output layer size       & $p = 3 n_M $ \\
Input layer size        & $m = 3 n_M L$\\
Number of hidden layers & 1 \\
Size of the hidden layer & $q$ \\
Activation function $\phi$ &  Hyperbolic tangent \\
Training algorithm &  RTRL or UORO\\
Optimization method & Stochastic grad. descent \\ 
                    & with gradient clipping \\
Weights initialization & Gaussian \\
Input data normalization & Yes (online)\\
Cross-validation metric & RMSE (Eq. \ref{eq:RMSE_def}) \\
Nb. of runs for cross-val. & $n_{cv}=50$\\
Nb. of runs for evaluation & $n_{test}=300$ \\
\hline
\end{tabular}
\caption{Configuration of the RNN forecasting the motion of the external markers. $n_M$ refers to the number of external markers and $L$ to the SHL.}
\label{table:RNNs_configuration}
\end{table}

\subsection{Experimental design}

We compare RNNs trained with UORO with other prediction methods: RNNs trained with RTRL, LMS, and multivariate linear regression (Table \ref{table:models comparison}). We clipped the gradient estimate of the instantaneous loss (Eq. \ref{eq:loss function}) with respect to the parameter vector $\vec{\nabla}_{\theta} L_n $ for UORO, RTRL, and LMS when $ \| \vec{\nabla}_{\theta} L_n \|_2 > \tau$ with the same threshold value $\tau = 2.0$ for these three algorithms.

\begin{table*}[htb!]
\normalsize
\begin{center}
\begin{tabular}{llll}
\hline
Prediction    &  Mathematical                               & Development set        & Range of hyper-parameters \\
method        &  model                                      & partition              & for cross-validation     \\
\hline
\hline
RNN           & $x_{n+1} = \Phi(W_{a,n} x_n + W_{b,n} u_n)$ & Training 30s           & $\eta \in \{ 0.05, 0.1, 0.2\}$ \\
UORO          & $y_{n} = W_{c,n} x_n$                       & Cross-validation 30s   & $\sigma_{init} \in \{ 0.02, 0.05\}$ \\
              &                                             &                        & $L \in \{ 10, 30, 50, 70, 90\}$\\
              &                                             &                        & $q \in \{ 10, 30, 50, 70, 90 \}$ \\    
\hline
RNN           & $x_{n+1} = \Phi(W_{a,n} x_n + W_{b,n} u_n)$ & Training 30s           & $\eta \in \{ 0.02, 0.05, 0.1, 0.2\}$ \\
RTRL          & $y_{n} = W_{c,n} x_n$                       & Cross-validation 30s   & $\sigma_{init} \in \{ 0.01, 0.02, 0.05\}$ \\
              &                                             &                        & $L \in \{ 10, 25, 40, 55 \}$ \\
              &                                             &                        & $q \in \{ 10, 25, 40, 55 \}$ \\  
\hline
LMS           & $y_{n+1} = W_n u_n$                         & Training 30s           & $\eta \in \{ 0.002, 0.005, 0.01, $ \\
              &                                             & Cross-validation 30s   & \multicolumn{1}{r}{$ 0.02, 0.05, 0.1, 0.2\}$}\\
              &                                             &                        & $L \in \{ 10, 30, 50, 70, 90\}$\\
\hline
Linear        & $y_{n+1} = W u_n$                           & Training 54s           & $L \in \{ 10, 20, 30, 40, 50, $ \\
regression    &                                             & Cross-validation 6s    & \multicolumn{1}{r}{$ 60, 70, 80, 90 \}$} \\
\hline
\end{tabular}
\end{center}
\caption{Overview of the different forecasting methods compared in this study. The input vector \text{\normalsize $u_n$}, corresponding to the positions in the past, and the output vector \text{\normalsize $y_{n+1}$}, corresponding to the predicted positions, which appear in the second column, are defined in Eq. \ref{eq:RNN_in_out_def}. The fourth column describes the hyper-parameter range for cross-validation with grid search.  $\eta$ refers to the learning rate, $\sigma_{init}$ to the standard deviation of the initial Gaussian distribution of the synaptic weights, $L$ to the SHL (expressed in number of time steps), and $q$ to the number of hidden units. $W_n$ and $W$ are matrices used respectively in LMS and linear regression, and their size is $p \times (m+1)$.}
\label{table:models comparison}
\end{table*}

Learning is performed using only information from the sequence that is used for testing. Each time series is split into a training and development set of 1 min and the remaining test set. The training set comprises the data between 0s and 30s except in the case of linear regression as using more data is beneficial to offline methods. The hyper-parameters that minimize the root-mean-square error (Eq. \ref{eq:RMSE_def}) of the cross-validation set during the grid search process are selected for evaluation. The term $\delta_j(t_k)$ in Eq. \ref{eq:RMSE_def} designates the instantaneous prediction error at time $t_k$ due to marker $j$, defined in Eq. \ref{eq:instant_error}.

\begin{equation} \label{eq:instant_error}
\delta_j(t_k) = \| \vec{u}_j^{pred}(t_k) - \vec{u}_j^{true}(t_k) \|_2
\end{equation}

\begin{equation} \label{eq:RMSE_def}
 RMSE = \sqrt{\frac{1}{3(k_{max}-k_{min}+1)} \sum_{k = k_{min}}^{k_{max}} \sum_{j = 1}^{3}
 	\delta_j(t_k)^2
 	}
\end{equation}

To analyze the forecasting performance of each algorithm, we compute the RMSE, but also the normalized RMSE (Eq. \ref{eq:nRMSE_def}), mean average error (Eq. \ref{eq:MAE_def}), and maximum error (Eq. \ref{eq:max_error_def}) of the test set.  In Eq. \ref{eq:nRMSE_def}, $\vec{\mu}_j^{true}$ designates the mean 3D position of marker $j$ in the test set. Because the weights of the RNNs are initialized randomly, given each set of hyper-parameters, we average the RMSE of the cross-validation set over $n_{cv} = 50$ successive runs. Then, during performance evaluation, each metric is averaged over $n_{test} = 300$ runs.

\begin{equation} \label{eq:nRMSE_def}
 nRMSE = \frac{\sqrt{\sum_{k = k_{min}}^{k_{max}} \sum_{j =1}^3
 						\delta_j(t_k)^2}}
 				 {\sqrt{\sum_{k = k_{min}}^{k_{max}} \sum_{j =1}^3
 				 		\| \vec{\mu}_j^{true} - \vec{u}_j^{true}(t_k) \|_2^2}}
\end{equation}

\begin{equation} \label{eq:MAE_def}
 MAE = \frac{1}{3(k_{max}-k_{min}+1)} \sum_{k = k_{min}}^{k_{max}} \sum_{j =1}^3
	\delta_j(t_k) 
\end{equation} 

\begin{equation}\label{eq:max_error_def}
 e_{max} = \underset{k = k_{min}, ..., k_{max}}{max} \underset{j =1, 2, 3}{max}
 	\delta_j(t_k) 
\end{equation}

Furthermore, we examine the jitter of the test set, which measures the oscillation amplitude of the predicted signal (Eq. \ref{eq:jitter_def}). High fluctuations of the prediction signal result in difficulties concerning robot control during the treatment. The jitter $J$ is minimized when the prediction is constant, thus there is a trade-off between accuracy and jitter \cite{krilavicius2016predicting}.

\begin{equation} \label{eq:jitter_def}
\small
 J = \frac{1}{3(k_{max}-k_{min})} \sum_{k = k_{min}}^{k_{max}-1} \sum_{j =1}^3
	\| \vec{u}_j^{pred}(t_{k+1}) - \vec{u}_j^{pred}(t_k) \|_2 
\end{equation} 

We assume that given an RNN training method, each associated error measure $e_{i,h}$ (the MAE, RMSE, nRMSE, maximum error, or jitter) corresponding to sequence $i$ and horizon $h$ follows a Gaussian distribution $\mathcal{N}(\mu_{i,h},\,\sigma_{i,h}^{2})$. Indeed, each realization of the random variable $e_{i,h}$ depends on the run index $r \in [\![ 1, ..., n_{test} ]\!] $, and we denote that value $e_{i,h}^{(r)}$. This enables calculating the 95\% confidence interval $[ \overline{\mu}_{i,h} - \Delta \mu_{i,h}, \overline{\mu}_{i,h} + \Delta \mu_{i,h}]$ for $\mu_{i,h}$, where $\Delta \mu_{i,h}$ is defined in Eq. \ref{eq:conf interval mu_i,h}. \footnotemark

\footnotetext{We write $\overline{\mu}_{i,h}$ instead of $\mu_{i,h}$ and $\overline{\sigma}_{i,h}$ instead of $\sigma_{i,h}$ to designate estimators of these parameters given the $n_{test}$ runs.}

\begin{equation} \label{eq:std dev error}
\overline{\sigma}_{i,h}^2 = \frac{1}{n_{test} - 1} \sum_{r = 1}^{n_{test}} \left( e_{i,h}^{(r)} - \overline{\mu}_{i,h} \right)^2
\end{equation}

\begin{equation} \label{eq:conf interval mu_i,h}
\Delta \mu_{i,h} = 1.96 \frac{\overline{\sigma}_{i,h}}{\sqrt{n_{test}}}
\end{equation}

The mean of the error $e_{i,h}$ over a subset $I \subseteq [\![ 1, ..., 9 ]\!]$ of the 9 sequences\footnotemark and $h \in H = \{ h_{min}, ..., h_{max} \}$, denoted by $e_I$, follows a Gaussian distribution with mean $\mu_I$. The half-range of the 95\% confidence interval for $\mu_I$, denoted by $\Delta \mu_I$, can be calculated according to Eq. \ref{eq:conf interval mu}.

\footnotetext{When $I$ is the set $[\![ 1, ..., 9 ]\!]$, the confidence intervals calculated are those associated with the 9 records. Otherwise, we select $I$ as the set of indexes associated with the regular or irregular breathing sequences.}

\begin{equation} \label{eq:conf interval mu}
\Delta \mu_I = \frac{1}{|I| \, |H|} \sqrt{\sum_{i \in I} \sum_{h=h_{min}}^{h_{max}} (\Delta \mu_{i,h})^2}
\end{equation}

\section{Results}

\subsection{Prediction accuracy and oscillatory behavior of the predicted signal}
\label{section:accuracy and jitter}

\begin{table*}[tb!]
\normalsize
\setlength{\tabcolsep}{8pt}
\begin{center}
\begin{tabular}{lllll}
\hline
Error     &  Prediction & Average over   & Regular   & Irregular \\
type      &  method     & the 9 sequences  & breathing & breathing\protect\footnotemark \\
\hline \hline
MAE       & RNN UORO        & $0.845 \pm 0.001$ & $0.674 \pm 0.001$ & $0.916 \pm 0.001$ \\
(in mm)   & RNN RTRL        & $0.834 \pm 0.002$ & $0.684 \pm 0.002$ & $0.973 \pm 0.003$ \\
          & LMS             & 0.957 & 0.907 & 1.18 \\
          & Lin. regression & 4.45 & 3.23 & 6.57 \\
          & No prediction   & 3.27 & 2.89 & 3.43 \\
\hline               
RMSE      & RNN UORO        & $1.275 \pm 0.001$ & $1.030 \pm 0.001$ & $1.505 \pm 0.002$ \\
(in mm)   & RNN RTRL        & $1.419 \pm 0.005$ & $1.119 \pm 0.004$ & $1.721 \pm 0.005$ \\
          & LMS             & 1.370 & 1.247 & 1.818 \\
          & Lin. regression & 6.089 & 4.454 & 9.164 \\
          & No prediction   & 4.243 & 3.952 & 4.461 \\    
\hline
nRMSE     & RNN UORO        & $0.2824 \pm 0.0002$ & $0.2868 \pm 0.0004$ & $0.3211 \pm 0.0004$ \\
(no unit) & RNN RTRL        & $0.3027 \pm 0.0007$ & $0.2914 \pm 0.0008$ & $0.3688 \pm 0.0010$ \\
          & LMS             & 0.3116 & 0.2987 & 0.4198 \\
          & Lin. regression & 1.411 & 1.181 & 2.132 \\
          & No prediction   & 0.9312 & 1.006 & 0.9833 \\    
\hline
Max error & RNN UORO        & $8.81 \pm 0.01$ & $7.20 \pm 0.02$ & $12.34 \pm 0.02$ \\
(in mm)   & RNN RTRL        & $11.68 \pm 0.04$ & $10.01 \pm 0.04$ & $14.56 \pm 0.06$ \\
          & LMS             & 9.31 & 8.59 & 12.9 \\
          & Lin. regression & 30.6 & 23.2 & 49.0 \\
          & No prediction   & 14.8 & 13.9 & 18.2 \\   
\hline
Jitter    & RNN UORO        & $0.9672 \pm 0.0004$ & $0.7778 \pm 0.0002$ & $0.9973 \pm 0.0007$ \\
(in mm)   & RNN RTRL        & $0.7532 \pm 0.0015$ & $0.6494 \pm 0.0012$ & $0.8735 \pm 0.0014$ \\
          & LMS             & 1.596 & 1.646 & 1.724 \\
          & Lin. regression & 0.7767 & 0.6011 & 1.078 \\
          & No prediction   & 0.4395 & 0.3877 & 0.5045 \\                            
\hline
\end{tabular}
\end{center}
\caption{Comparison of the forecasting performance of each algorithm. Each error value corresponds to the average of a given performance measure of the test set over the sequences considered and the horizon values between 0.1s and 2.0s. The 95\% mean confidence intervals associated with the RNNs are calculated assuming that the error distribution is Gaussian (Eq. \ref{eq:conf interval mu}).}
\label{table:pred perf}
\end{table*}

\footnotetext{Sequence 201205111057-LACLARUAR-3-O-72 (cf \cite{krilavicius2016predicting}) has been removed from the sequences with abnormal respiratory motion when reporting performance measures in the last column, as it does not contain abrupt or sudden motion that typically makes forecasting difficult. In particular, this is why the nRMSE of UORO averaged over the 9 sequences is lower than nRMSE of UORO averaged over the regular or irregular breathing sequences.}

\begin{figure*}[tb!]
    \centering
    \includegraphics[width=.38\textwidth]{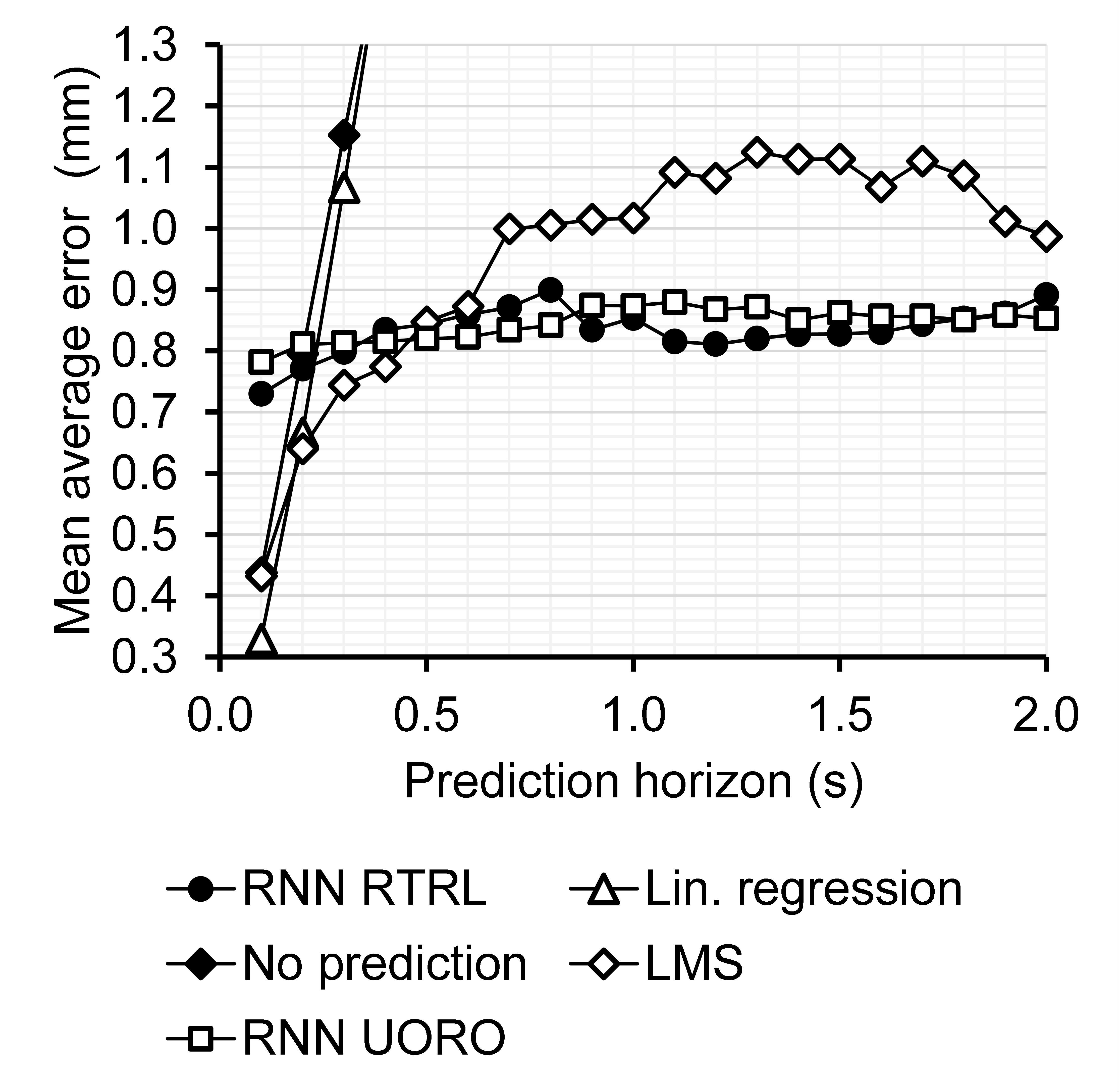}%
    \qquad
    \includegraphics[width=.38\textwidth]{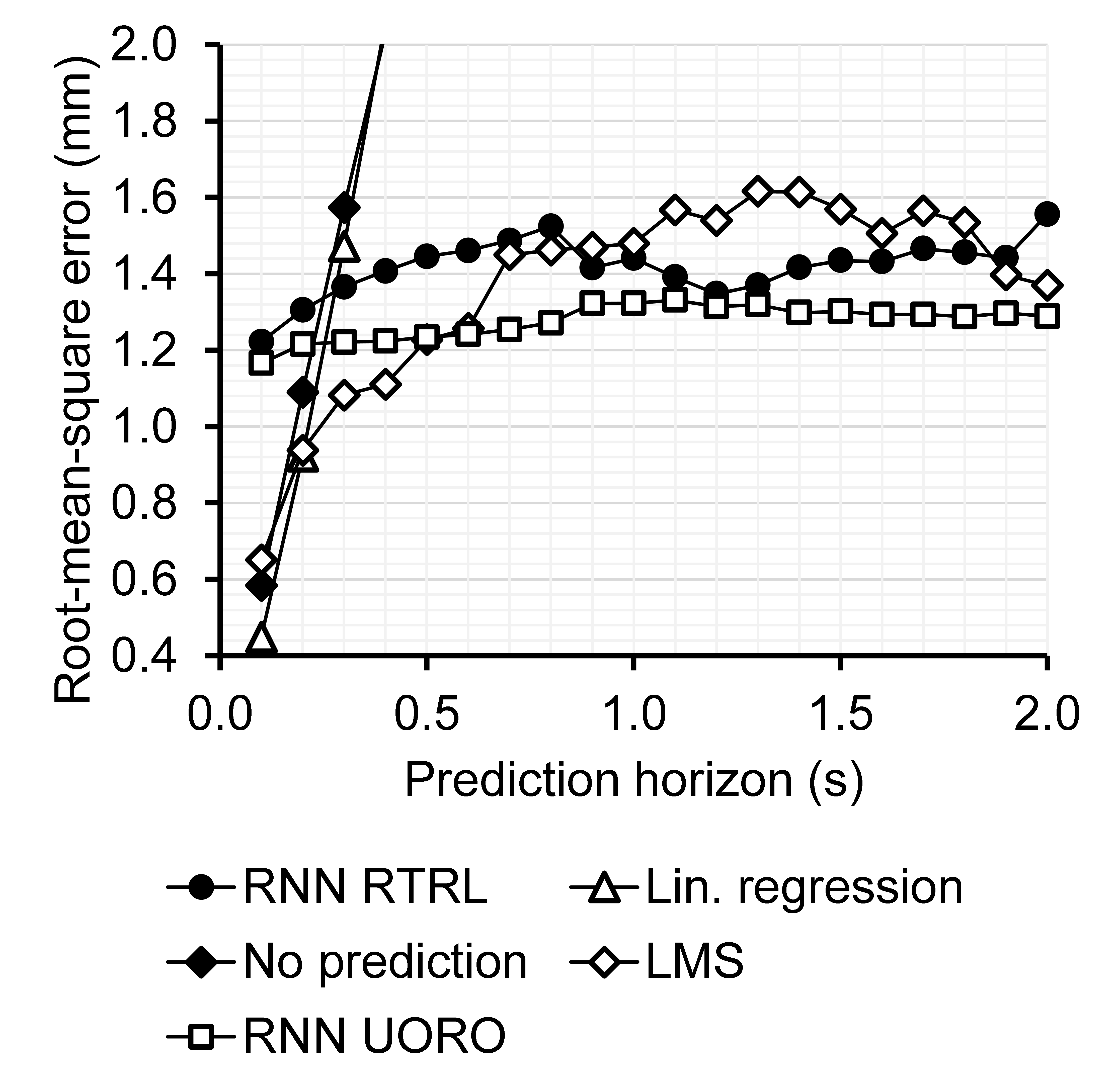}%
    \qquad
    \includegraphics[width=.38\textwidth]{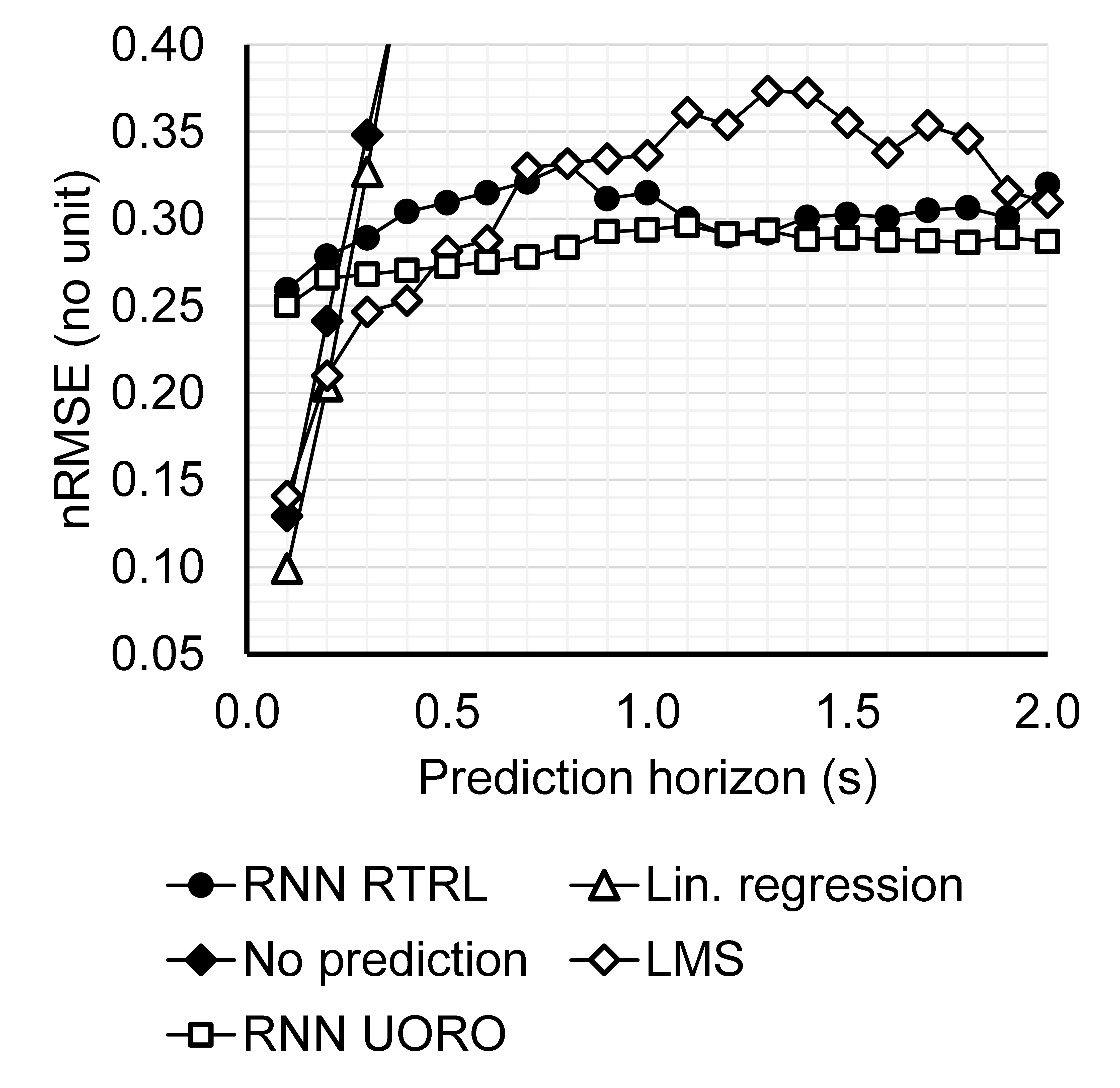}%
    \qquad
    \includegraphics[width=.38\textwidth]{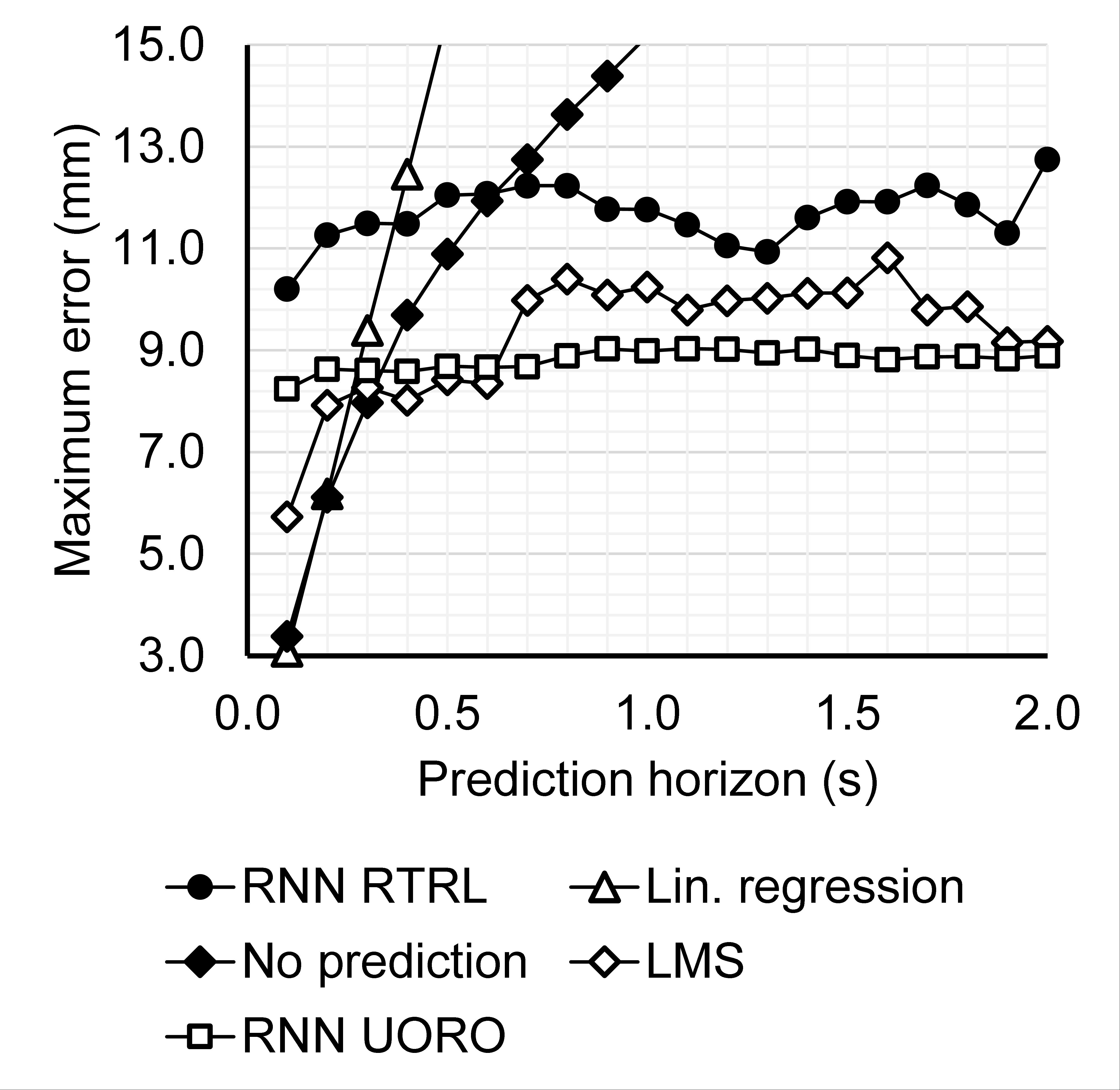}%
    \qquad
    \includegraphics[width=.38\textwidth]{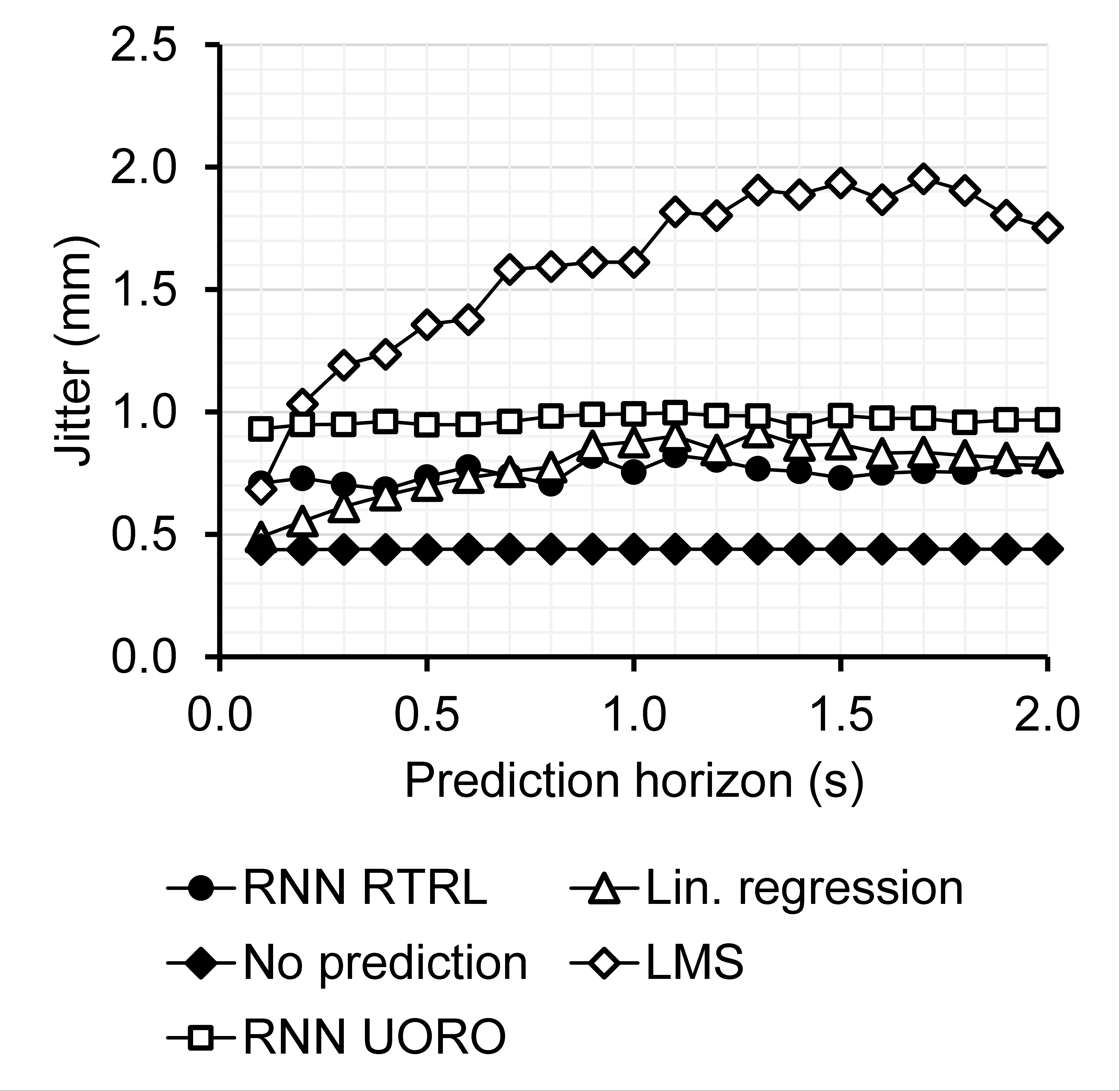}%
    \caption{Forecasting performance of each algorithm as a function of the prediction horizon. Each point corresponds to the average of one performance measure of the test set across the 9 sequences. }
    \label{fig:pred perf}
\end{figure*}

\begin{figure}[htb!]
	\centering
		\includegraphics[width=\columnwidth]{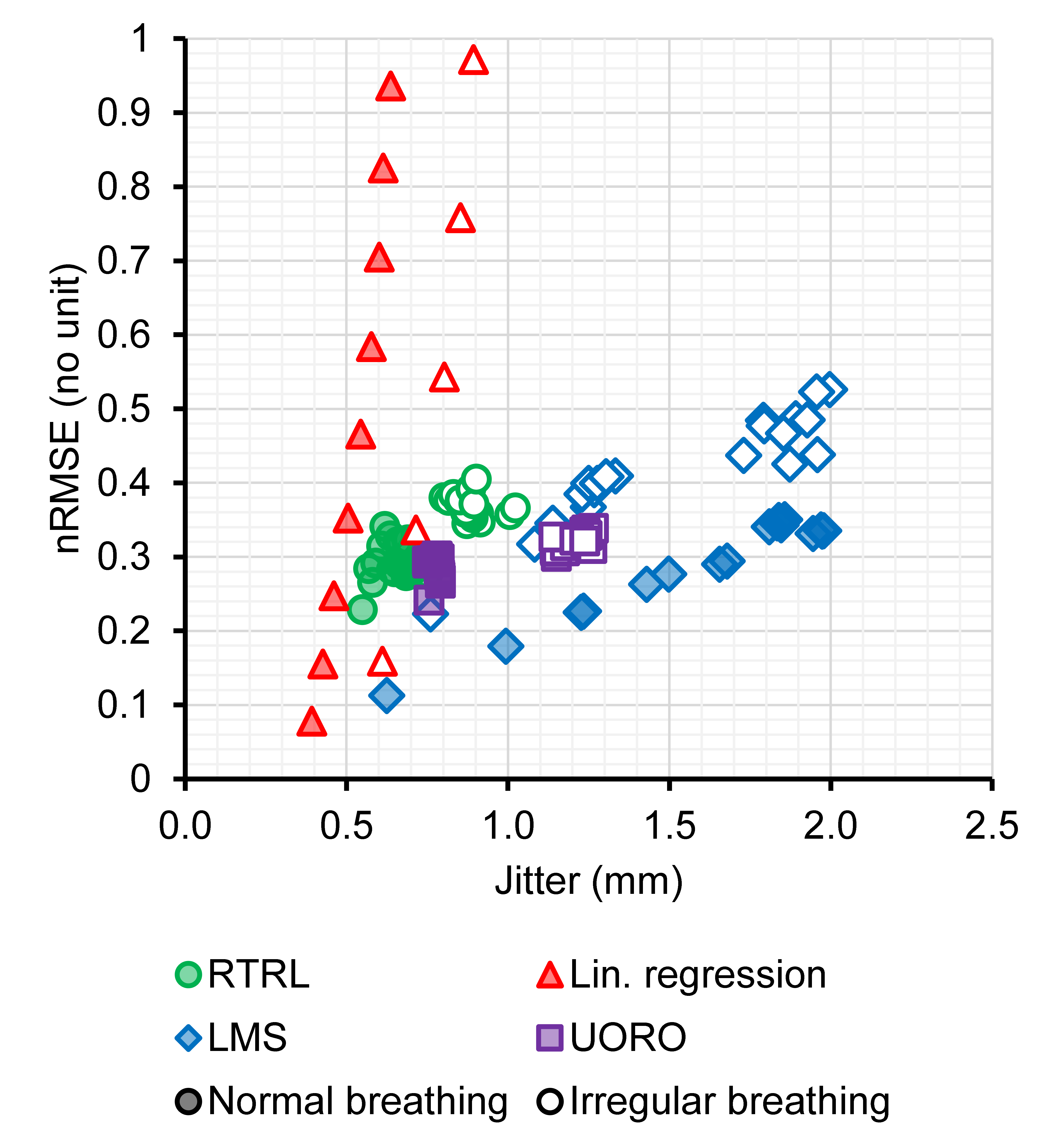}
	\caption{Prediction performance of each algorithm in terms of nRMSE and jitter. Each point corresponds to the mean of the nRMSE and jitter of a given algorithm of the test set over the regular or irregular breathing sequences for a single horizon value. Datapoints corresponding to linear regression with high horizon values have not been displayed for readability as they correspond to high nRMSE values.}
	\label{fig:nRMSE vs jitter}
\end{figure}

\begin{figure*}[htb!]
    \centering
    \subfloat[\normalsize Prediction with an RNN trained with UORO]{{\includegraphics[width=.85\textwidth]{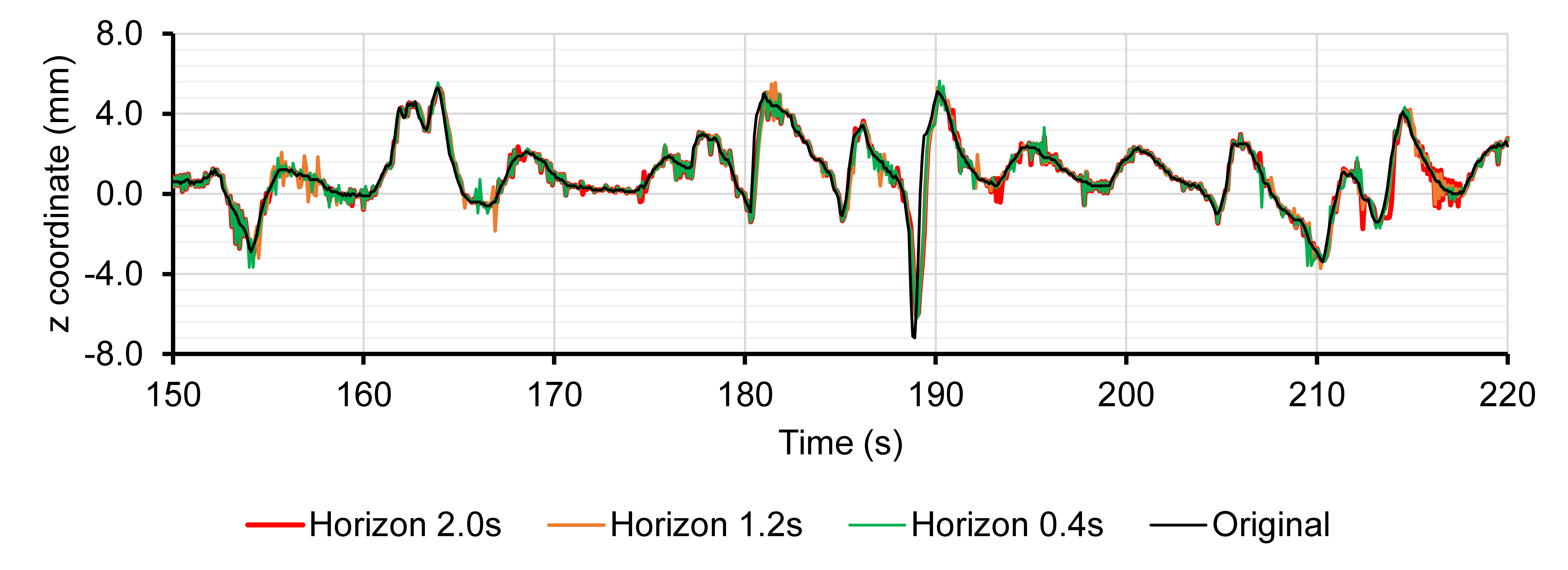} }}%
    \quad
    \subfloat[\normalsize Prediction with an RNN trained with RTRL]{{\includegraphics[width=.85\textwidth]{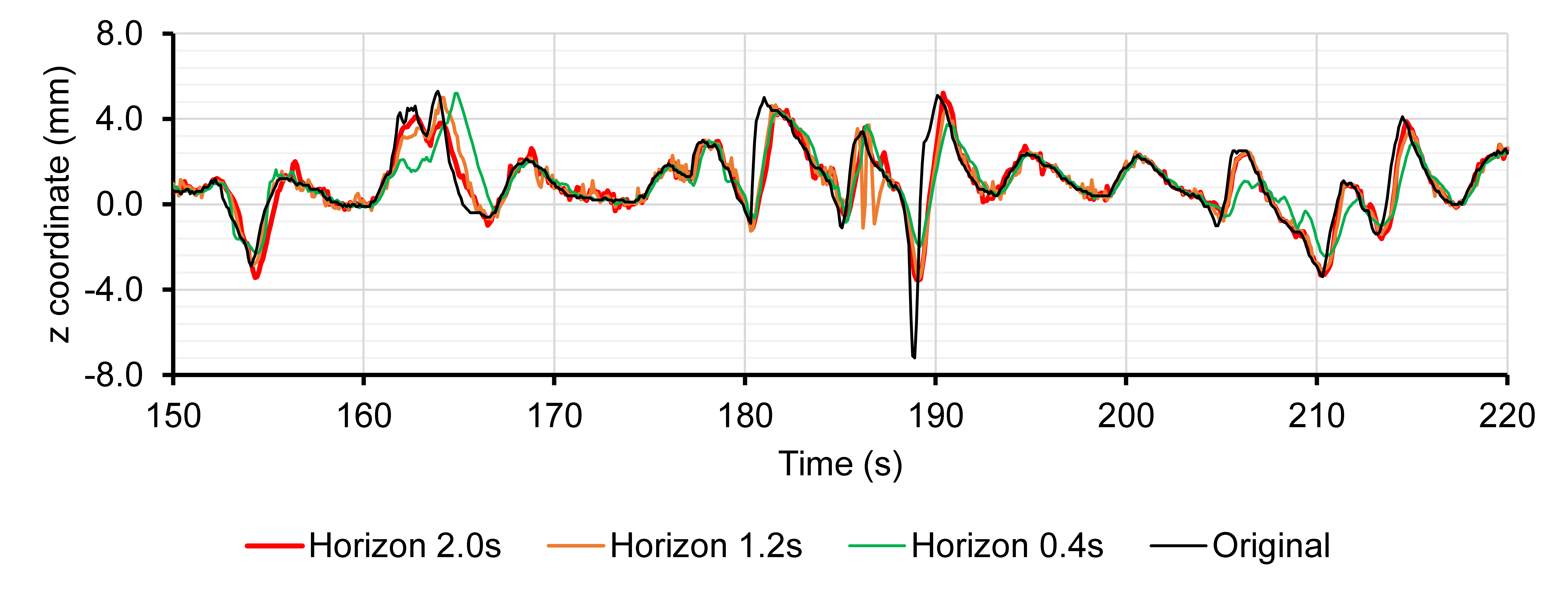} }}%
    \quad
    \subfloat[\normalsize Prediction with LMS]{{\includegraphics[width=.85\textwidth]{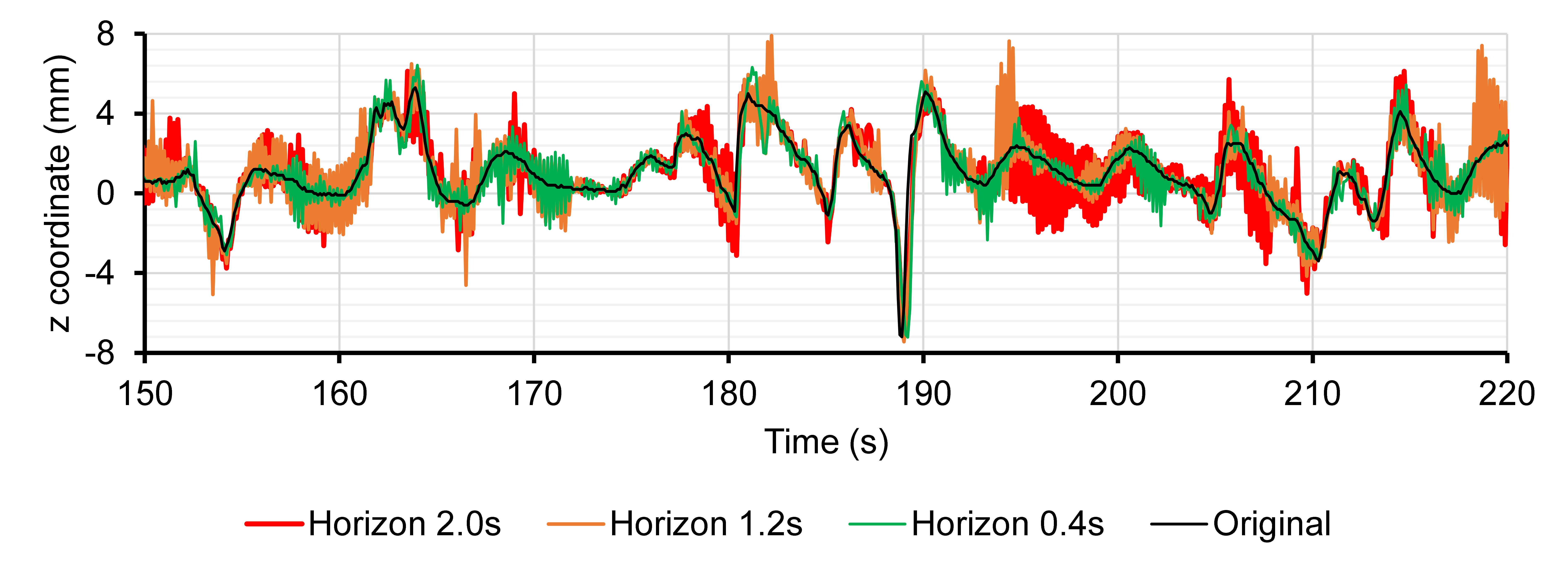} }}%
    \caption{Comparison between RTRL, UORO, and LMS regarding the prediction of the position of the z coordinate (spine axis) of marker 3 in sequence 1 (person talking)}%
    \label{fig:coordz_marker3_seq1}%
\end{figure*}

UORO achieves the lowest RMSE, nRMSE and maximum error averaged over all the sequences and horizons (cf Table \ref{table:pred perf}). It is relatively robust to irregular motion, as its nRMSE only increases by 10.6\% between regular and irregular breathing. LMS is subject to high jitter values (cf also Fig. \ref{fig:pred perf}, Fig. \ref{fig:nRMSE vs jitter}, Fig. \ref{fig:coordz_marker3_seq1}, and Fig. \ref{fig:coordz_marker3_seq5}). The high maximum errors corresponding to RTRL, relative to UORO and LMS, can be observed in Fig. \ref{fig:loss function}. The narrow 95\% confidence intervals associated with the performance measures reported in Table \ref{table:pred perf} indicate that selecting $n_{test} = 300$ runs is sufficient for providing accurate results.

The graphs representing the performance of each algorithm as a function of the horizon value $h$ appear to have irregular and changeable local variations, especially in the case of RTRL and LMS, because the set of hyper-parameters automatically selected by cross-validation is different for each horizon value (Fig. \ref{fig:pred perf}). These instabilities may also be caused by the relatively low number of breathing records in our dataset. However, it can be observed that the prediction errors and jitter of the test set corresponding to each algorithm globally tend to increase with $h$. 

Linear regression achieves the lowest RMSE and nRMSE for $h \leq 0.2s$ as well as the lowest MAE and maximum error for $h = 0.1s$. The RMSE corresponding to linear regression for $h=0.2s$ is equal to 0.92mm. LMS gives the lowest RMSE for $ 0.3s \leq h \leq 0.5s$, the lowest MAE for $ 0.2s \leq h \leq 0.4s$, the lowest nRMSE for $h = 0.3s$ and $h = 0.4s$, and the lowest maximum error for $ 0.4s \leq h \leq 0.6s$. The RMSE corresponding to LMS for $h=0.5s$ is equal to 1.23mm. UORO outperforms the other algorithms in terms of RMSE for $h \geq 0.6s$ and maximum error for $h \geq 0.7s$. The RMSE given by UORO is rather constant and stays below 1.33mm across all the horizon values considered. RTRL and UORO both have a lower prediction MAE than LMS for $h \geq 0.5s$. Our analysis of the influence of the latency on the relative performance of linear filters, adaptive filters, and ANNs agrees with the review of Verma et al. \cite{verma2010survey} (cf section \ref{section:intro pred in radiotherapy}).

The jitter associated with RTRL and UORO respectively increases from 0.71mm and 0.94mm for $h=0.1s$ to 0.78mm and 0.96mm for $h=2.0s$. However, the jitter associated with linear regression and LMS increases more significantly with $h$. The jitter corresponding to linear regression is the lowest among the four prediction methods for $h \leq 0.6s$.

The performance of each algorithm as a function of the horizon in the cases of normal breathing and abnormal breathing is detailed in Appendix \ref{appendix:pred perf}. The local unsteadiness of the variations of each performance measure with $h$ in Figs. \ref{fig:pred perf regular} and \ref{fig:pred perf irregular} is more pronounced than in Fig. \ref{fig:pred perf} because both situations involve averaging results over fewer respiratory traces. However, it still appears that the prediction errors globally tend to increase with $h$ in both cases. UORO performs better than the other algorithms for lower horizon values in the scenario of abnormal breathing. Indeed, it achieves the lowest RMSE and nRMSE for $h \geq 0.3s$, and the lowest MAE for $h \geq 0.2s$ (RTRL and UORO achieve comparable performance for high horizons in terms of MAE).

\subsection{Influence of the hyper-parameters on prediction accuracy}

\begin{figure*}[ht!]%
    \centering
    \subfloat[\normalsize Prediction nRMSE of the cross-validation set as a function of the learning rate]{{\includegraphics[width=.40\textwidth]{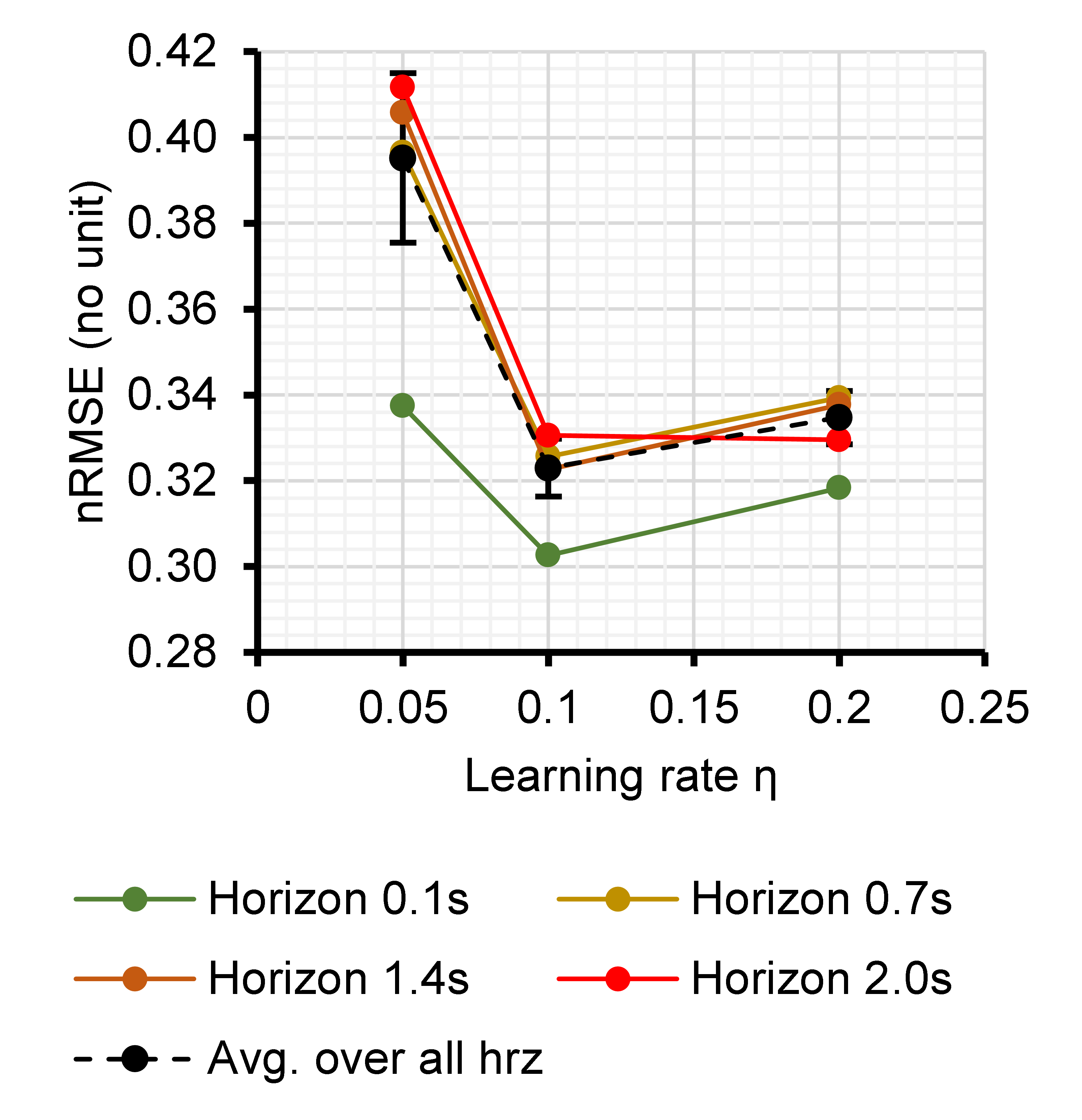} \label{fig:learn rate influence}}}%
    \qquad
    \subfloat[\normalsize Prediction nRMSE of the cross-validation set as a function of the standard deviation of the Gaussian distribution of the initial synaptic weights]{{\includegraphics[width=.40\textwidth]{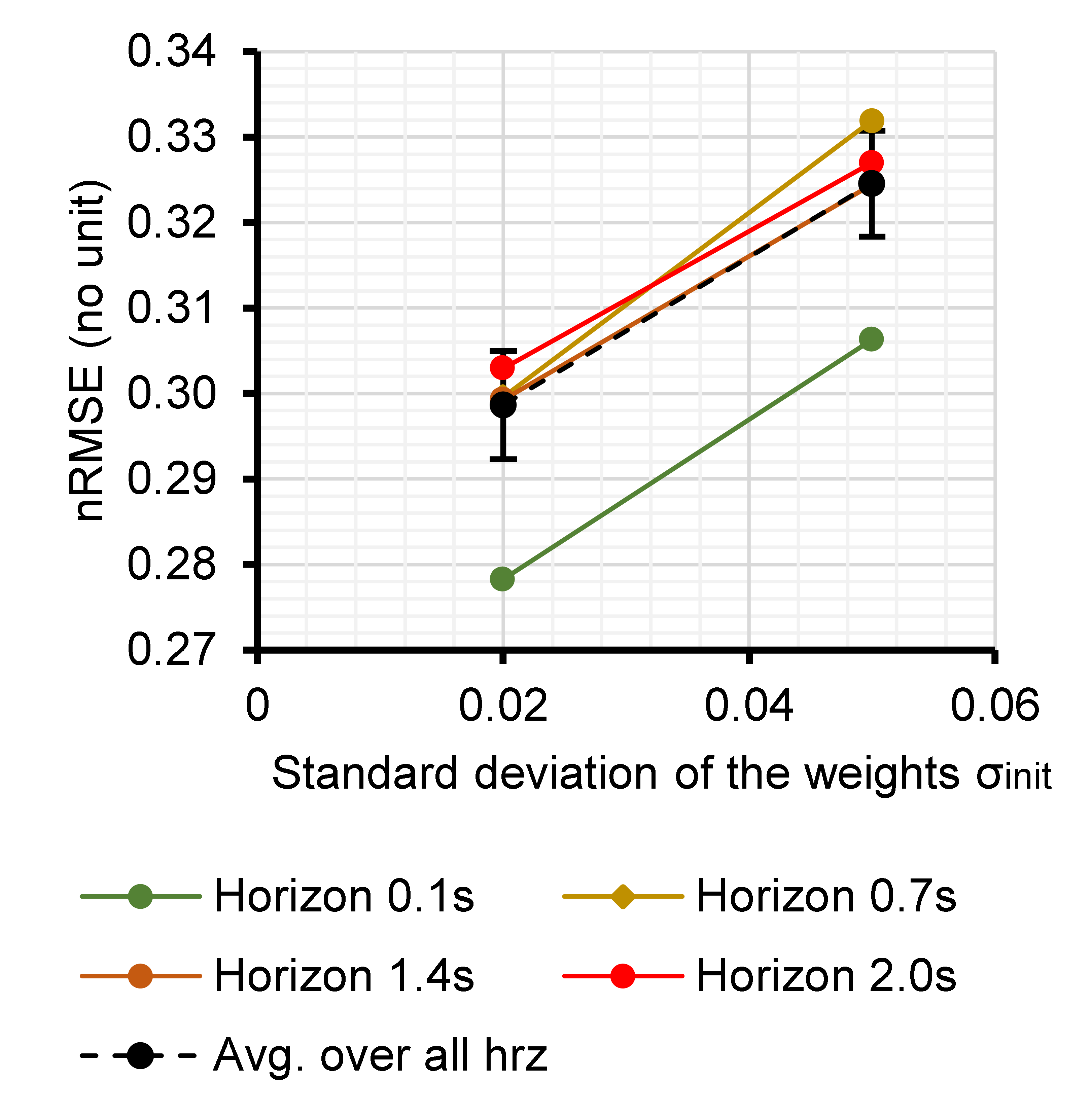} \label{fig:init std weights influence}}}%
    \qquad
    \subfloat[\normalsize Prediction nRMSE of the cross-validation set as a function of the signal history length]{{\includegraphics[width=.40\textwidth]{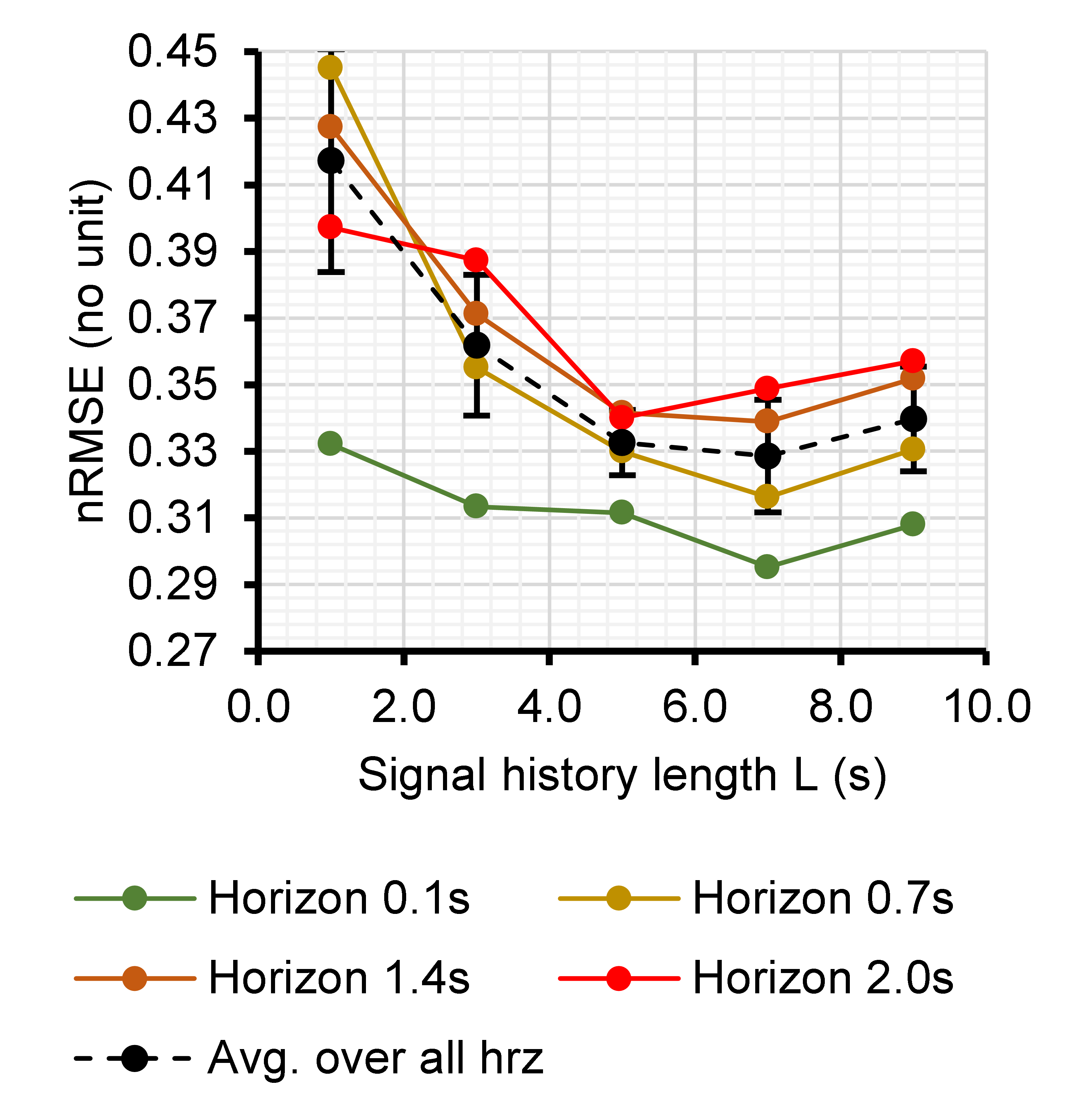} \label{fig:SHL influence}}}%
    \qquad
    \subfloat[\normalsize Prediction nRMSE of the cross-validation set as a function of the number of hidden units]{{\includegraphics[width=.40\textwidth]{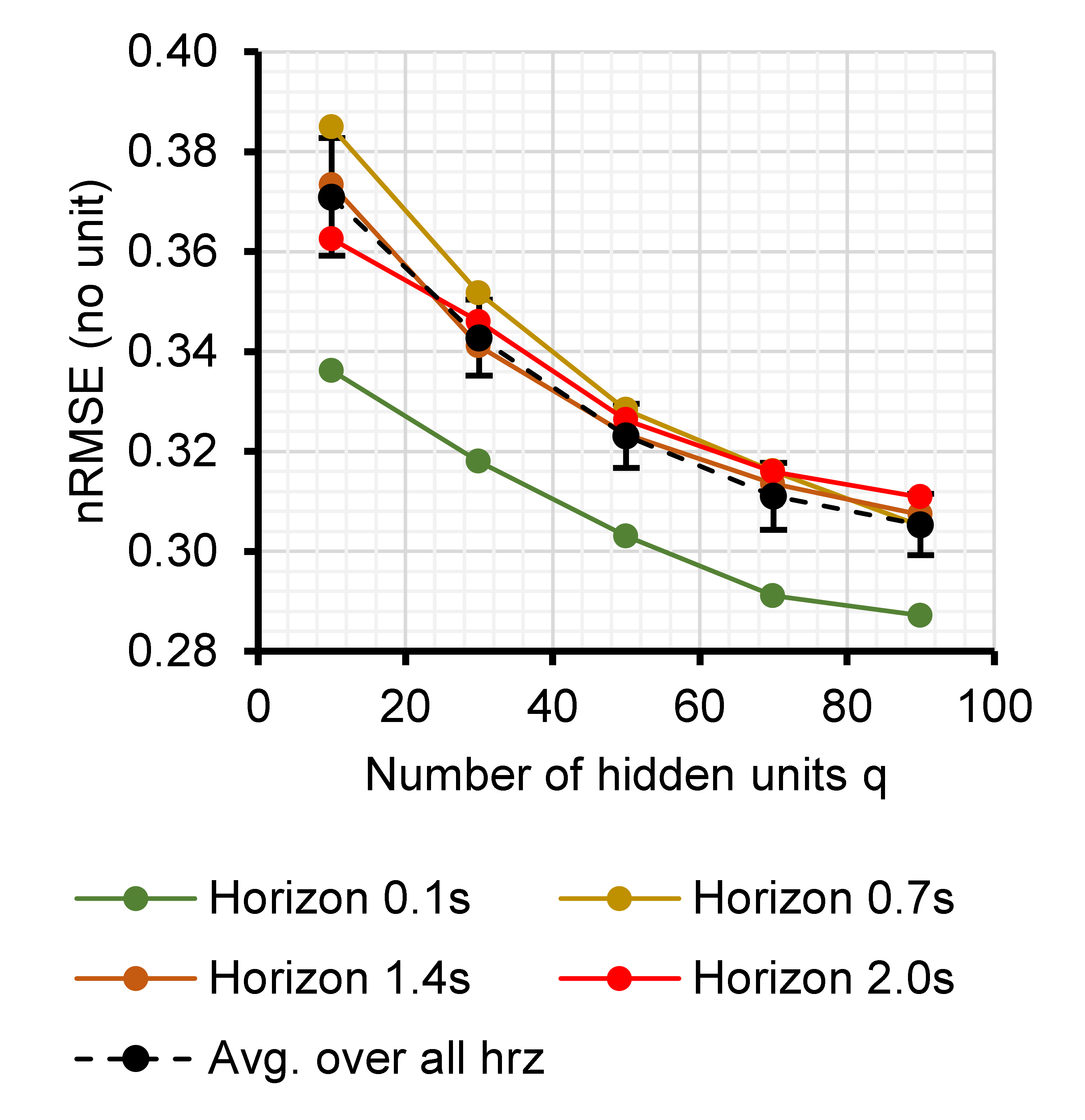} \label{fig:nb hidden neurons influence}}}%
    \caption{Prediction nRMSE of UORO of the cross-validation set as a function of each RNN hyper-parameter, for different horizon values. Given one hyper-parameter, each color point of the associated graph corresponds to the minimum of the nRMSE over every possible combination of the other hyper-parameters in the cross-validation range (Table \ref{table:models comparison}). Each nRMSE measure is averaged over the 9 sequences and 50 runs. The black dotted curves correspond to the nRMSE minimum averaged over the horizon values between 0.1s and 2.0s, and the associated error bars correspond to its standard deviation over these horizon values.}%
    \label{fig:hyperparam influence}
\end{figure*}

\begin{figure}[htb!]
\centering
\includegraphics[width=.75\columnwidth]{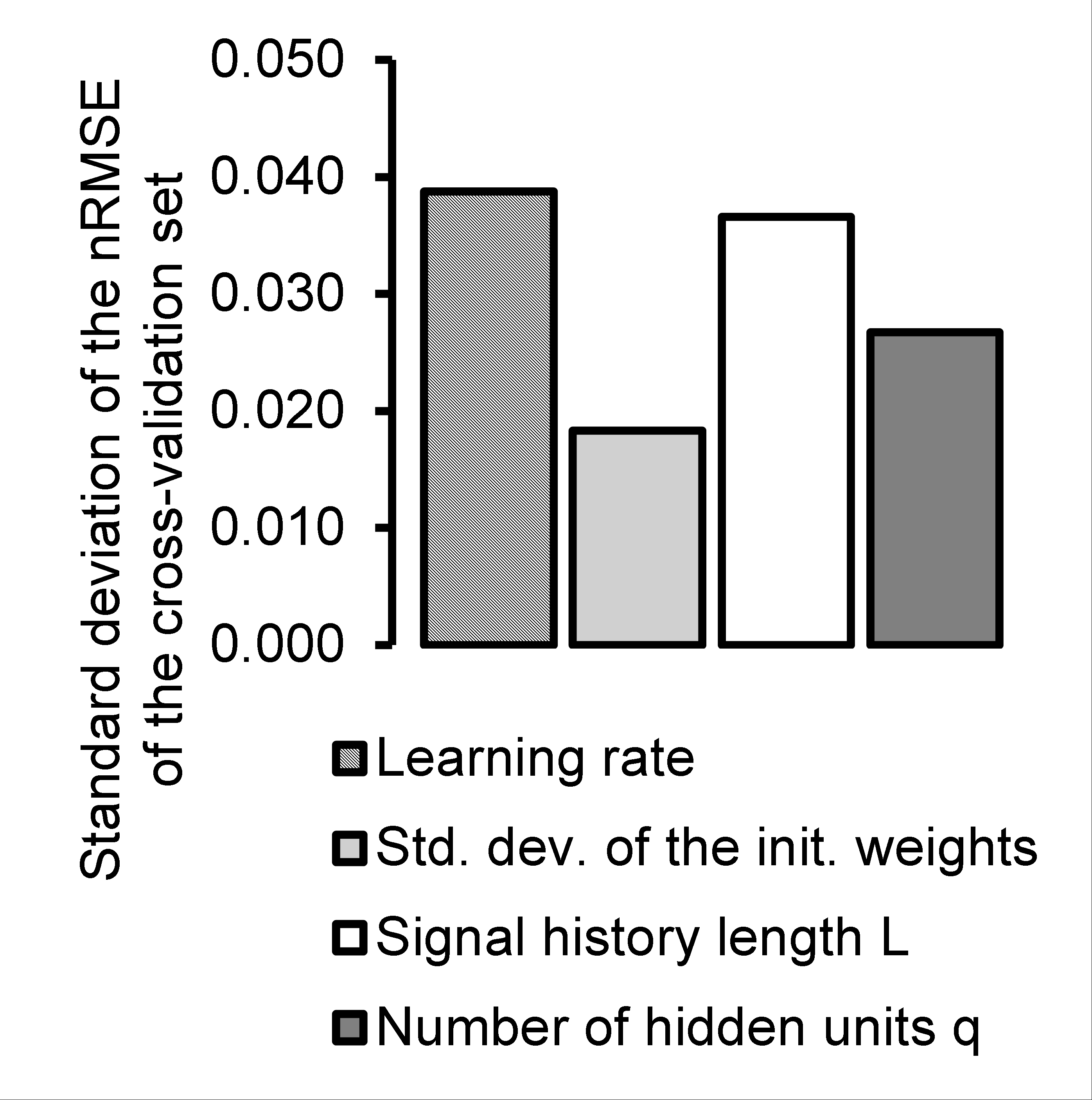}
\caption{Standard deviation of the nRMSE of the cross-validation set (black dotted curves in Fig. \ref{fig:hyperparam influence}) for each hyper-parameter. A hyper-parameter corresponding to a high standard deviation value has a high influence on the prediction error.} 
\label{fig:relative param influence}
\end{figure}

Fig. \ref{fig:hyperparam influence} also shows that the prediction nRMSE of the cross-validation set tends to increase as the horizon value $h$ increases. On average over the 9 sequences and all the horizon values, $\eta = 0.1$ and $L = 7.0s$ give the best prediction results. However, for $h=2.0s$, a higher learning rate $\eta = 0.2$ and a lower value of the SHL $L= 5.0s$ give better results (Figs. \ref{fig:learn rate influence}, \ref{fig:SHL influence}). In other words, when performing prediction with a high look-ahead time, it seems better to make the RNN more dependent on the recent inputs, and quickly correct the synaptic weights when large prediction errors occur. In our experimental setting, $\sigma_{init} = 0.02$ and $q=90$ hidden units correspond to the lowest nRMSE of the cross-validation set (Figs. \ref{fig:init std weights influence}, \ref{fig:nb hidden neurons influence}). The nRMSE of the cross-validation set decreases as the number of hidden units increases, therefore we may achieve higher accuracy with more hidden units. However, that would consequently increase the computing time (Fig. \ref{fig:prediction time}). With the hyper-parameters $L = 7.0s$ and $q = 90$ corresponding to the highest accuracy on average over the 9 sequences and all the horizon values, our shallow network already has 65,700 parameters to learn. Similarly, it has been reported in \cite{POHL2021101941} that increasing the number of hidden units of a vanilla RNN with a single hidden layer trained to predict breathing signals using RTRL led to a decrease of the forecasting MAE. Fig. \ref{fig:hyperparam influence} displays the nRMSE averaged over the 9 sequences, and the general aforementioned recommendations are not optimal for each sequence. Therefore, we recommend using cross-validation to determine the best hyper-parameter set for each breathing record. The learning rate and SHL appear to be the most important hyper-parameters to tune (Fig. \ref{fig:relative param influence}). Appropriately selecting them resulted in a decrease of the mean cross-validation nRMSE of 18.2\% (from 0.395 to 0.323) and 21.3\% (from 0.417 to 0.329), respectively.

\subsection{Time performance}

\begin{figure}[htb!]
\centering
\includegraphics[width=\columnwidth]{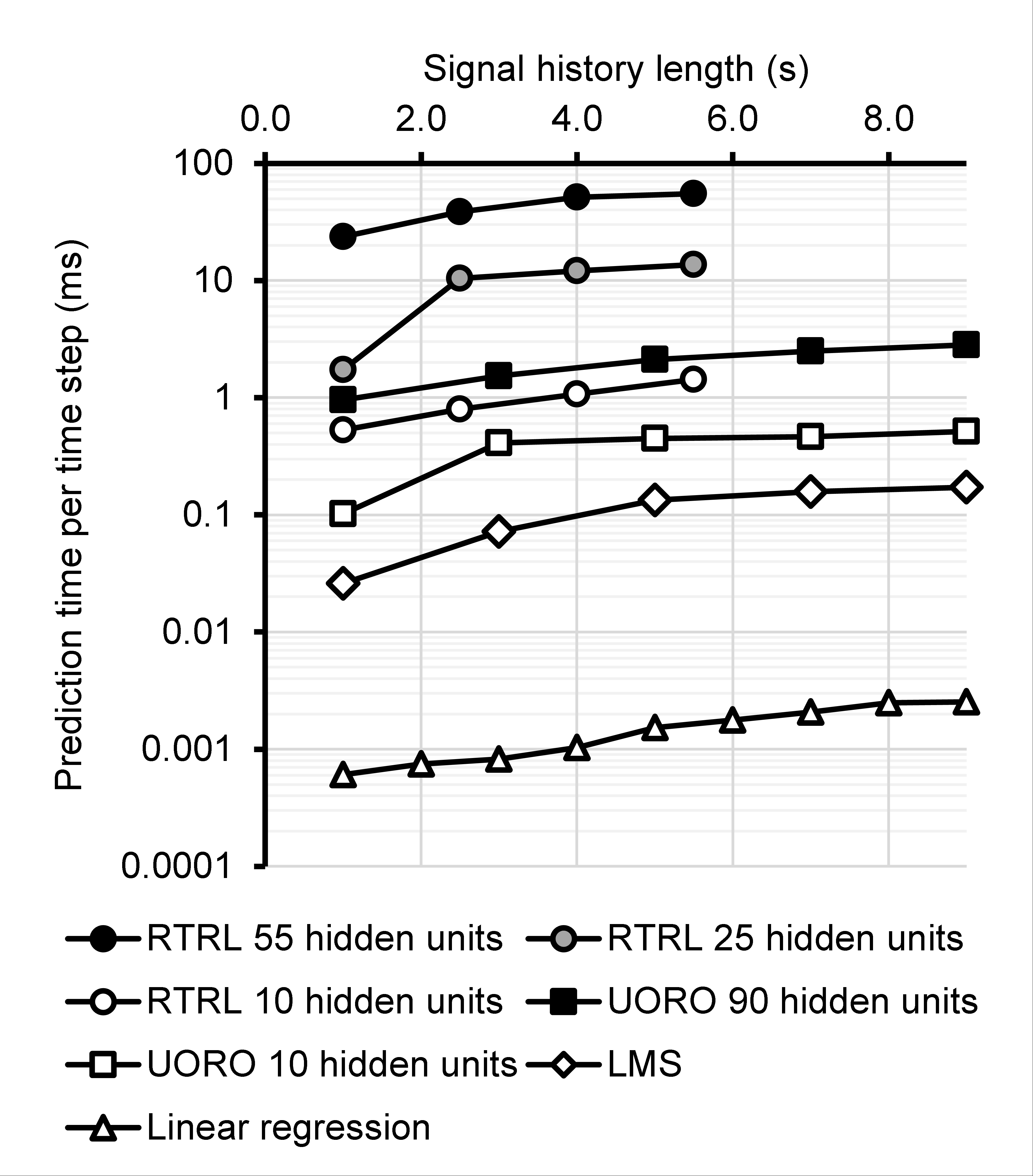}
\caption{Calculation time (Dell Intel Core i9-9900K 3.60GHz CPU 32Gb RAM with Matlab)} 
\label{fig:prediction time}
\end{figure}

UORO has a prediction time per time step equal to 2.8ms for 90 hidden neurons and an SHL of 9.0s, whereas RTRL requires 55ms to perform a single prediction using 55 hidden units with an SHL of 5.5s (Dell Intel Core i9-9900K 3.60GHz CPU 32Gb RAM with Matlab, Fig. \ref{fig:prediction time}). The complexity $\mathcal{O}(q^3 (q+L))$ and resulting high computing time of RTRL is the reason why we performed cross-validation for RTRL with fewer hidden units and lower SHL values than UORO, which has a complexity $\mathcal{O}(q(q+L))$ (Table \ref{table:models comparison}).

\section{Discussion}

\subsection{Significance of our results relative to the dataset used}
\label{section: comments on dataset}

One drawback of our study is the number of sequences used and their duration, which are low in comparison with some other studies related to forecasting in radiotherapy (cf section \ref{section:intro pred in radiotherapy} and Table \ref{table:comparison_with litterature}). Therefore, our numerical results might appear to lack a certain degree of confidence. However, the dataset used is representative of a large variety of breathing patterns including shifts, drifts, slow motion, sudden irregularities, as well as resting and non-perturbed motion. In addition, our results are consistent with previous studies that claim that linear prediction, linear adaptive filters, and ANNs achieve high performance respectively for low, intermediate, and high horizon values (cf section \ref{section:accuracy and jitter}). The algorithms studied in our work are online algorithms that do not need a high amount of prior data for making accurate predictions, as demonstrated by the high performance that we achieved with only one minute of training. Because of the reasons mentioned above, we think that the results presented in our study have a significantly high level of confidence and would generalize well to larger datasets.

The online availability of the dataset used is a particular strength of our study, as it enables reproducibility of our results. Most of the previous studies about the prediction of breathing signals for radiotherapy rely on datasets that are not publicly available (cf section \ref{section:intro pred in radiotherapy} and Table \ref{table:comparison_with litterature}), which makes performance comparison difficult.

Laughing and talking are situations where prediction is difficult, and are controlled in a clinical setting. However, evaluating performance with such difficult scenarios gives information about other situations that will sometimes happen during treatment, such as yawning, hiccuping, and coughing. Detecting these anomalies and turning off the irradiation beam when they occur is currently the standard clinical approach. Distinguishing between normal and irregular breathing enabled us to objectively study and quantify the robustness of the algorithms compared (cf Table \ref{table:pred perf} and Appendix \ref{appendix:pred perf}). Since irregular breathing sequences comprise almost half of our entire dataset, the numerical error measures averaged over the nine sequences should be higher than one can expect in more realistic scenarios.

\subsection{Comparison with previous works}

Table \ref{table:comparison_with litterature} compares the performance of UORO in our work with the results previously reported in the literature. Comparison with the previous research is complex because the datasets are different. In particular, the frequency, amplitude, and regularity of the signals vary from study to study. Furthermore, the response time, as well as the partition of the data into development and test set are usually arbitrarily selected, thus they also differ between the studies. 

The prediction errors in our research might appear relatively large, but this is due to the low sampling frequency (10Hz), the high amplitude of the breathing signals, and the high proportion of irregular patterns in our dataset (cf section \ref{section: comments on dataset}). Furthermore, the breathing records that we use have a relatively low duration and therefore our RNN has fewer data available for training. When taking these circumstances into account, it appears that the errors reported in our study are consistent with the findings of the previous related works.

Our purpose is to examine the extent to which RNNs can efficiently learn to adaptively predict respiratory motion with little data. We do not aim to build a generalized model with a high amount of data. All the RNN-based models reported in Table \ref{table:comparison_with litterature} may benefit from adaptive retraining with UORO.

Teo et al. studied breathing records with a frequency of 7.5 Hz and reported lower errors using an MLP with one hidden layer trained first with backpropagation and retrained online \cite{teo2018feasibility}. The RMSE that they achieved was 26\% lower than that of UORO in our research. Our higher errors are partly due to the amplitude of the breathing signals in our dataset, which are approximately 3 times higher. Mafi et al. and Jiang et al. also reported similar but lower prediction errors using RTRL to train a 1-layer RNN \cite{mafi2020real} and a 1-layer non-linear auto-regressive exogenous model (NARX) \cite{jiang2019prediction}, respectively. However, the former do not provide information concerning motion amplitude, and the latter use signals with amplitudes nearly twice lower and a higher sampling rate, equal to 30 Hz. Moreover, our results demonstrate that UORO has more benefits than RTRL in practice. The RMSE error that UORO achieved is approximately 2 to 4 times lower than the RMSEs reported by Sharp et al., who used a multilayer perceptron (MLP) with one hidden layer and breathing records of the same frequency (10Hz) with similar amplitudes \cite{sharp2004prediction}. Furthermore, the nRMSE error of UORO in our work is approximately 1 to 2 times lower than those corresponding to the 3-layer MLP in the study of Jöhl et al.,  even though they use breathing records with a relatively higher sampling frequency (25 Hz) and lower signal amplitudes. They claim that linear filters are the most appropriate to forecast respiratory motion, but our results indicate that this is true only when the response time is low with respect to the sampling frequency (Section \ref{section:accuracy and jitter}). The RMSE that we found is within the range reported in the first study of Jiang et al., who predicted the position of an internal marker using an RNN with 1 hidden layer trained with BPTT with a higher frequency (30 Hz) \cite{kai2018prediction}. 

\begin{table*}[htb]
\setlength{\tabcolsep}{1.5pt}
\begin{center}
\begin{tabular}{llllllllll}
\hline
First                          & Network & Training          & Breathing      & Sampling & Amount of       & Signal      & Response & Prediction       \\
author                         &         & method            & data           & rate     & data            & amplitude   & time     & error            \\
\hline
\hline
Sharp \cite{sharp2004prediction}& 1-layer & -                 & 1 implanted    & 10 Hz    & 14 records      & 9.1mm       & 1) 200ms & 1) RMSE 2.6mm   \\
                                & MLP     &                   & marker         &          & 48s to 352s     & to 31.6mm   & 2) 1s    & 2) RMSE 5.3mm  \\
Sun \cite{sun2017respiratory}  & 1-layer & Levenberg-Marq.   & RPM data       & 30 Hz    & data from       & Rescaling   & 500ms    & Max error 0.65   \\
                               & MLP     & \& adapt. boosting& (Varian)       &          & 138 scans       & between -1  &          & RMSE 0.17        \\
                               &         &                   &                &          &                 & and 1       &          & nRMSE 0.28       \\                                                              
Jiang \cite{kai2018prediction}   & 1-layer & BPTT              & 1 implanted    & 30 Hz    & 7 records of    & -           & 1.0s     & RMSE from        \\
                               & RNN     &                   & marker         &          & 40s to 70s      &             &          & 0.48mm to 1.37mm \\
Teo \cite{teo2018feasibility}  & 1-layer & Backprop. \&      & Cyberknife     & 7.5 Hz   & 27 records      & 2mm         & 650      & MAE 0.65mm       \\                               
                               & MLP     & adapt. training  & Synchrony      &          & of 1 min        & to 16mm     &          & RMSE 0.95mm      \\
                               &         &                   &                &          &                 & (dvlpmt set)&          & Max error 3.94mm \\                                                              
Jiang \cite{jiang2019prediction}& 1-layer & RTRL              & 1 implanted    & 30 Hz   & 7 records       & 2.5mm   & 1) 600ms & 1) RMSE 0.97mm       \\                                                              
                               & NARX    &                   & marker         &         &               & to 26.5mm            & 2) 1.0s  & 2) RMSE 1.18mm      \\
Lin \cite{lin2019towards}      & 3-layer & -                 & RPM data       & 30 Hz    & 1703 records    & Rescaling   & 280ms    & MAE 0.112        \\                                                              
                               & LSTM    &                   & (Varian)       &          & of 2 to 5 min   & between -1  & 500ms    & RMSE 0.139       \\
                               &         &                   &                &          &                 & and 1       &          & Max error 1.811  \\                                                              
Yun \cite{yun2019deep}         & 3-layer & adapt. training   & tumor 3D       & 25 Hz    & 158 records     & 0.6mm       & 280ms    & RMSE 0.9mm       \\                                                              
                               & LSTM    &                   & center of mass &          & of 8 min        & to 51.2mm   &          &                  \\
Jöhl \cite{johl2020performance}& 3-layer & Levenberg-Marq.   & Cyberknife SNR & 25 Hz   & 95 records of      & up to    & 160ms    & 1) nRMSE 0.38       \\                                                              
                               & MLP    &                    & 1) 30dB 2) 20dB &          & 11 to 131 min & 12mm            &          & 2) nRMSE 0.66      \\
Mafi \cite{mafi2020real}       & RNN     & RTRL              & Cyberknife     & 7.5 Hz   & 43 records of   & -           & 665ms    & MAE 0.54mm       \\                                                              
                               & -FCL    &                   & Synchrony      &          & 2.2s to 6.4s    &             &          & RMSE 0.57mm      \\

Lee \cite{lee2021geometric}    & LSTM    & BPTT              & RPM data       & 30 Hz    & 550 records     & 11.9mm      & 210ms    & RMSE 0.28mm      \\
                               & -FCL    &                   & (Varian)       &          & 91s to 188s     & to 25.9mm   &          &                  \\                               
\hline
Our                            & 1-layer & UORO              & 3 external     & 10 Hz    & 9 records       & 6mm         & 0.1s     & MAE 0.85mm       \\
work                           & RNN     &                   & markers        &          & 73s to 222s     & to 40mm     & to 2.0s  & Max error 8.8mm  \\
                               &         &                   & (Polaris)      &          &                 & (SI         &          & RMSE 1.28mm      \\
                               &         &                   &                &          &                 & direction)  &          & nRMSE 0.28       \\                               
\hline 
\end{tabular}
\end{center}
\caption{Comparison of our work with some of the previous studies about time-series forecasting with ANNs for respiratory motion compensation in radiotherapy (cf Section \ref{section:intro pred in radiotherapy}). In this table, the term "RNN" designates a vanilla RNN, as opposed to LSTMs. "LSTM-FCL" or "RNN-FCL" designates a combination of LSTM/RNN layers and fully connected layers. A field with " - " indicates that the information is not available in the corresponding research article. The results corresponding to our study are the performance measures averaged over the horizon values between 0.1s and 2.0s in Table \ref{table:pred perf}. \protect\footnotemark}
\label{table:comparison_with litterature}
\end{table*}

\footnotetext{RPM stands for "real-time position management". See also footnote \ref{footnote abbreviations}.}

\section{Conclusions}

This is the first study about RNNs trained with UORO applied to the forecast of the position of external markers on the chest and abdomen for safe radiotherapy, to the extent of our knowledge. This method can mitigate the latency of treatment systems due to robot control and radiation delivery preparation. This will in turn help decrease irradiation to healthy tissues and avoid lung radiation therapy side effects such as radiation pneumonitis or pulmonary fibrosis. The dataset used and our code are accessible online \cite{krilavicius2015predicting, pohl_michel_2021_5506965}. 

Online processing is suitable for breathing motion prediction during the radiotherapy treatment as it enables adaptation to each patient's individual respiratory patterns varying over time. Up to now, the only online learning algorithm for RNNs that has been evaluated in the context of respiratory motion prediction is RTRL \cite{mafi2020real, POHL2021101941}, but UORO has the advantage of being much faster, as they respectively have a theoretical complexity of $\mathcal{O}(q^4)$ and $\mathcal{O}(q^2)$, where $q$ is the number of hidden units. We derived an efficient implementation of UORO in the case of vanilla RNNs that uses closed-form expressions for quantities appearing in the loss gradient calculation, in contrast to the original article \citep{tallec2017unbiased} describing UORO in the general case. We could efficiently train RNNs using only one minute of breathing data per sequence, as dynamic training can be implemented with limited data. 

Most previous research in respiratory motion forecasting focused on univariate signals, but we undertook 3D marker position prediction, as it will help better estimate the 3D tumor motion via correspondence modeling. In addition, prediction was performed simultaneously for the three markers so that the RNN discovers and uses information from the correlation between their motion. We suggest using such multi-dimensional input and output framework (Eq. \ref{eq:RNN_in_out_def}), as it may improve the 3D forecasting accuracy in comparison to independently predicting univariate coordinate signals as in \cite{kai2018prediction, jiang2019prediction}. To the best of our knowledge, the comparison of the different prediction filters in our study takes into consideration the most extensive range of response times $h$ among the previous studies in the literature about respiratory motion prediction. Also, our study compares the highest number of forecasting quality metrics (MAE, RMSE, nRMSE, maximum error, and jitter) for each algorithm as h varies, among the works on breathing motion forecasting. Using different metrics helps better characterize the behavior of each algorithm.

UORO achieved the lowest prediction RMSE for horizon values $h \geq 0.6s$, with an average value over 9 breathing sequences not exceeding 1.4mm. These sequences last from 73s to 222s, correspond to a sampling rate of 10Hz and marker position amplitudes varying from 6mm to 40mm in the SI direction. Moreover, UORO achieved the lowest maximum error for $h \geq 0.7s$ with an average value over the 9 sequences not exceeding 9.1mm. The average of the RMSE and maximum error over the sequences corresponding to regular breathing were respectively lower than 1.1mm and 7.7mm. The nRMSE of UORO only increased by 10.6\% when performing the evaluation with the sequences corresponding to irregular breathing instead of regular breathing, which indicates good robustness to sudden changes in respiratory patterns. The calculation time per time step of UORO is equal to 2.8ms for 90 hidden units and an SHL of 9.0s (Dell Intel Core i9-9900K 3.60GHz CPU 32Gb RAM with Matlab). UORO has a much better time performance than RTRL, whose calculation time per time step is equal to 55.2ms for 55 hidden units and an SHL of 5.5s.

Linear regression was the most efficient prediction algorithm for low look-ahead time values, with an RMSE lower than 0.9mm for $h \leq 0.2s$. LMS gave the best prediction results for intermediate look-ahead values, with an RMSE lower than 1.2mm for $h \leq 0.5s$. These observations regarding the influence of the horizon agree with those in \cite{verma2010survey}. The errors reported in our study may be higher than in clinical scenarios due to the relatively high proportion of records corresponding to irregular breathing in our dataset.

Gradient clipping was used to ensure numerical stability and we selected a clipping threshold $\tau = 2.0$. The learning rate and SHL were the hyper-parameters with the strongest influence on the prediction performance of UORO. To the best of our knowledge, our work is the first to examine the influence of the horizon on hyper-parameter optimization among the works about respiratory motion forecasting during radiotherapy. Concerning UORO, we found that a learning rate  $\eta=0.1$ and SHL of 7.0s gave the best results on average, except with high horizon values close to $h=2.0s$, for which a higher learning rate $\eta=0.2$ and lower SHL of 5.0s led to better performance. The prediction error decreased as the number of hidden units increased. That fact has also been observed previously with RTRL in the context of respiratory motion forecasting \cite{POHL2021101941}. 

LSTM networks or gated recurrent units (GRUs) could be used instead of a vanilla RNN structure, as that could lead to higher prediction accuracy. Furthermore, UORO could be used to dynamically retrain in real-time the last hidden layer of a deep RNN that predicts respiratory waveform signals, as a form of transfer learning. This could improve the robustness of that RNN to unseen examples corresponding to irregular breathing patterns. Moreover, tumor position forecasting in lung radiotherapy will benefit from the development of new efficient online learning algorithms for deep RNNs. Using more data acquired from other institutions or synthesized via generative models \cite{pastor2021semi} may be beneficial to subsequent studies.

%

\section*{Acknowledgments}

We thank Prof. Masaki Sekino, Prof. Ichiro Sakuma, and Prof. Hitoshi Tabata (The University of Tokyo, Graduate School of Engineering) for their insightful comments that helped improve the quality of this research. We also thank Dr. Stephen Wells (Nikon) who proofread the article.

\section*{Ethical approval}

The authors did not perform experiments involving human participants or animals.

\section*{Funding} 

This research did not receive any specific grant from funding agencies in the public, commercial, or not-for-profit sectors.

\section*{Declaration of competing interests}

The authors declare that they have no conflict of interest.

\bibliographystyle{spbasic}      
\bibliography{bibliography}

\begin{thebibliography}{57}
\providecommand{\natexlab}[1]{#1}
\providecommand{\url}[1]{{#1}}
\providecommand{\urlprefix}{URL }
\expandafter\ifx\csname urlstyle\endcsname\relax
  \providecommand{\doi}[1]{DOI~\discretionary{}{}{}#1}\else
  \providecommand{\doi}{DOI~\discretionary{}{}{}\begingroup
  \urlstyle{rm}\Url}\fi
\providecommand{\eprint}[2][]{\url{#2}}

\bibitem[{Aicher et~al.(2020)Aicher, Foti, and Fox}]{aicher2020adaptively}
Aicher C, Foti NJ, Fox EB (2020) Adaptively truncating backpropagation through
  time to control gradient bias. In: Uncertainty in Artificial Intelligence,
  PMLR, pp 799--808

\bibitem[{Azizmohammadi et~al.(2019)Azizmohammadi, Martin, Miro, and
  Duong}]{azizmohammadi2019model}
Azizmohammadi F, Martin R, Miro J, Duong L (2019) Model-free cardiorespiratory
  motion prediction from {X}-ray angiography sequence with {LSTM} network. In:
  2019 41st Annual International Conference of the IEEE Engineering in Medicine
  and Biology Society (EMBC), IEEE, pp 7014--7018

\bibitem[{Benzing et~al.(2019)Benzing, Gauy, Mujika, Martinsson, and
  Steger}]{benzing2019optimal}
Benzing F, Gauy MM, Mujika A, Martinsson A, Steger A (2019) Optimal
  kronecker-sum approximation of real time recurrent learning. In:
  International Conference on Machine Learning, PMLR, pp 604--613

\bibitem[{Bohnstingl et~al.(2020)Bohnstingl, Wo{\'z}niak, Maass, Pantazi, and
  Eleftheriou}]{bohnstingl2020online}
Bohnstingl T, Wo{\'z}niak S, Maass W, Pantazi A, Eleftheriou E (2020) Online
  spatio-temporal learning in deep neural networks. arXiv preprint
  arXiv:200712723

\bibitem[{Chang et~al.(2021)Chang, Dang, Dai, Sun et~al.}]{chang2021real}
Chang P, Dang J, Dai J, Sun W, et~al. (2021) Real-time respiratory tumor motion
  prediction based on a temporal convolutional neural network: Prediction model
  development study. Journal of Medical Internet Research 23(8):e27235

\bibitem[{Choi et~al.(2014)Choi, Chang, Kim, Park, Song, and
  Kang}]{choi2014performance}
Choi S, Chang Y, Kim N, Park SH, Song SY, Kang HS (2014) Performance
  enhancement of respiratory tumor motion prediction using adaptive support
  vector regression: Comparison with adaptive neural network method.
  International journal of imaging systems and technology 24(1):8--15

\bibitem[{Ehrhardt et~al.(2013)Ehrhardt, Lorenz et~al.}]{ehrhardt20134d}
Ehrhardt J, Lorenz C, et~al. (2013) 4D modeling and estimation of respiratory
  motion for radiation therapy, vol~10. Springer

\bibitem[{Fan et~al.(2020)Fan, Yu, Zhao, and Yu}]{fan2020respiratory}
Fan Q, Yu X, Zhao Y, Yu S (2020) A respiratory motion prediction method based
  on improved relevance vector machine. Mobile Networks and Applications
  25(6):2270--2279

\bibitem[{Goodband et~al.(2008)Goodband, Haas, and
  Mills}]{goodband2008comparison}
Goodband JH, Haas OC, Mills J (2008) A comparison of neural network approaches
  for on-line prediction in {IGRT}. Medical physics 35(3):1113--1122

\bibitem[{Jaderberg et~al.(2017)Jaderberg, Czarnecki, Osindero, Vinyals,
  Graves, Silver, and Kavukcuoglu}]{jaderberg2017decoupled}
Jaderberg M, Czarnecki WM, Osindero S, Vinyals O, Graves A, Silver D,
  Kavukcuoglu K (2017) Decoupled neural interfaces using synthetic gradients.
  In: International Conference on Machine Learning, PMLR, pp 1627--1635

\bibitem[{Jaeger(2002)}]{jaeger2002tutorial}
Jaeger H (2002) Tutorial on training recurrent neural networks, covering BPPT,
  RTRL, EKF and the "echo state network" approach, vol~5. GMD-Forschungszentrum
  Informationstechnik Bonn

\bibitem[{Jiang et~al.(2019)Jiang, Fujii, and Shiinoki}]{jiang2019prediction}
Jiang K, Fujii F, Shiinoki T (2019) Prediction of lung tumor motion using
  nonlinear autoregressive model with exogenous input. Physics in Medicine \&
  Biology 64(21):21NT02

\bibitem[{J{\"o}hl et~al.(2020)J{\"o}hl, Ehrbar, Guckenberger, Kl{\"o}ck,
  Meboldt, Zeilinger, Tanadini-Lang, and Schmid~Daners}]{johl2020performance}
J{\"o}hl A, Ehrbar S, Guckenberger M, Kl{\"o}ck S, Meboldt M, Zeilinger M,
  Tanadini-Lang S, Schmid~Daners M (2020) Performance comparison of prediction
  filters for respiratory motion tracking in radiotherapy. Medical physics
  47(2):643--650

\bibitem[{Kai et~al.(2018)Kai, Fujii, and Shiinoki}]{kai2018prediction}
Kai J, Fujii F, Shiinoki T (2018) Prediction of lung tumor motion based on
  recurrent neural network. In: 2018 IEEE International Conference on
  Mechatronics and Automation (ICMA), IEEE, pp 1093--1099

\bibitem[{Khankan et~al.(2017)Khankan, Althaqfi
  et~al.}]{khankan2017demystifying}
Khankan A, Althaqfi S, et~al. (2017) Demystifying {C}yberknife stereotactic
  body radiation therapy for interventional radiologists. The Arab Journal of
  Interventional Radiology 1(2):55

\bibitem[{Krauss et~al.(2011)Krauss, Nill, and Oelfke}]{krauss2011comparative}
Krauss A, Nill S, Oelfke U (2011) The comparative performance of four
  respiratory motion predictors for real-time tumour tracking. Physics in
  Medicine \& Biology 56(16):5303

\bibitem[{Krilavicius et~al.(2015)Krilavicius, Zliobaite, Simonavicius, and
  Jarusevicius}]{krilavicius2015predicting}
Krilavicius T, Zliobaite I, Simonavicius H, Jarusevicius L (2015) Predicting
  respiratory motion for real-time tumour tracking in radiotherapy.
  \eprint{1508.00749}

\bibitem[{Krilavicius et~al.(2016)Krilavicius, Zliobaite, Simonavicius, and
  Jaruevicius}]{krilavicius2016predicting}
Krilavicius T, Zliobaite I, Simonavicius H, Jaruevicius L (2016) Predicting
  respiratory motion for real-time tumour tracking in radiotherapy. In: 2016
  IEEE 29th International Symposium on Computer-Based Medical Systems (CBMS),
  IEEE, pp 7--12

\bibitem[{Lee et~al.(2021)Lee, Cho, Lee, Jeong, Kwak, Jung, Kim, Yoon, Song,
  Lee et~al.}]{lee2021geometric}
Lee M, Cho MS, Lee H, Jeong C, Kwak J, Jung J, Kim SS, Yoon SM, Song SY, Lee
  Sw, et~al. (2021) Geometric and dosimetric verification of a recurrent neural
  network algorithm to compensate for respiratory motion using an articulated
  robotic couch. Journal of the Korean Physical Society 78(1):64--72

\bibitem[{Lee and Motai(2014)}]{lee2014prediction}
Lee SJ, Motai Y (2014) Prediction and classification of respiratory motion.
  Springer

\bibitem[{Lee et~al.(2011)Lee, Motai, and Murphy}]{lee2011respiratory}
Lee SJ, Motai Y, Murphy M (2011) Respiratory motion estimation with hybrid
  implementation of extended {Kalman} filter. IEEE Transactions on Industrial
  Electronics 59(11):4421--4432

\bibitem[{Lee et~al.(2013)Lee, Motai, Weiss, and Sun}]{lee2013customized}
Lee SJ, Motai Y, Weiss E, Sun SS (2013) Customized prediction of respiratory
  motion with clustering from multiple patient interaction. ACM Transactions on
  Intelligent Systems and Technology (TIST) 4(4):1--17

\bibitem[{Lin et~al.(2019)Lin, Shi, Wang, Chan, Tang, and Ji}]{lin2019towards}
Lin H, Shi C, Wang B, Chan MF, Tang X, Ji W (2019) Towards real-time
  respiratory motion prediction based on long short-term memory neural
  networks. Physics in Medicine \& Biology 64(8):085010

\bibitem[{Mafi and Moghadam(2020)}]{mafi2020real}
Mafi M, Moghadam SM (2020) Real-time prediction of tumor motion using a dynamic
  neural network. Medical \& biological engineering \& computing 58(3):529--539

\bibitem[{Marschall et~al.(2020)Marschall, Cho, and
  Savin}]{marschall2020unified}
Marschall O, Cho K, Savin C (2020) A unified framework of online learning
  algorithms for training recurrent neural networks. Journal of Machine
  Learning Research 21(135):1--34

\bibitem[{Mass{\'e} and Ollivier(2020)}]{masse2020convergence}
Mass{\'e} PY, Ollivier Y (2020) Convergence of online adaptive and recurrent
  optimization algorithms. arXiv preprint arXiv:200505645

\bibitem[{McClelland et~al.(2013)McClelland, Hawkes, Schaeffter, and
  King}]{mcclelland2013respiratory}
McClelland JR, Hawkes DJ, Schaeffter T, King AP (2013) Respiratory motion
  models: a review. Medical image analysis 17(1):19--42

\bibitem[{Menick et~al.(2020)Menick, Elsen, Evci, Osindero, Simonyan, and
  Graves}]{menick2020practical}
Menick J, Elsen E, Evci U, Osindero S, Simonyan K, Graves A (2020) A practical
  sparse approximation for real time recurrent learning. arXiv preprint
  arXiv:200607232

\bibitem[{Michel(2022)}]{pohl_michel_2021_5506965}
Michel P (2022) {Time series forecasting with UORO, RTRL, LMS, and linear
  regression: Fourth release}. \doi{10.5281/zenodo.5506964},
  \urlprefix\url{https://doi.org/10.5281/zenodo.5506964}

\bibitem[{Mujika et~al.(2018)Mujika, Meier, and
  Steger}]{mujika2018approximating}
Mujika A, Meier F, Steger A (2018) Approximating real-time recurrent learning
  with random kronecker factors. arXiv preprint arXiv:180510842

\bibitem[{Murphy and Pokhrel(2009)}]{murphy2009optimization}
Murphy MJ, Pokhrel D (2009) Optimization of an adaptive neural network to
  predict breathing. Medical physics 36(1):40--47

\bibitem[{Murray(2019)}]{murray2019local}
Murray JM (2019) Local online learning in recurrent networks with random
  feedback. ELife 8:e43299

\bibitem[{Nabavi et~al.(2020)Nabavi, Abdoos, Moghaddam, and
  Mohammadi}]{nabavi2020respiratory}
Nabavi S, Abdoos M, Moghaddam ME, Mohammadi M (2020) Respiratory motion
  prediction using deep convolutional long short-term memory network. Journal
  of Medical Signals and Sensors 10(2):69

\bibitem[{{National Cancer Institute - Surveillance, Epidemiology and End
  Results Program}(2021)}]{NIC2020LungBronchusCancer}
{National Cancer Institute - Surveillance, Epidemiology and End Results
  Program} (2021) Cancer stat facts: Lung and bronchus cancer.
  \url{https://seer.cancer.gov/statfacts/html/lungb.html}, [Online; accessed
  26-April-2021]

\bibitem[{Ollivier et~al.(2015)Ollivier, Tallec, and
  Charpiat}]{ollivier2015training}
Ollivier Y, Tallec C, Charpiat G (2015) Training recurrent networks online
  without backtracking. \eprint{1507.07680}

\bibitem[{Pascanu et~al.(2013)Pascanu, Mikolov, and
  Bengio}]{pascanu2013difficulty}
Pascanu R, Mikolov T, Bengio Y (2013) On the difficulty of training recurrent
  neural networks. In: International conference on machine learning, pp
  1310--1318

\bibitem[{Pastor-Serrano et~al.(2021)Pastor-Serrano, Lathouwers, and
  Perk{\'o}}]{pastor2021semi}
Pastor-Serrano O, Lathouwers D, Perk{\'o} Z (2021) A semi-supervised
  autoencoder framework for joint generation and classification of breathing.
  Computer Methods and Programs in Biomedicine 209:106312

\bibitem[{Pohl et~al.(2021)Pohl, Uesaka, Demachi, and
  Chhatkuli}]{POHL2021101941}
Pohl M, Uesaka M, Demachi K, Chhatkuli RB (2021) Prediction of the motion of
  chest internal points using a recurrent neural network trained with real-time
  recurrent learning for latency compensation in lung cancer radiotherapy.
  Computerized Medical Imaging and Graphics p 101941,
  \urlprefix\url{https://doi.org/10.1016/j.compmedimag.2021.101941}

\bibitem[{Remy et~al.(2021)Remy, Ahumada, Labine, C{\^o}t{\'e}, Lachaine, and
  Bouchard}]{remy2021potential}
Remy C, Ahumada D, Labine A, C{\^o}t{\'e} JC, Lachaine M, Bouchard H (2021)
  Potential of a probabilistic framework for target prediction from surrogate
  respiratory motion during lung radiotherapy. Physics in Medicine \& Biology
  66(10):105002

\bibitem[{Romaguera et~al.(2020)Romaguera, Plantef{\`e}ve, Romero, H{\'e}bert,
  Carrier, and Kadoury}]{romaguera2020prediction}
Romaguera LV, Plantef{\`e}ve R, Romero FP, H{\'e}bert F, Carrier JF, Kadoury S
  (2020) Prediction of in-plane organ deformation during free-breathing
  radiotherapy via discriminative spatial transformer networks. Medical image
  analysis 64:101754

\bibitem[{Roth et~al.(2018)Roth, Kanitscheider, and Fiete}]{roth2018kernel}
Roth C, Kanitscheider I, Fiete I (2018) Kernel {RNN} learning ({KeRNL}). In:
  International Conference on Learning Representations

\bibitem[{Salehinejad et~al.(2017)Salehinejad, Sankar, Barfett, Colak, and
  Valaee}]{salehinejad2017recent}
Salehinejad H, Sankar S, Barfett J, Colak E, Valaee S (2017) Recent advances in
  recurrent neural networks. arXiv preprint arXiv:180101078

\bibitem[{Sarudis et~al.(2017)Sarudis, Karlsson~Hauer, Nyman, and
  B{\"a}ck}]{sarudis2017systematic}
Sarudis S, Karlsson~Hauer A, Nyman J, B{\"a}ck A (2017) Systematic evaluation
  of lung tumor motion using four-dimensional computed tomography. Acta
  Oncologica 56(4):525--530

\bibitem[{Schweikard et~al.(2004)Schweikard, Shiomi, and
  Adler}]{schweikard2004respiration}
Schweikard A, Shiomi H, Adler J (2004) Respiration tracking in radiosurgery.
  Medical physics 31(10):2738--2741

\bibitem[{Sharp et~al.(2004)Sharp, Jiang, Shimizu, and
  Shirato}]{sharp2004prediction}
Sharp GC, Jiang SB, Shimizu S, Shirato H (2004) Prediction of respiratory
  tumour motion for real-time image-guided radiotherapy. Physics in Medicine \&
  Biology 49(3):425

\bibitem[{Sun et~al.(2017)Sun, Jiang, Ren, Dang, You, and
  Yin}]{sun2017respiratory}
Sun W, Jiang M, Ren L, Dang J, You T, Yin F (2017) Respiratory signal
  prediction based on adaptive boosting and multi-layer perceptron neural
  network. Physics in Medicine \& Biology 62(17):6822

\bibitem[{Takao et~al.(2016)Takao, Miyamoto, Matsuura, Onimaru, Katoh, Inoue,
  Sutherland, Suzuki, Shirato, and Shimizu}]{takao2016intrafractional}
Takao S, Miyamoto N, Matsuura T, Onimaru R, Katoh N, Inoue T, Sutherland KL,
  Suzuki R, Shirato H, Shimizu S (2016) Intrafractional baseline shift or drift
  of lung tumor motion during gated radiation therapy with a real-time
  tumor-tracking system. International Journal of Radiation Oncology* Biology*
  Physics 94(1):172--180

\bibitem[{Tallec and Ollivier(2017{\natexlab{a}})}]{tallec2017unbiased}
Tallec C, Ollivier Y (2017{\natexlab{a}}) Unbiased online recurrent
  optimization. arXiv preprint arXiv:170205043

\bibitem[{Tallec and Ollivier(2017{\natexlab{b}})}]{tallec2017unbiasing}
Tallec C, Ollivier Y (2017{\natexlab{b}}) Unbiasing truncated backpropagation
  through time. \eprint{1705.08209}

\bibitem[{Teo et~al.(2018)Teo, Ahmed, Kawalec, Alayoubi, Bruce, Lyn, and
  Pistorius}]{teo2018feasibility}
Teo TP, Ahmed SB, Kawalec P, Alayoubi N, Bruce N, Lyn E, Pistorius S (2018)
  Feasibility of predicting tumor motion using online data acquired during
  treatment and a generalized neural network optimized with offline patient
  tumor trajectories. Medical physics 45(2):830--845

\bibitem[{Verma et~al.(2010)Verma, Wu, Langer, Das, and
  Sandison}]{verma2010survey}
Verma P, Wu H, Langer M, Das I, Sandison G (2010) Survey: real-time tumor
  motion prediction for image-guided radiation treatment. Computing in Science
  \& Engineering 13(5):24--35

\bibitem[{Wang et~al.(2021)Wang, Li, Li, Dai, Xiao, Bai, He, Liu, and
  Bai}]{wang2021real}
Wang G, Li Z, Li G, Dai G, Xiao Q, Bai L, He Y, Liu Y, Bai S (2021) Real-time
  liver tracking algorithm based on {LSTM} and {SVR} networks for use in
  surface-guided radiation therapy. Radiation Oncology 16(1):1--12

\bibitem[{Wang et~al.(2018)Wang, Liang, Zhu, and Xie}]{wang2018feasibility}
Wang R, Liang X, Zhu X, Xie Y (2018) A feasibility of respiration prediction
  based on deep {Bi-LSTM} for real-time tumor tracking. IEEE Access
  6:51262--51268

\bibitem[{Wang et~al.(2020)Wang, Yu, Sivanagaraja, and Veluvolu}]{wang2020fast}
Wang Y, Yu Z, Sivanagaraja T, Veluvolu KC (2020) Fast and accurate online
  sequential learning of respiratory motion with random convolution nodes for
  radiotherapy applications. Applied Soft Computing 95:106528

\bibitem[{Williams and Zipser(1989)}]{williams1989learning}
Williams RJ, Zipser D (1989) A learning algorithm for continually running fully
  recurrent neural networks. Neural computation 1(2):270--280

\bibitem[{Yu et~al.(2020)Yu, Wang, Liu, Sun, Kuang, and Sun}]{yu2020rapid}
Yu S, Wang J, Liu J, Sun R, Kuang S, Sun L (2020) Rapid prediction of
  respiratory motion based on bidirectional gated recurrent unit network. IEEE
  Access 8:49424--49435

\bibitem[{Yun et~al.(2019)Yun, Rathee, and Fallone}]{yun2019deep}
Yun J, Rathee S, Fallone B (2019) A deep-learning based {3D} tumor motion
  prediction algorithm for non-invasive intra-fractional tumor-tracked
  radiotherapy (nifte{RT}) on {Linac-MR}. International Journal of Radiation
  Oncology, Biology, Physics 105(1):S28

\end{thebibliography}

\appendix

\section{Appendix : Notes on the derivation of UORO for standard RNNs}%
\label{appendix:UORO_alg}

The derivation of UORO for RNNs in the general case is described in \citep{tallec2017unbiased}. In this section, we provide details concerning the calculation of several quantities involved in the computation of the loss gradient $\nabla_{\theta} L_n$ in the case of vanilla RNNs defined by Eqs. \ref{eq:state_vanilla} and \ref{eq:measurement_vanilla}.

\subsection{Calculation of $\nabla_x L_{n+1}$}
\label{Calculation of nabla_x L_n+1}

In this section we compute the quantity \footnotemark:
\begin{equation} \label{eq:delta_x definition}
\frac{\partial L_{n+1}}{\partial x} = \frac{\partial L_{n+1}}{\partial y} \frac{\partial F_{out}}{\partial x}
\end{equation}

\footnotetext{
Note: with the notations of Tallec et al., Eq. \ref{eq:delta_x definition} can be rewritten as:
$\delta s = \frac{\partial l_{t+1}}{\partial o} \frac{\partial F_{out}}{\partial s}$
}

We recall that $e_n = y_n^* - y_n$ where $y_n^*$ is the exact signal and $y_n$ the predicted signal. We compute first the left factor:

\begin{align}
\frac{\partial L_{n+1}}{\partial y} &= \frac{\partial}{\partial y} \left[ \frac{1}{2} e_{n+1}^T e_{n+1}  \right] \\
&= \frac{\partial}{\partial e_{n+1}} \left[ \frac{1}{2} e_{n+1}^T e_{n+1} \right] \frac{\partial e_{n+1}}{\partial y} \\
&= e_{n+1}^T (-I_p) \\
&= - e_{n+1}^T \label{eq:UORO aux 1}
\end{align}

Furthermore, is straightforward that:
\begin{equation}
\frac{\partial F_{out}}{\partial x} = W_{c,n} \label{eq:UORO aux 2}
\end{equation}

By combining Eqs. \ref{eq:UORO aux 1} and \ref{eq:UORO aux 2}, we can rewrite Eq. \ref{eq:delta_x definition} as:
\begin{equation} \label{eq:delta_x computation result}
\frac{\partial L_{n+1}}{\partial x} = - e_{n+1}^T W_{c,n}
\end{equation}

We have proven the formula for $\nabla_x L_{n+1}$ appearing in line 19 of Algorithm \ref{alg:RNN-UORO}.

\subsection{Calculation of $\delta \theta$}
\label{section: calculation of delta theta}

In this section, we compute the quantity:
\begin{equation} \label{eq:delta theta def}
\delta \theta = \frac{\partial L_{n+1}}{\partial y} \frac{\partial F_{out}}{\partial \theta}
\end{equation} 

The left factor has already been computed previously (Eq. \ref{eq:UORO aux 1}). What remains to compute is the right factor. We write the parameter (line) vector as:

\begin{equation}
\theta_n = [W_{a,n}^{unrolled}, W_{b,n}^{unrolled}, W_{c,n}^{unrolled}]
\end{equation}

where $W_{a,n}^{unrolled}$ (resp. $W_{b,n}^{unrolled}$, $W_{c,n}^{unrolled}$) is a line vector containing the elements of $W_{a,n}$ (resp. $W_{b,n}$, $W_{c,n}$). We thus have:

\begin{align}
\delta \theta &= - e_{n+1}^T \left[ 0_{1 \times (|\theta|-pq)}, \frac{\partial F_{out}}{\partial W_{c,n}^{unrolled}} \right] \\
&= \left[ 0_{1 \times (|\theta|-pq)}, - e_{n+1}^T \frac{\partial F_{out}}{\partial W_{c,n}^{unrolled}} \right] \label{eq: delta theta aux}
\end{align}

We need to calculate the quantity $e_{n+1}^T (\partial F_{out} / \partial W_{c,n}^{unrolled})$. We have $F_{out}(x_{n+1}, \theta_n) = y_{n+1} = W_{c,n} x_{n+1}$ (Eqs. \ref{eq:RNN_general_eqs} and \ref{eq:measurement_vanilla}), so the $k^{th}$ component of $F_{out}(x_{n+1}, \theta_n)$ is simply calculated as:

\begin{equation}
y_{n+1,k} = \sum_{l=1}^q W_{c,n}^{k,l} x_{n+1,l}
\end{equation}

Thus, for $(i,j) \in [\![ 1, ..., p ]\!] \times [\![ 1, ..., q ]\!]$, we have:
\begin{equation} \label{eq:delta theta calculation aux 1}     
	\frac{\partial y_{n+1,k}}{\partial W_{c,n}^{i,j}} = 
    \begin{cases}
      x_{n+1,j} & \text{if $k=i$}\\
      0 & \text{otherwise}
    \end{cases} 
\end{equation} 

Therefore:
\begin{align}
e_{n+1}^T \frac{\partial F_{out}}{\partial W_{c,n}^{i,j}} &= e_{n+1}^T  \begin{bmatrix} 
0 \\
\vdots \\
x_{n+1,j} \\
\vdots \\
0
\end{bmatrix}
\leftarrow i^{th} \mbox{ row} \\
&= e_{n+1, i} x_{n+1,j}
\end{align}

This can be rewritten as:
\begin{equation}
e_{n+1}^T \frac{\partial F_{out}}{\partial W_{c,n}} = e_{n+1} x_{n+1}^T 
\end{equation}

Therefore:
\begin{equation}
e_{n+1}^T \frac{\partial F_{out}}{\partial W_{c,n}^{unrolled}} = \text{reshape($e_{n+1} x_{n+1}^T$, $1 \times pq$)} 
\end{equation}

We plug this expression into Eq. \ref{eq: delta theta aux} to obtain finally: 
\begin{equation}\label{eq:delta theta calculation result}
\delta \theta = [ 0_{1 \times (|\theta|-pq)}, \text{reshape($- e_{n+1} x_{n+1}^T$, $1 \times pq$)}] 
\end{equation} 

This justifies the update of $\delta \theta$ in line 18 of Algorithm \ref{alg:RNN-UORO} \footnotemark.

\footnotetext{Note: with the notations of Tallec et al., Eqs. \ref{eq:delta theta def} and \ref{eq:delta theta calculation result} can be rewritten as:
$ \delta \theta = (\partial l_{t+1} / \partial o) (\partial F_{out} / \partial \theta) = [ 0_{1 \times (|\theta|-pq)}, \text{reshape($e_{t+1} s_{t+1}^T$, $1 \times pq$)}] $}

\subsection{Calculation of $\delta \theta_g$}

In this section, we detail the calculation of the following quantity:

\begin{equation}
\delta \theta_g = \nu ^T \frac{\partial F_{st}}{\partial \theta}
\end{equation}

In this equation, $\nu$ represents a column vector of size $q$ with random values in $\{ -1, 1\}$ (cf line 20 of Algorithm \ref{alg:RNN-UORO}). $F_{st}(x_n, u_n, \theta_n)$ does not depend on the quantity $W_{c,n}$ so we can write:

\begin{align}
\delta \theta_g &= \nu^T \left[ \frac{\partial F_{st}}{\partial W_{a,n}^{unrolled}}, \frac{\partial F_{st}}{\partial W_{b,n}^{unrolled}}, 0_{1 \times pq} \right] \\
&= \left[ \nu^T \frac{\partial F_{st}}{\partial W_{a,n}^{unrolled}}, \nu^T \frac{\partial F_{st}}{\partial W_{b,n}^{unrolled}}, 0_{1 \times pq} \right] \label{eq:delta theta_g aux 3}
\end{align} 

We focus first on the calculation of the component $\nu^T \partial F_{st}/\allowbreak\partial W_{a,n}^{unrolled}$. We recall that $\Phi$ is the non-linear activation function defined in Eq. \ref{eq:non_linearity} and we define $z_n$ as the following column vector:
\begin{equation} \label{eq:z_n definition}
z_n = W_{a,n} x_n + W_{b,n} u_n
\end{equation} 

The state equation can be rewritten as:
\begin{equation} \label{eq:state equation rewritten}
F_{st}(x_n, u_n, \theta_n) = \Phi(z_n)
\end{equation}

We select $(i,j) \in [\![ 1, ..., q ]\!]^2$. We deduce from the previous equation that:
\begin{equation} \label{eq:delta theta_g aux 2}
\frac{\partial F_{st}}{\partial W_{a,n}^{i,j}} = \frac{\partial \Phi}{\partial z} \frac{\partial z_n}{\partial W_{a,n}^{i,j}}
\end{equation} 

The calculation of the left factor directly comes from the definition of $\Phi$ in Eq. \ref{eq:non_linearity}:
\begin{equation} \label{eq:partial Phi over partial z}
\frac{\partial \Phi}{\partial z} = 
\begin{bmatrix}
   \phi'(z_{n,1}) & & 0\\ 
   & \ddots & \\
  0 &  &  \phi'(z_{n,q}) 
 \end{bmatrix}
\end{equation}

The right factor can simply be calculated using an approach similar to that leading to Eq. \ref{eq:delta theta calculation aux 1}:
\begin{equation}
\frac{\partial z_n}{\partial W_{a,n}^{i,j}} = \begin{bmatrix} 
0 \\
\vdots \\
x_{n,j} \\
\vdots \\
0
\end{bmatrix}
\leftarrow i^{th} \mbox{ row}
\end{equation}

Eq. \ref{eq:delta theta_g aux 2} can thus be rewritten as:

\begin{align}
\frac{\partial F_{st}}{\partial W_{a,n}^{i,j}}
&= \begin{bmatrix}
   \phi'(z_{n,1}) &  &  & & 0\\ 
   & \ddots &  & & \\ 
   &  &  \ddots & & \\
   &  &         & \ddots & \\
  0 &  &   &  & \phi'(z_{n,q}) 
 \end{bmatrix} \begin{bmatrix} 
0 \\
\vdots \\
x_{n,j} \\
\vdots \\
0
\end{bmatrix}
\leftarrow i^{th} \mbox{ row} \\
&= \begin{bmatrix} 
0 \\
\vdots \\
\phi'(z_{n,i}) x_{n,j} \\
\vdots \\
0
\end{bmatrix}
\leftarrow i^{th} \mbox{ row} \label{eq:delta theta_g aux 1}
\end{align} 

We define for $j \in [\![ 1, ..., q ]\!]$ the following matrix:
\begin{equation} \label{eq: def partial F_st over partial W_a,n^j}
\frac{\partial F_{st}}{\partial W_{a,n}^j} = 
\left[ \frac{\partial F_{st}}{\partial W_{a,n}^{1,j}}, ..., \frac{\partial F_{st}}{\partial W_{a,n}^{q,j}} \right]
\end{equation}

Using Eq. \ref{eq:delta theta_g aux 1}, we can compute it the following way:
\begin{align}
\frac{\partial F_{st}}{\partial W_{a,n}^j} &= 
 \begin{bmatrix}
   \phi'(z_{n,1}) x_{n,j} & & 0\\ 
   & \ddots &  \\ 
  0 &  & \phi'(z_{n,q}) x_{n,j}
 \end{bmatrix} \\
 &= x_{n,j} \begin{bmatrix}
   \phi'(z_{n,1}) & & 0\\ 
   & \ddots &  \\ 
  0 &  & \phi'(z_{n,q})
 \end{bmatrix} \label{calculation partial F_st over partial W_a,n^j}
\end{align}

Therefore,
\begin{align}
\nu^T \frac{\partial F_{st}}{\partial W_{a,n}^j} &= \nu^T x_{n,j} \begin{bmatrix}
   \phi'(z_{n,1}) & & 0\\ 
   & \ddots &  \\ 
  0 &  & \phi'(z_{n,q})
 \end{bmatrix} \\
&= x_{n,j} [\nu_1 \phi'(z_{n,1}), ..., \nu_q \phi'(z_{n,q})] \\
&= x_{n,j} [\nu * \phi'(z_n)]^T 
\end{align}
where $*$ refers to element-wise multiplication.

We can finally compute the quantity $\nu^T \partial F_{st} / \partial W_{a,n}^{unrolled}$ appearing in Eq. \ref{eq:delta theta_g aux 3} as:
\begin{align}
\nu^T \frac{\partial F_{st}}{\partial W_{a,n}^{unrolled}}
&= \left[ \nu^T \frac{\partial F_{st}}{\partial W_{a,n}^1}, ..., \nu^T \frac{\partial F_{st}}{\partial W_{a,n}^q} \right] \\
&= [x_{n,1}(\nu * \phi'(z_n))^T, ..., x_{n,q}(\nu * \phi'(z_n))^T] \\
&= \text{reshape} \big( \big[ x_{n,1}(\nu * \phi'(z_n)), ..., \nonumber \\
& \phantomrel{=} \hphantom{\text{reshape} \big( \big[} x_{n,q}(\nu * \phi'(z_n)) \big], 1 \times q^2 \big) \\
&= \text{reshape} \left( (\nu * \phi'(z_n)) x_n^T, 1 \times q^2 \right) \label{eq:delta theta_g aux 4}
\end{align}

The reshaping operation, which was also used in Eq. \ref{eq:delta theta calculation result}, enables writing expressions with simple matrix operations that can be quickly performed with appropriate linear algebra libraries. In a similar way, we can compute the quantity $\nu^T \partial F_{st} / \partial W_{b,n}^{unrolled}$ as:
\begin{equation} \label{eq:delta theta_g aux 5}
\nu^T \frac{\partial F_{st}}{\partial W_{b,n}^{unrolled}} = \text{reshape} \left( (\nu * \phi'(z_n)) u_n^T, 1 \times q(m+1) \right)
\end{equation}


To summarize, we can compute the quantity $\delta \theta_g$ using Eqs. \ref{eq:delta theta_g aux 3}, \ref{eq:delta theta_g aux 4} and \ref{eq:delta theta_g aux 5}. The quantity $\delta \theta_g^{aux} = \nu * \phi'(z_n)$ appears both in Eqs. \ref{eq:delta theta_g aux 4} and \ref{eq:delta theta_g aux 5} so it can be computed beforehand to improve time performance. In this section, we justified the lines 22, 23, and 24 in Algorithm \ref{alg:RNN-UORO}.

\onecolumn


\section{Appendix : Predicted motion for sequence 5 (regular breathing)}%
\label{appendix:coordz_marker3_seq5}

\begin{figure}[hbt!]
	
    \centering
    \subfloat[\normalsize Prediction with an RNN trained with UORO]{{\includegraphics[width=.80\textwidth]{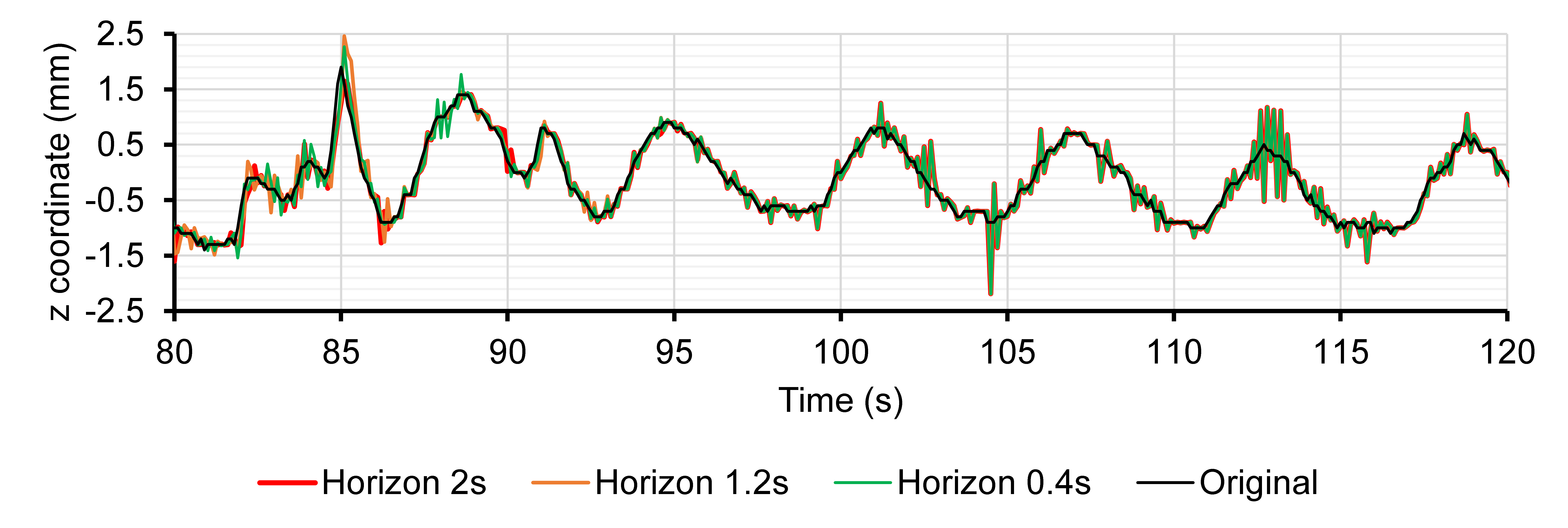} }}%
    \quad
    \subfloat[\normalsize Prediction with an RNN trained with RTRL]{{\includegraphics[width=.80\textwidth]{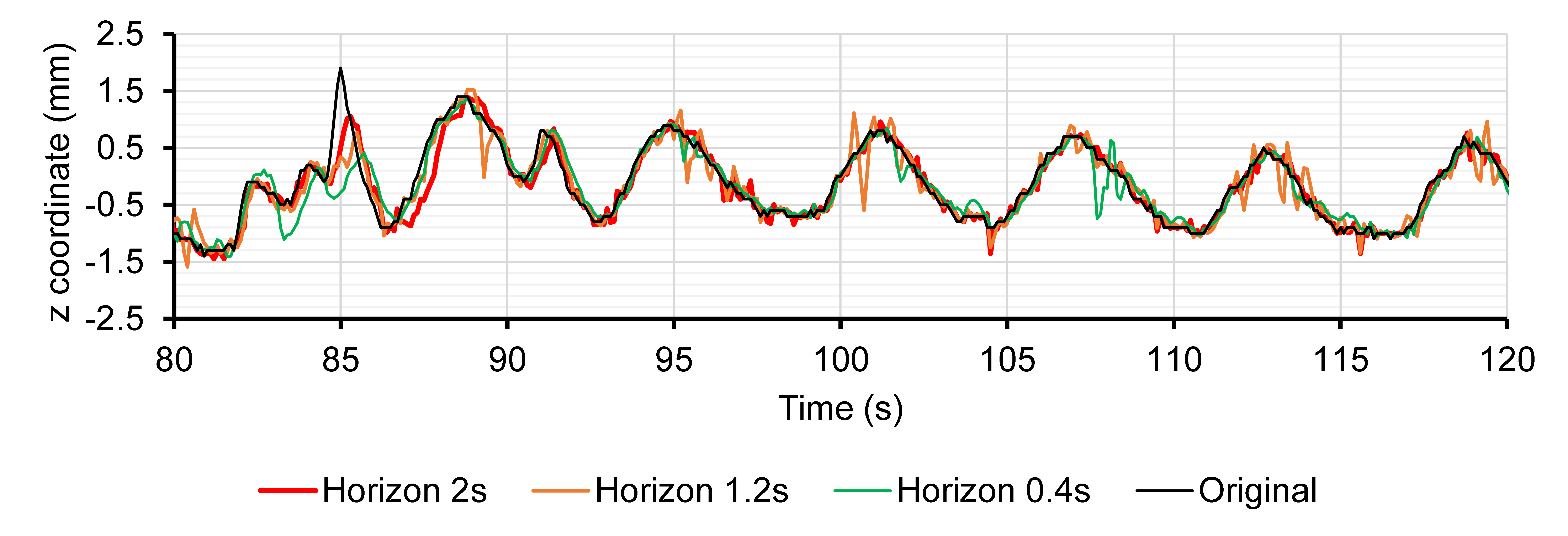} }}%
    \quad
    \subfloat[\normalsize Prediction with LMS]{{\includegraphics[width=.80\textwidth]{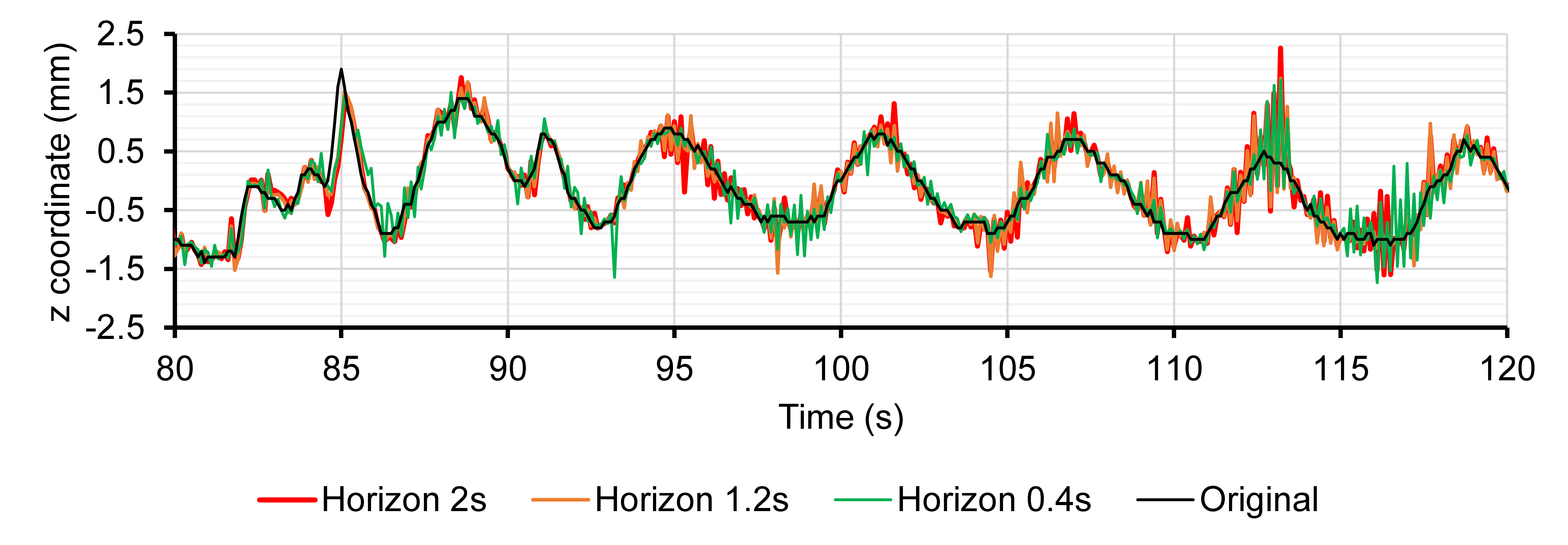} }}%
    \caption{Comparison between RTRL, UORO, and LMS regarding the prediction of the position of the z coordinate (spine axis) of marker 3 in sequence 5 (normal breathing)}%
    \label{fig:coordz_marker3_seq5}%
\end{figure}

\clearpage

\section{Appendix : Loss functions for sequence 1 (irregular breathing) and 5 (regular breathing)}%
\label{appendix:loss functions}

\begin{figure}[hbt!]
    \centering
    \subfloat[\normalsize Sequence 1 - UORO]{\includegraphics[width=.40\textwidth]{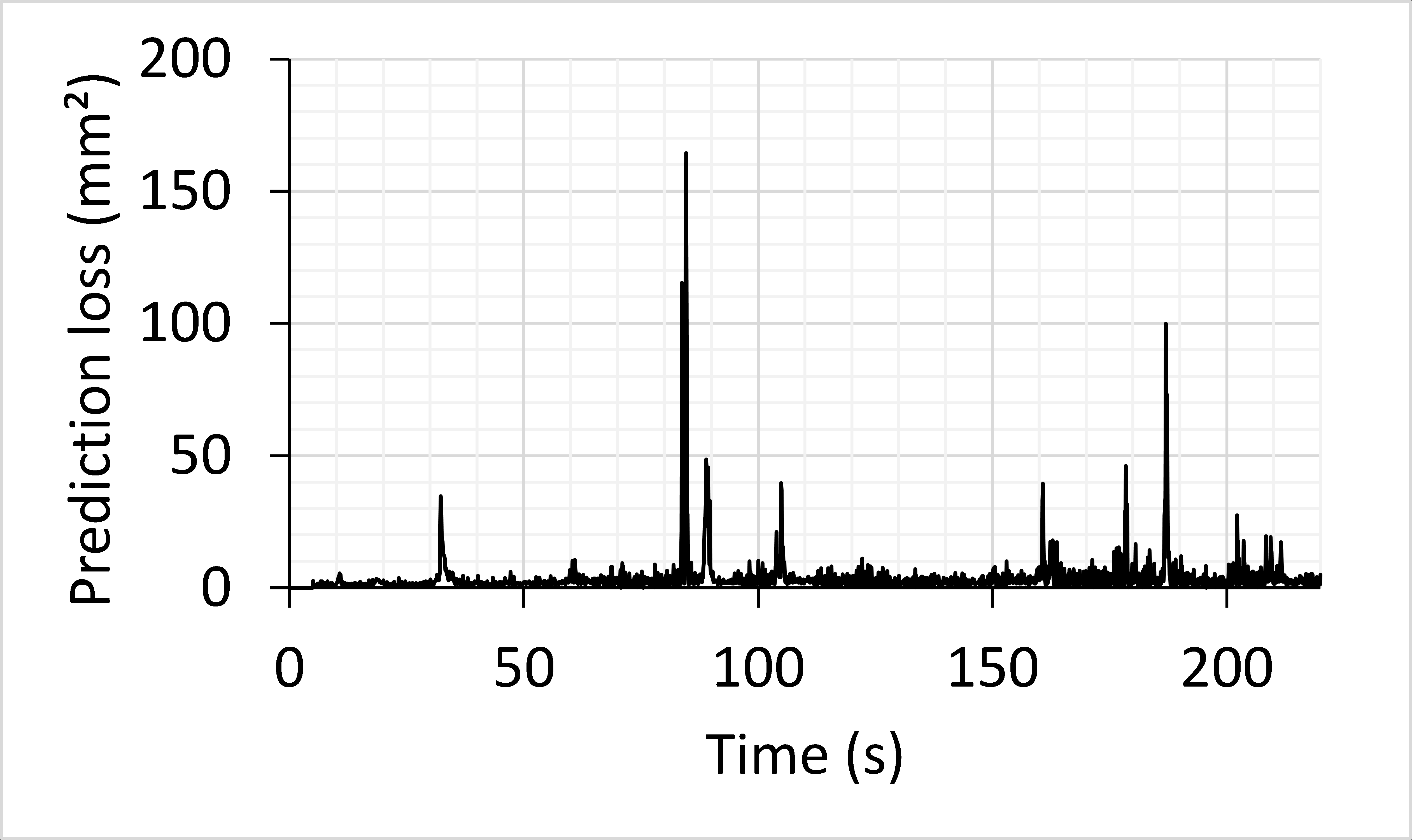} }%
    \quad
    \subfloat[\normalsize Sequence 5 - UORO]{\includegraphics[width=.40\textwidth]{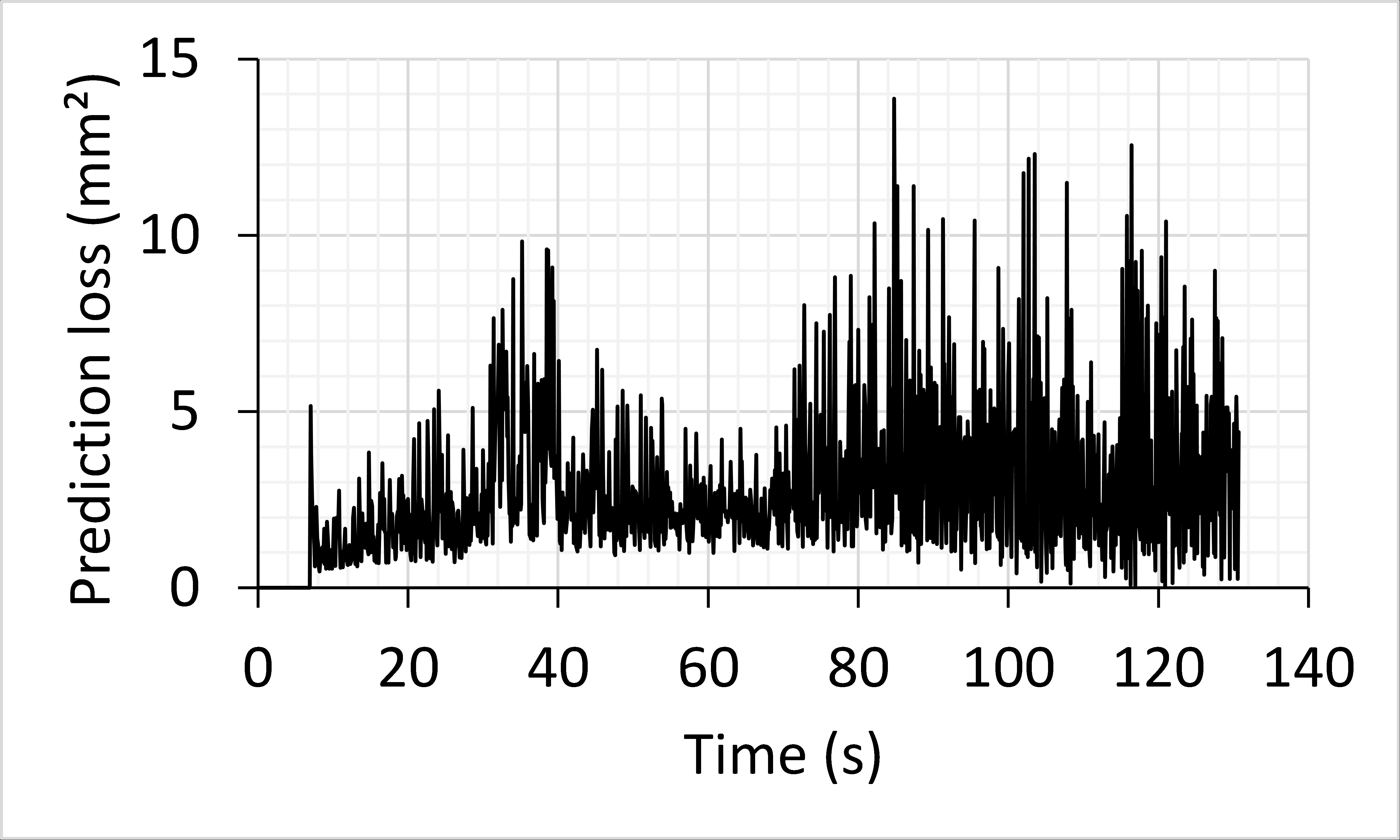} }%
    \quad
    \subfloat[\normalsize Sequence 1 - RTRL]{\includegraphics[width=.40\textwidth]{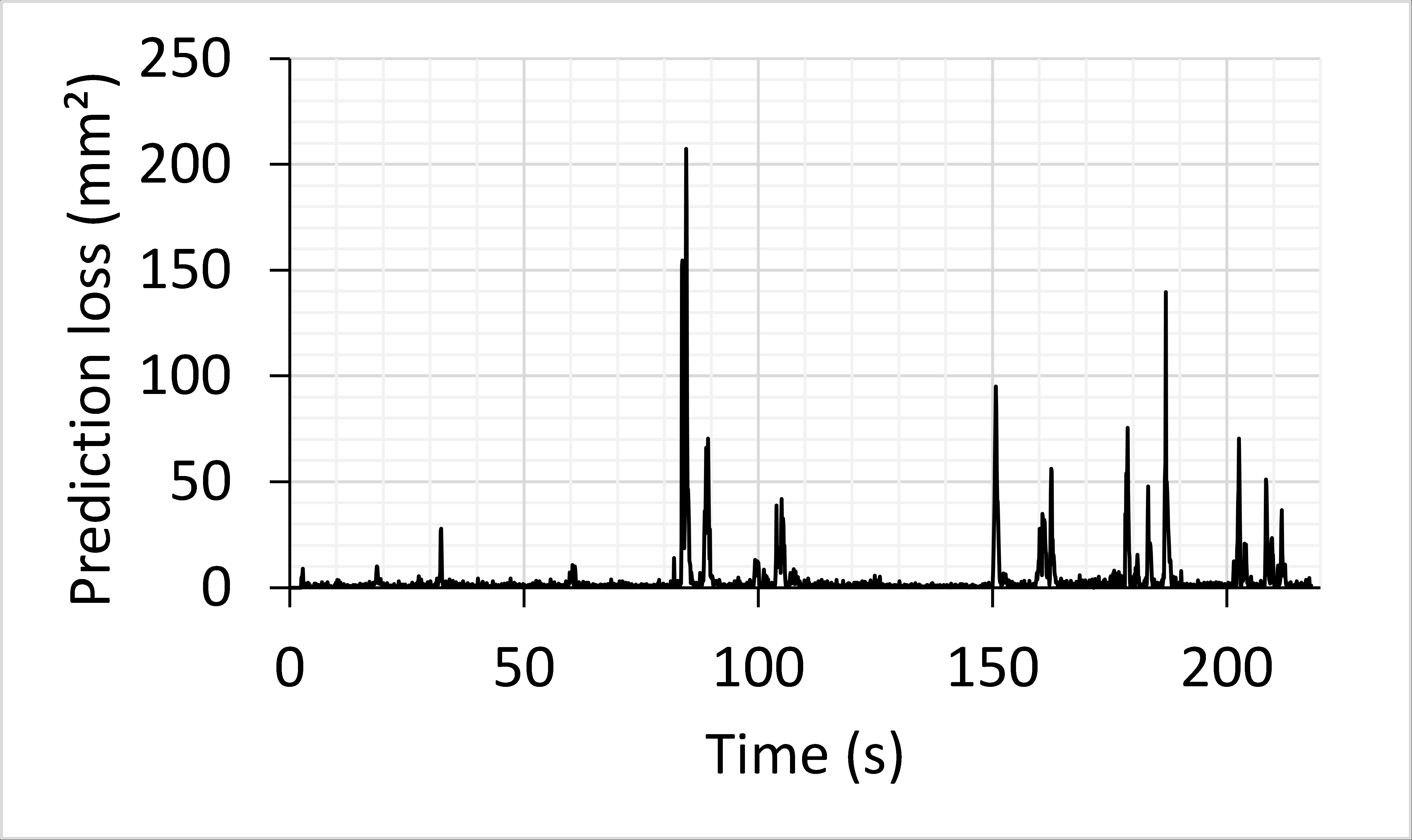} }%
    \quad
    \subfloat[\normalsize Sequence 5 - RTRL]{\includegraphics[width=.40\textwidth]{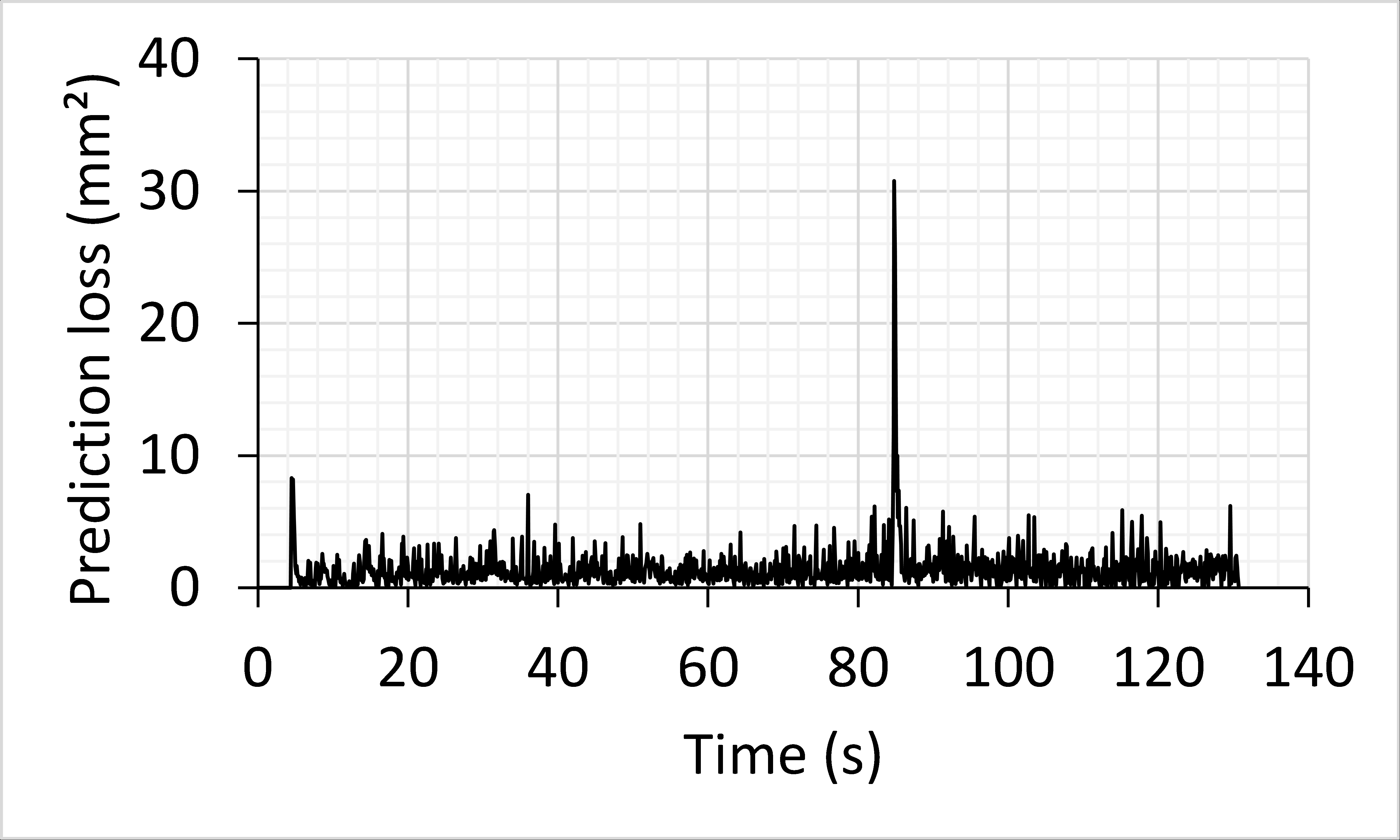} }%
    \quad
    \subfloat[\normalsize Sequence 1 - LMS]{\includegraphics[width=.40\textwidth]{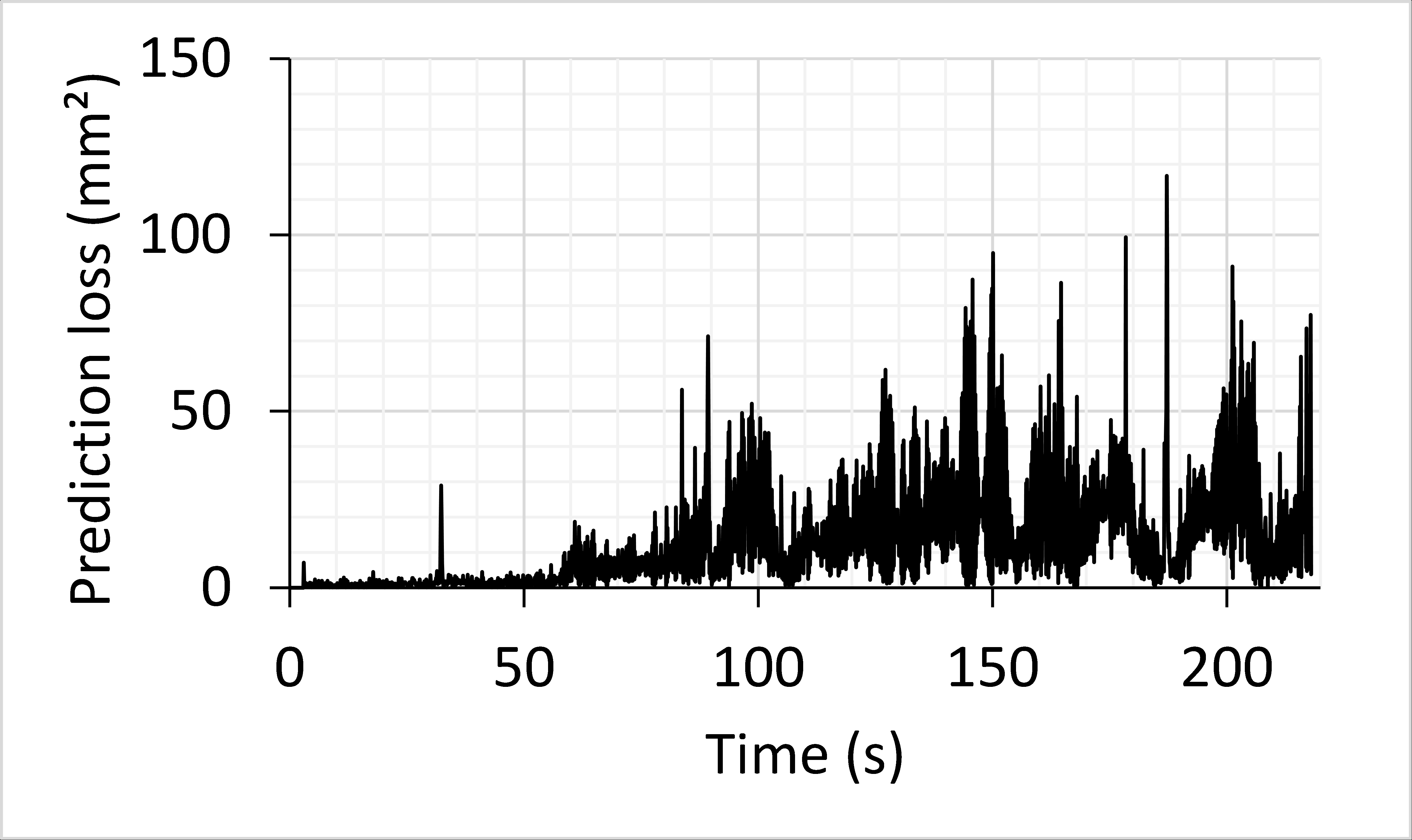} }%
    \quad
    \subfloat[\normalsize Sequence 5 - LMS]{\includegraphics[width=.40\textwidth]{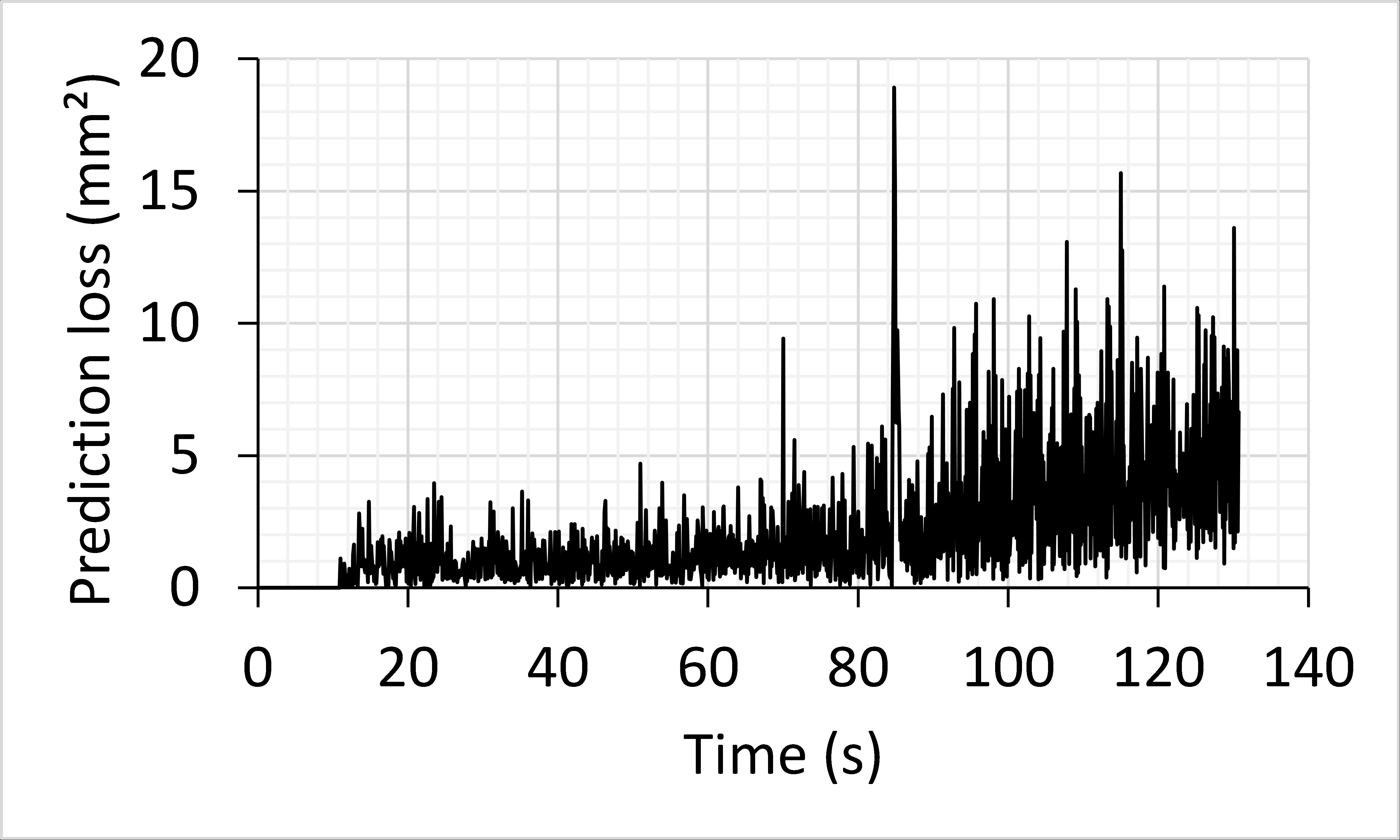} }%
    \quad                  
    \caption{Prediction instantaneous square loss (cf Eq. \ref{eq:loss function}) for sequence 1 (person talking) and sequence 5 (normal breathing). The horizon value is h=2.0s and the loss is averaged over 300 runs. }
    \label{fig:loss function}
\end{figure}

\clearpage

\section{Appendix : Prediction performance for regular and irregular breathing sequences}%
\label{appendix:pred perf}

\begin{figure*}[htb!]
    \centering
    \includegraphics[width=.35\textwidth]{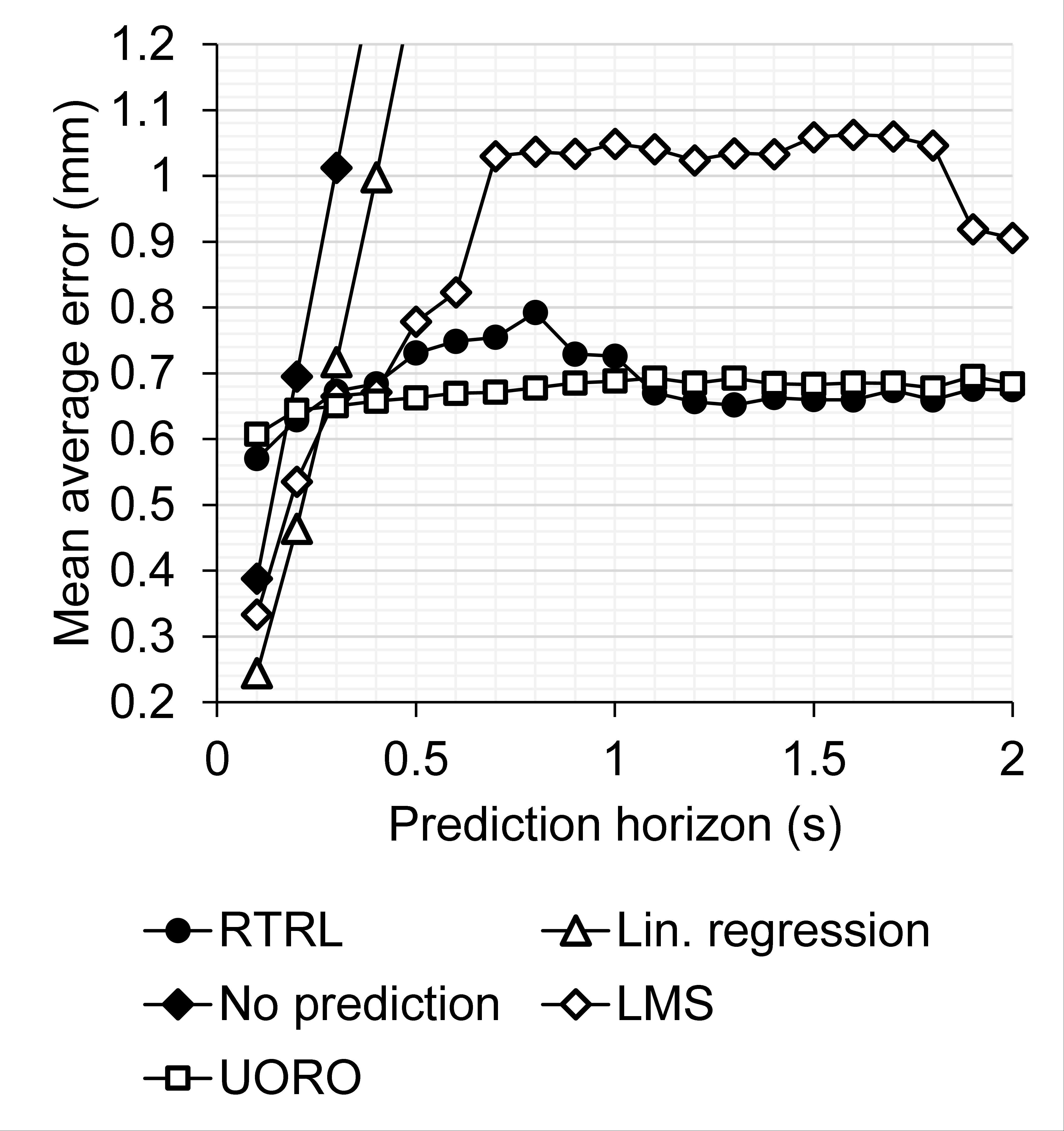}%
    \qquad
    \includegraphics[width=.35\textwidth]{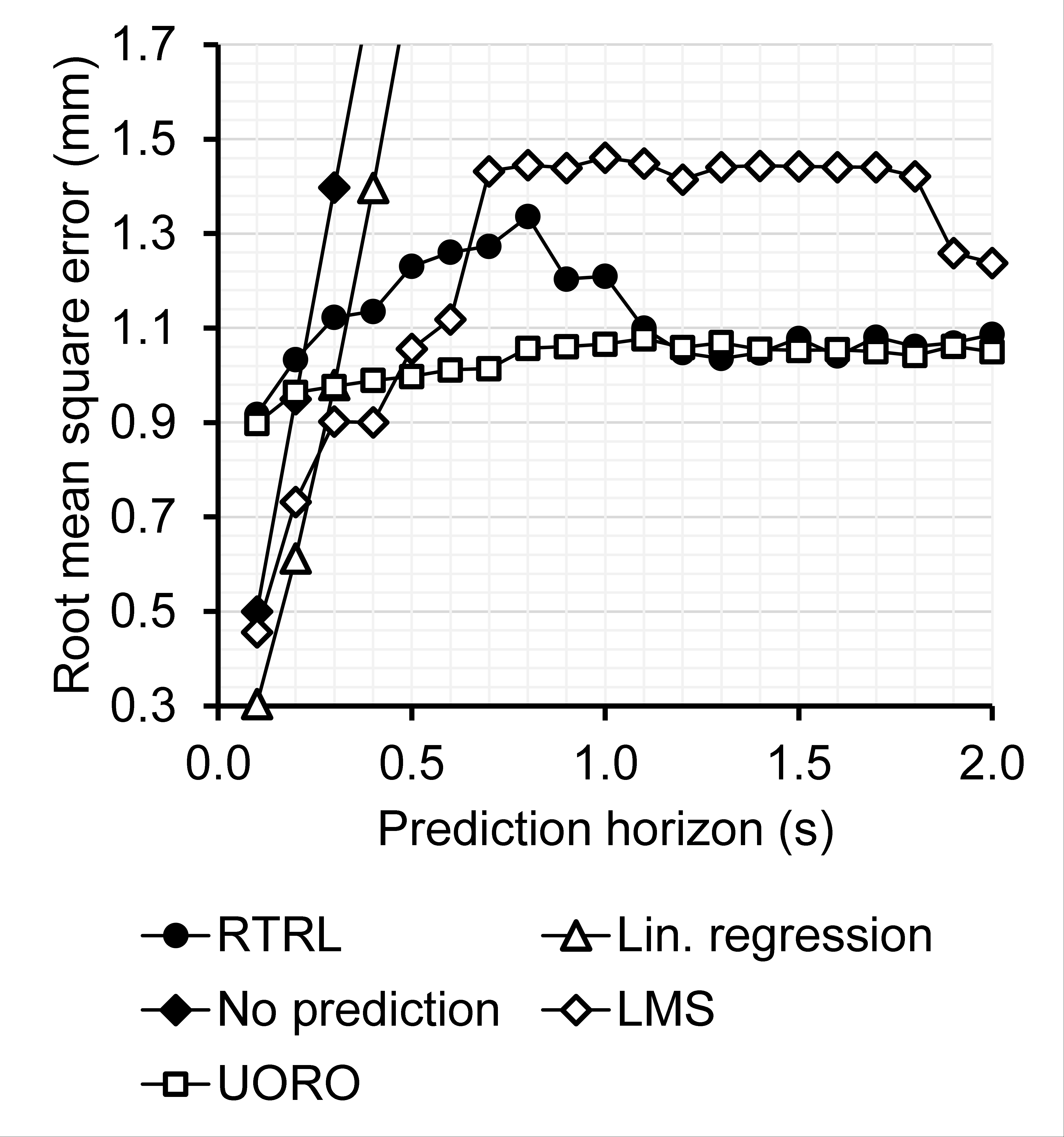}%
    \qquad
    \includegraphics[width=.35\textwidth]{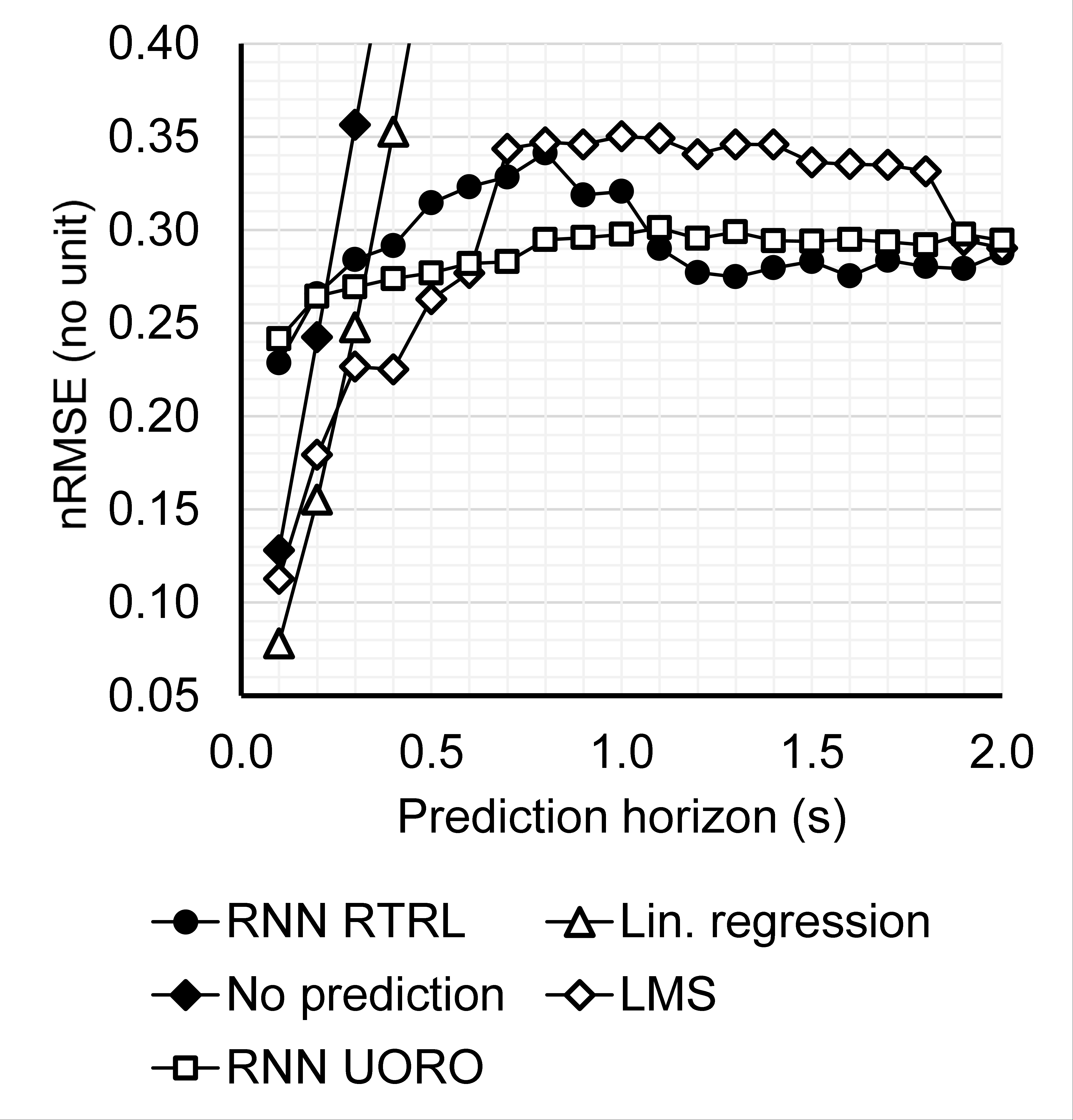}%
    \qquad
    \includegraphics[width=.35\textwidth]{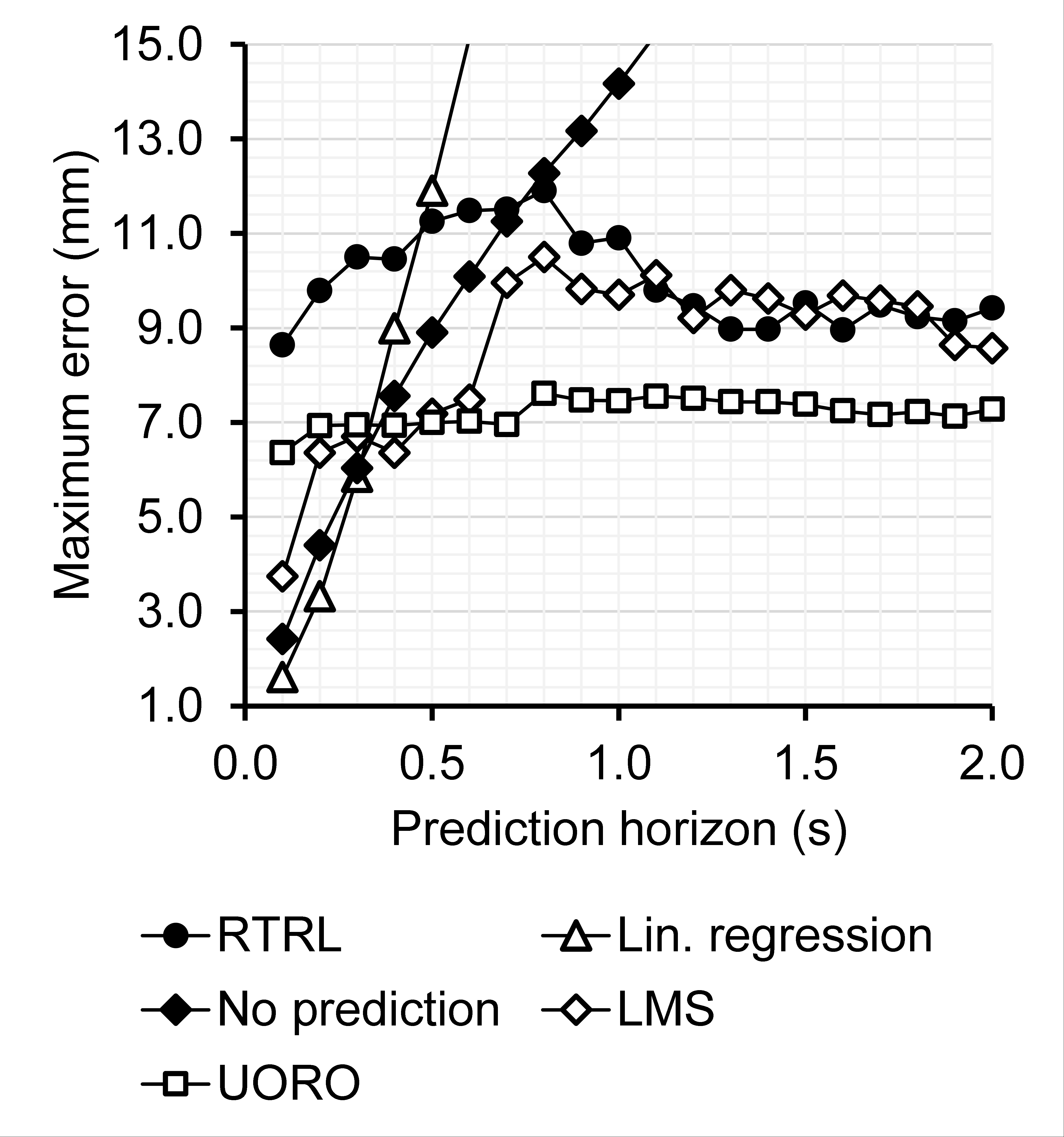}%
    \qquad
    \includegraphics[width=.35\textwidth]{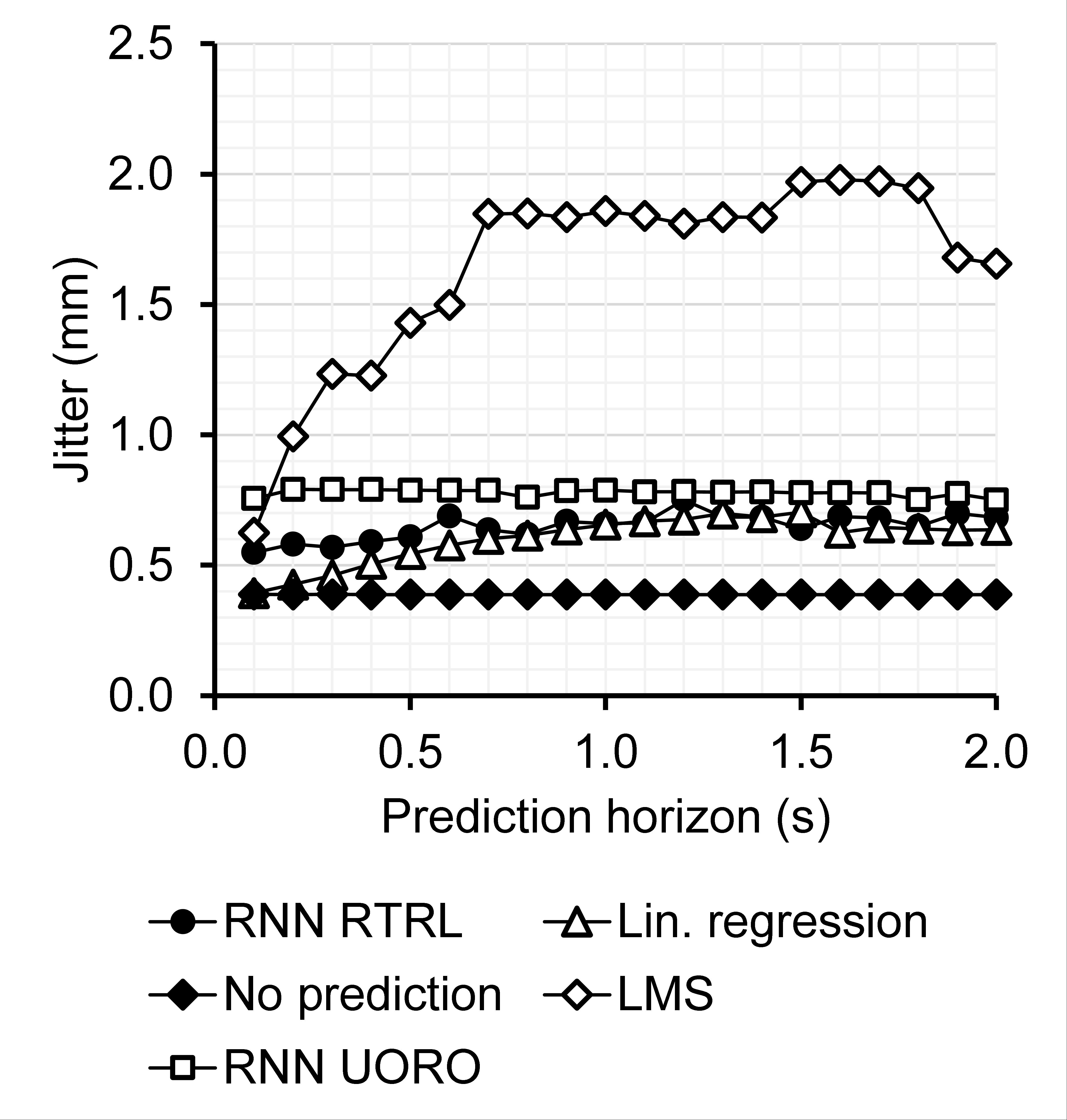}%
    \caption{Forecasting performance of each algorithm as a function of the prediction horizon. Each point corresponds to the average of one performance measure of the test set across the sequences associated with regular breathing. }
    \label{fig:pred perf regular}
\end{figure*}

\clearpage

\begin{figure*}[htb!]
    \centering
    \includegraphics[width=.35\textwidth]{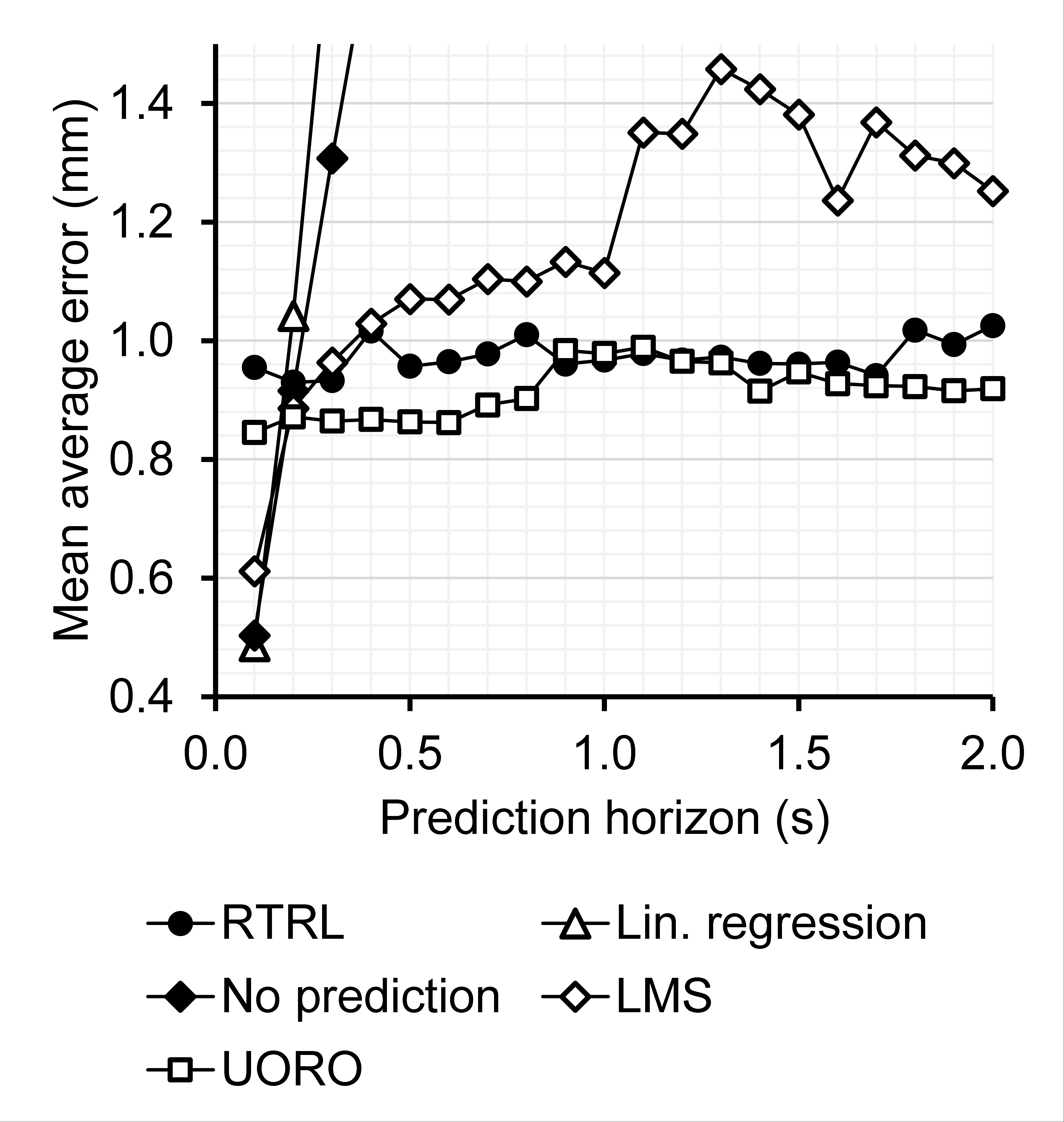}%
    \qquad
    \includegraphics[width=.35\textwidth]{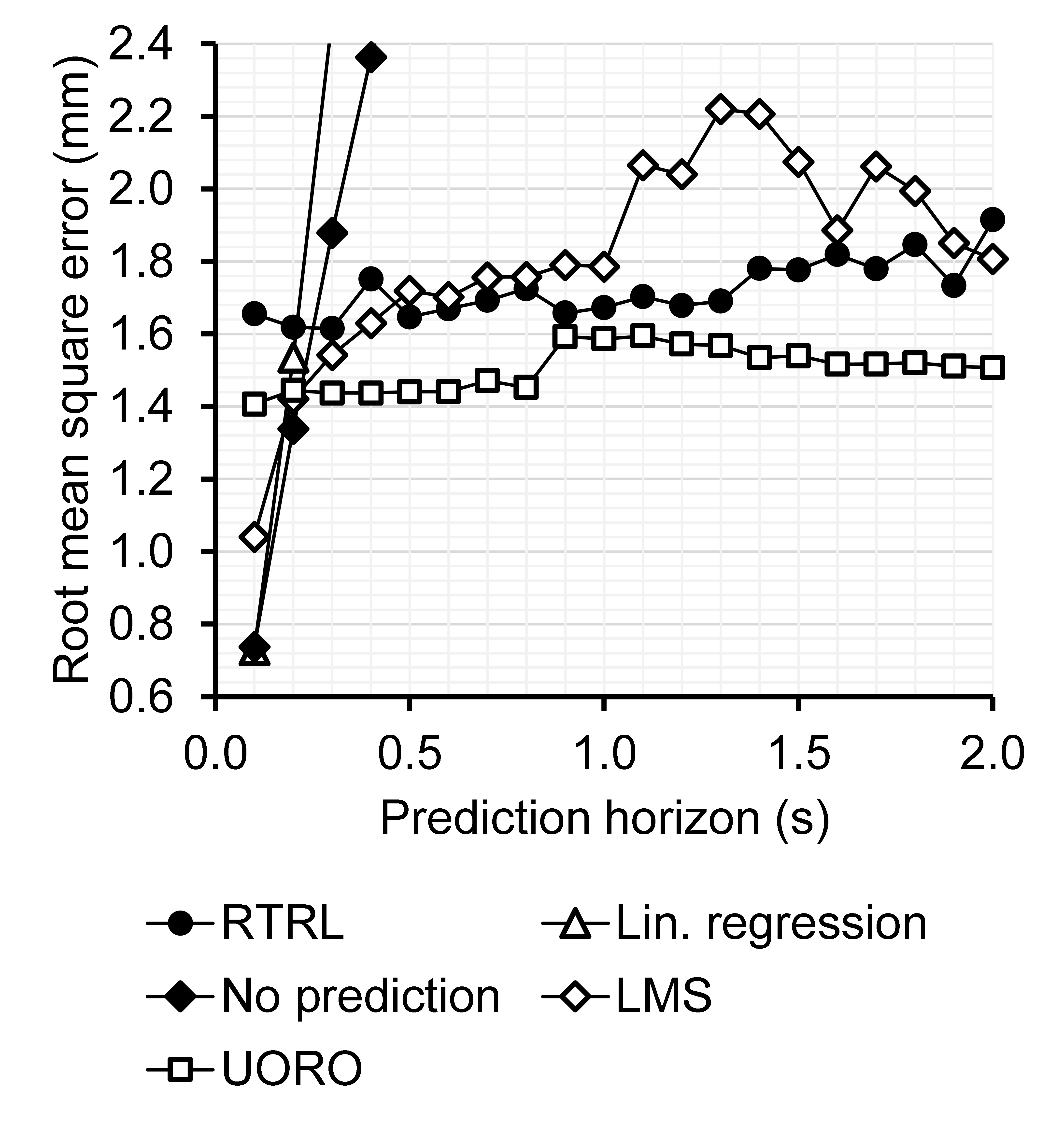}%
    \qquad
    \includegraphics[width=.35\textwidth]{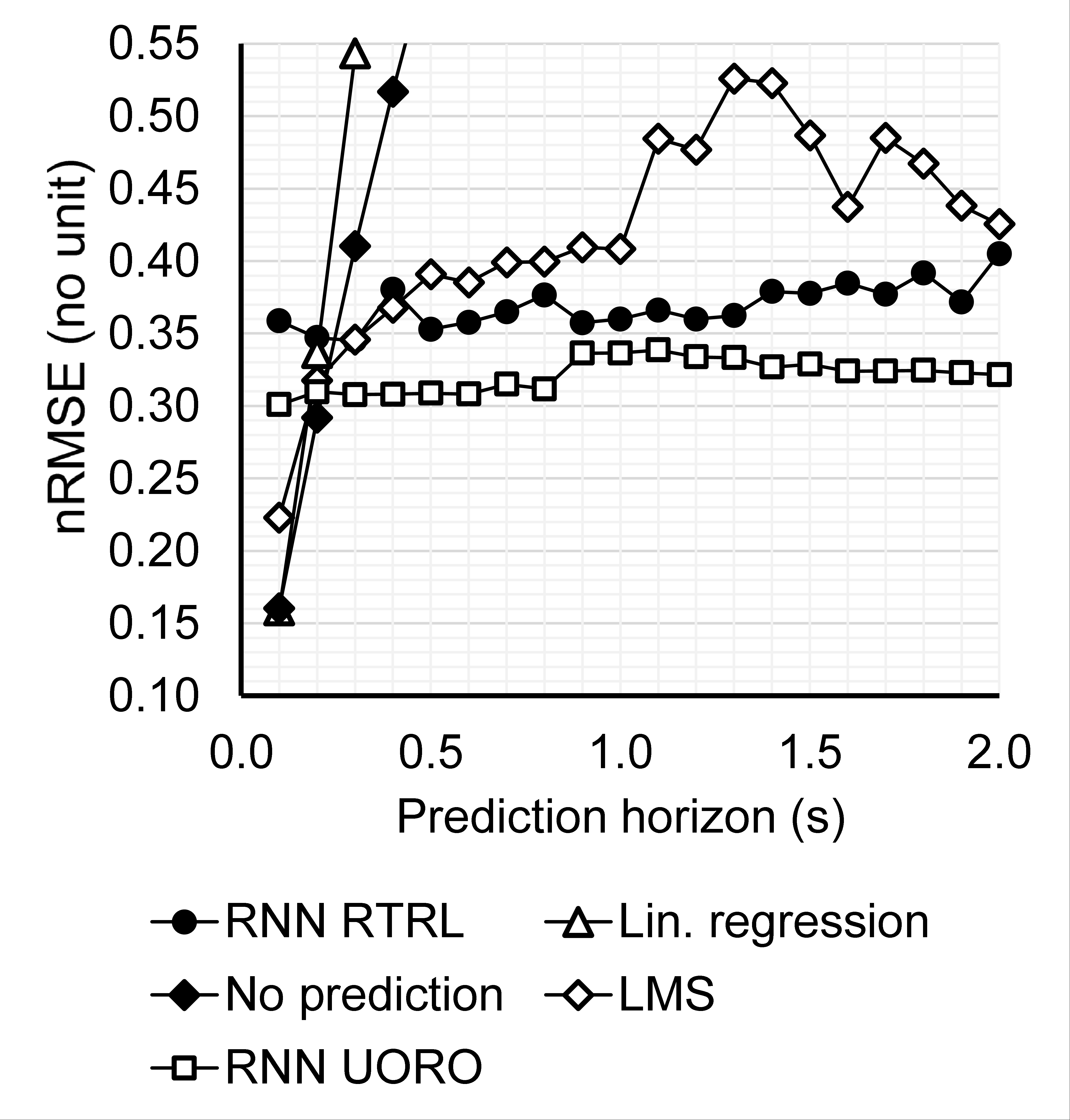}%
    \qquad
    \includegraphics[width=.35\textwidth]{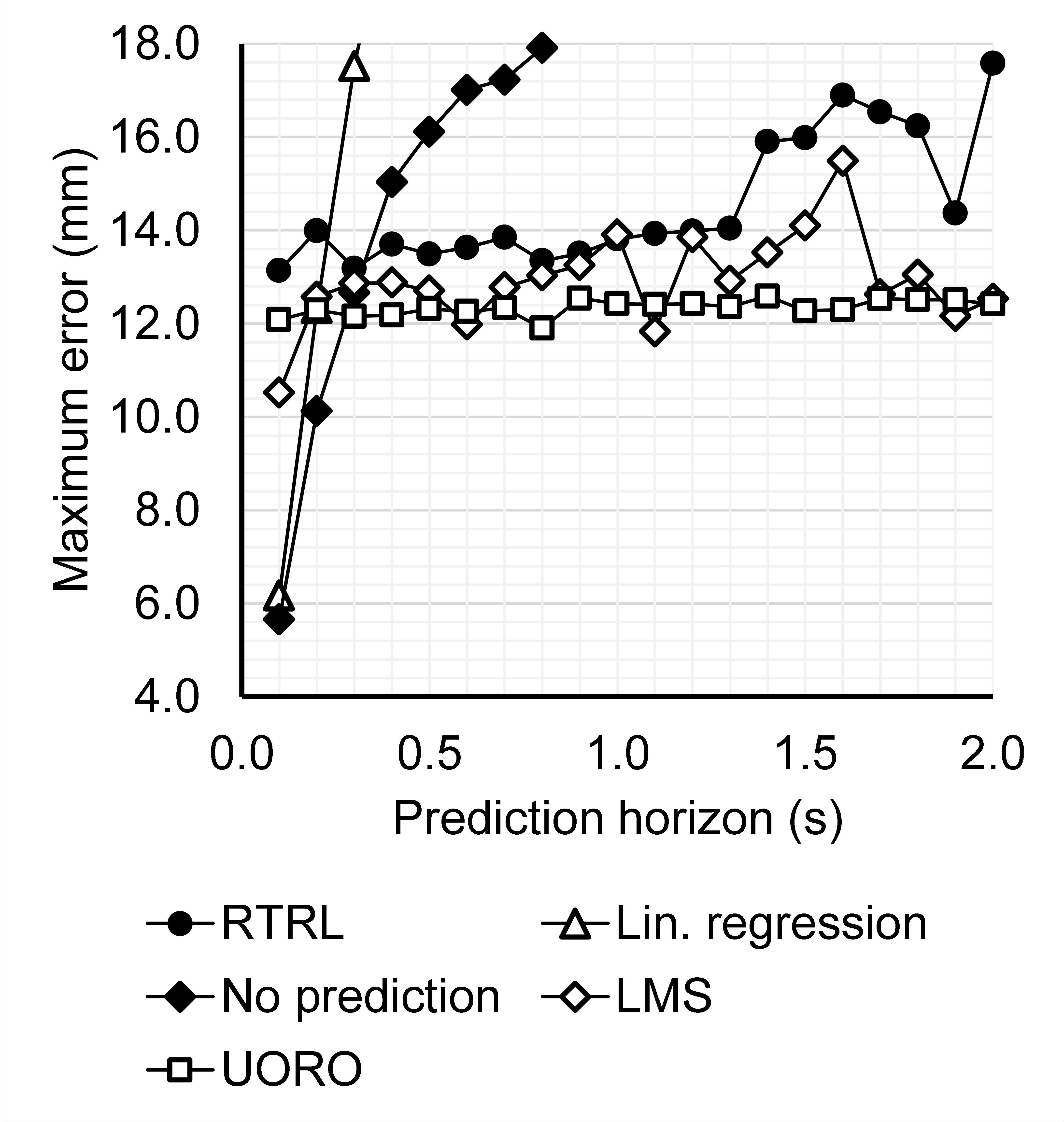}%
    \qquad
    \includegraphics[width=.35\textwidth]{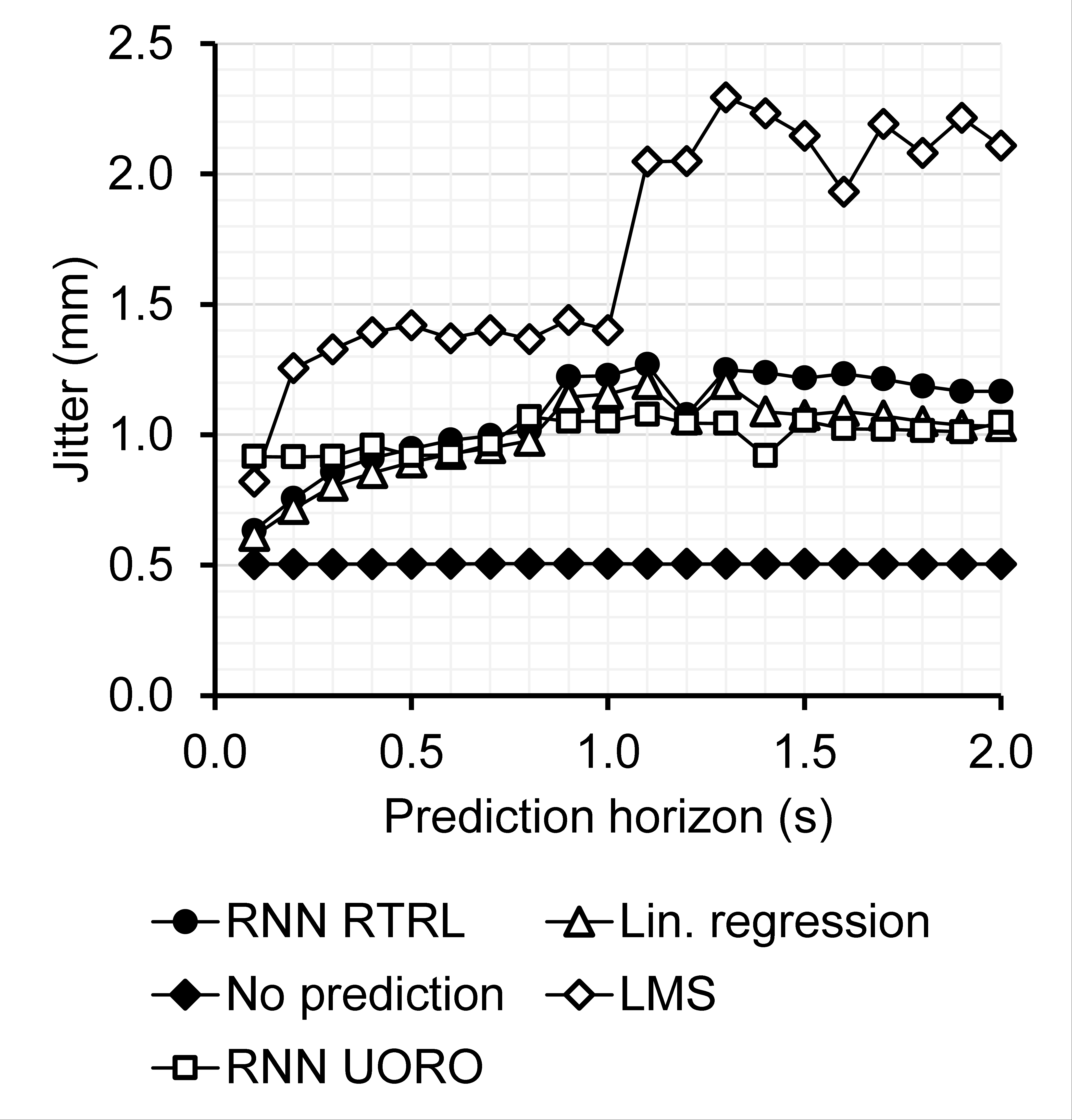}%
    \caption{Forecasting performance of each algorithm as a function of the prediction horizon. Each point corresponds to the average of one performance measure of the test set across the records associated with irregular breathing. \protect\footnotemark}
    \label{fig:pred perf irregular}
\end{figure*}

\footnotetext{Sequence 201205111057-LACLARUAR-3-O-72 (cf \cite{krilavicius2016predicting}) has been removed from the sequences associated with abnormal respiratory motion for plotting the performance graphs, as that sequence does not contain abrupt or sudden motion that typically makes forecasting difficult. }

\end{document}